\documentclass[structabstract,oldversion]{aa}

\usepackage{graphicx}
\usepackage{txfonts}
\usepackage{natbib}
\usepackage{subfigure}
\usepackage{xspace}
\usepackage{epsfig}
\usepackage{epsf}
\usepackage{comment}

\usepackage{hhline}
\usepackage{multirow}
\usepackage{textcomp} %pour le symbole \textborn
\usepackage{amssymb}  %pour le symbole \dashrightarrow

\newcommand{\floor}[1]{\lfloor #1 \rfloor}
\begin{document}
%\title{Computational issues connected to chemodynamical modelling used in galaxy formation and evolution}
%\title{Supernovae explosion in dSph simulations: dynamical and chemical improvement and limitations}
\title{Computational issues in chemo-dynamical modelling of the  formation and evolution of galaxies}

\author{Yves Revaz\inst{1} \and Alexis Arnaudon\inst{1,2} \and Matthew Nichols\inst{1} \and Vivien Bonvin\inst{1}  \and Pascale Jablonka\inst{1,3}}

\institute{Laboratoire d'Astrophysique, \'Ecole Polytechnique F\'ed\'erale de Lausanne (EPFL), 1290 Sauverny, Switzerland 
\and Department of Mathematics, Imperial College, London SW7 2AZ, UK. 
\and GEPI, Observatoire de Paris, CNRS UMR 8111, Universit\'e Paris Diderot, F-92125, Meudon, Cedex, France}

\date{Received -- -- 20--/ Accepted -- -- 20--}

\abstract
    {
      Chemo-dynamical N-body simulations are an essential tool for understanding the formation and evolution of galaxies.
      As the number of observationally determined stellar abundances continues to climb, these simulations are able to provide new constraints on the early star formaton history and chemical evolution inside both the Milky Way and Local Group dwarf galaxies.
      Here, we aim to reproduce the low $\alpha$-element scatter observed in metal-poor stars.
      We first demonstrate that as stellar particles inside simulations drop below a mass threshold, increases in the resolution produce an unacceptably large scatter as one particle is no longer a good approximation of an entire stellar population.
      This threshold occurs at around $10^3\,\rm{M_\odot}$, a mass limit easily reached in current (and future) simulations.
      By simulating the Sextans and Fornax dwarf spheroidal galaxies we show that this increase in scatter at high resolutions arises from stochastic supernovae explosions.
      In order to reduce this scatter down to the observed value, we show the necessity of introducing a metal mixing scheme into particle-based simulations.
      The impact of the method used to inject the metals into the surrounding gas is also discussed.
      We finally summarise the best approach for accurately reproducing the scatter in simulations of both Local Group dwarf galaxies and in the Milky Way.
}

\keywords{supernovae feedback -- galaxies evolution -- N-body simulation -- chemical evolution -- dSphs and spiral galaxies} 

   \titlerunning{Computational issues in chemo-dynamical modelling of galaxies}
   \authorrunning{Revaz, Arnaudon, Nichols, Bonvin, Jablonka}
   \maketitle

%\tableofcontents

%%%%%%%%%%%%%%%%%%%%%%%%%%%%%%%%%%%%%%%%%%%%%%%%%%%%%%%%%%%%%%%%%%%%%%%%%%%%%%%%%%%%%%%%%%%%%%%%%%%%%%%%%%%

\section{Introduction}
\label{introduction}

%%%%%%%%%%%%%%%%%%%%%%%%%%%%%%%%%%%%%%%%%%%%%%%%%%%%%%%%%%%%%%%%%%%%%%%%%%%%%%%%%%%%%%%%%%%%%%%%%%%%%%%%%%%

Over the last decade new observing facilities have allowed accurate measurements
of elemental abundances in a large number of individual stars not only in the
Milky Way but also within its satellites, specifically the Local Group dwarf
spheroidal galaxies (dSphs) and ultra-faint dwarfs (UFDs).

These measurements cover galaxies of very different mass and star formation
histories providing crucial constraints on models of galaxy evolution \citep[see][for a review]{tolstoy2009}.  Not only
should these models reproduce the observed dynamical properties of galaxies, but
they should also be able to account for their chemical properties.  Unfortunately,
these chemical constraints are often neglected despite the fact that they
can lead to erroneous conclusions.

While the mean metallicities at a given mass or luminosity is normally close to
observations, the abundance ratios and their corresponding scatter is often poorly reproduced.
This discrepancy suggests difficulties in accurately reproducing the number of supernovae or their associated feedback 
which are the dominant source of observable metals.

The yield of $\alpha$-elements in Type II supernovae (SNeII) strongly depends on
the progenitors mass.  \citet{tsujimoto1995} and \citet{woosley1995} found that
the ratio $[\rm{Mg}/\rm{Fe}]$ in SNII ejecta decreases by two orders of
magnitude between stars with masses above $50\,\rm{M_{\odot}}$ and stars lighter
than $10\,\rm{M_{\odot}}$.  As the lifetime of massive SNII ranges from
$\sim$$3$ to $30\,\rm{Myr}$, one naturally expects that the local ISM polluted by
massive stars is different from that polluted later by lighter stars even if the
stars originated in the same star-forming region.  It was pointed out long ago by
\citet{audouze95} that a small number of exploding supernovae with varied masses
sampling an initial mass function (IMF) introduce some scatter in abundance ratios.  Low scatter
in the observed abundances must then be a sign of continual
pollution by successive supernovae and/or the sign of efficient mixing, which must
be reproduced in simulations.

In the Milky Way, above $[\rm{Fe}/\rm{H}]>-3.5$, using a semi-analytical model,
\citet{karlsson2005a} found that the probability of finding a star enriched by less than ten supernovae is very low, explaining a possible mixing
of the different yields of stars with varied abundances.
However, despite this mixing, stars with $[\rm{Mg}/\rm{Fe}]<0$  are still predicted to exist \citep{karlsson2005b}.
The problem naturally worsens for dwarf galaxies, where the rate of exploding supernovae is much lower.
To efficiently mix the ISM, not only dynamical effects like ISM turbulence but also the motion of stars before they explode need to be considered.
Both are present with a certain degree of precision inside N-body simulations.
Recently, focusing on a small volume ($\sim 32\,\rm{pc}^3$) \citet{feng2014} showed that the star-to-star variation in abundances within an open cluster
may be considerably reduced owing to the mixing of inhomogeneous gas during the process of star formation. 
Similar conclusions have been recently obtained by \citet{petit2015}.
%{\color{red} mention also \citet{petit2014}(not accepted yet)

Different chemo-dynamical simulations have been performed since the 1990s to study
the metal enrichment in cosmological contexts
\citep{steinmetz1994,mosconi2001,lia2002,scannapieco2005,oppenheimer2008,wiersma2009,few2012},
in spiral galaxies
\citep{raiteri1996,berczik1999,friedli1994,friedli1995,lia2002,samland2003,stinson2006,kobayashi2011,tissera2012,few2012,vandevoort2014},
in ellipticals
\citep{kawata2003,kobayashi2004,kobayashi2005,martinez-serrano2008}, and in dwarf
irregular, spheroidal or elliptical galaxies
\citep{carraro2001,ricotti2005,marcolini2006,kawata2006,valcke2008,marcolini2008,revaz2009,okamoto2010,sawala2010,schroyen2011,revaz2012,schroyen2013}.
However, while a few authors present $[\alpha/{\rm Fe}]$ vs $[{\rm Fe}/{\rm H}]$
diagrams, very little discussion exists on the scatter of $\alpha$-elements.

Interestingly, it appears that in early works the scatter in abundance ratios was
not an issue, because of the artificial mixing induced by the poor resolution used
at that time \citep{raiteri1996,berczik1999}.  As pointed out by
\citet{mosconi2001}, increasing the resolution of Smoothed-Particle hydrodynamics (SPH) models increases the
scatter in the chemical properties.  This is indeed witnessed in recent
high-resolution models of dSphs \citep{sawala2010,revaz2012}.  In their zoom-in
simulation of a Milky Way-like galaxy, \citet{vandevoort2014} predicted a subsolar
$[{\rm Mg}/{\rm Fe}]$ population that is not observed.  Similar cases also exist
in elliptical models \citep{kawata2003,kawata2006,marinacci2014}.

This scatter in abundance ratios can be reduced by fixing the size of the region
in which metals are expelled \citep{kobayashi2011}, 
because on average, particles receive more ejecta with varied yields.
A similar mixing is achieved by increasing the number of SPH neighbours
as the volume that receives ejecta is increased, according to \citet{revaz2012}.
\citet{wiersma2009} introduced a natural SPH smoothing over the volume into which metals are ejected.

An alternative solution consists in
inserting a diffusion mechanism for the heavy elements through the ISM.
Following a suggestion from \citet{groom1997}, \citet{carraro1998} added a diffusive equation in 
their SPH implementation. However, the impact of this term on the distribution of abundances was not studied.
\citet{martinez-serrano2008} proposed an implementation of the classical diffusion
equation in their SPH code. A similar approach has been followed by
\citet{greif2009} but only applied to simulation of supernovae remnants.  This
technique was improved by \citet{shen2010} in the context of the intergalactic
medium, where the diffusion coefficient results from a turbulent mixing model.
Similar diffusivity has been used in the spiral galaxy simulations of \citet{brook2012}.
and has resulted in a considerable reduction of the oxygen abundance dispersion, thereby overcorrecting the problem.

It is clear that the impact of the
different possible schemes used to distribute stellar ejecta into the ISM needs to
be thoroughly examined.  This is particularly important for the SPH technique,
where metals issued from a supernovae are usually distributed among neighbouring
gas particles, using a weight proportional to the SPH kernel.  Consequently
particles at different radii receive differing amount of metals which may generate
an artificial scatter.  Other schemes could help to reduce this scatter, however, the
final impact on stellar abundances is not straightforward and detailed examinations are
needed.

A supplementary source of scatter comes from the modelling of the initial mass
function (IMF) as resolutions increase.  Numerical techniques usually assume that
a stellar particle represents a single stellar population (SSP).  However, with
the mass in the highest resolution simulations approaching that of small clusters,
serious concerns arise. These questions include: 
How to deal with stellar particles of masses smaller than what is required to ensure a
full sampling of the IMF? and What happens if the stellar particle mass goes below
$1000\,\rm{M_{\odot}}$, approaching the mass of the most massive stars?  It is
crucial to answer these questions and estimate the extent to which a poor IMF
sampling may bias the final stellar abundances along with the star formation history.

The aim of this paper is to examine how chemo-dynamical N-body simulations can
reproduce the observed scatter of [$\alpha$/Fe] at galactic scales in dSph and
Milky Way-like galaxies.  
We start by exposing observational facts concerning the
stellar abundance ratio scatter seen in observations.
We then describe the schemes used in our code
\texttt{GEAR}, including recent improvements concerning the IMF sampling,
spreading of elements, and mixing schemes.  We apply our method to very
simple cases, where only one or two supernovae explode and pollute an initially
homogeneous box.  In addition to the verification that our code correctly
reproduces the Sedov-Talor solution, these tests allow us to understand the
bias that metal spreading techniques induce and to test the efficiency
of mixing schemes.  In a following step, we simulate Fornax and Sextans like dSphs.
Based on the prediction of star formation histories and abundance scatters, the
limitations of different IMF sampling schemes and the necessity of mixing schemes
are discussed.  Finally, we show that our choice of parameters for dSphs also
reproduces the star formation rate and $\alpha$-element abundances in a Milky
Way-like galaxy.

\section{Observational facts}\label{facts}

\begin{figure*}  
  \resizebox{1.0\hsize}{!}{\includegraphics[angle=0]{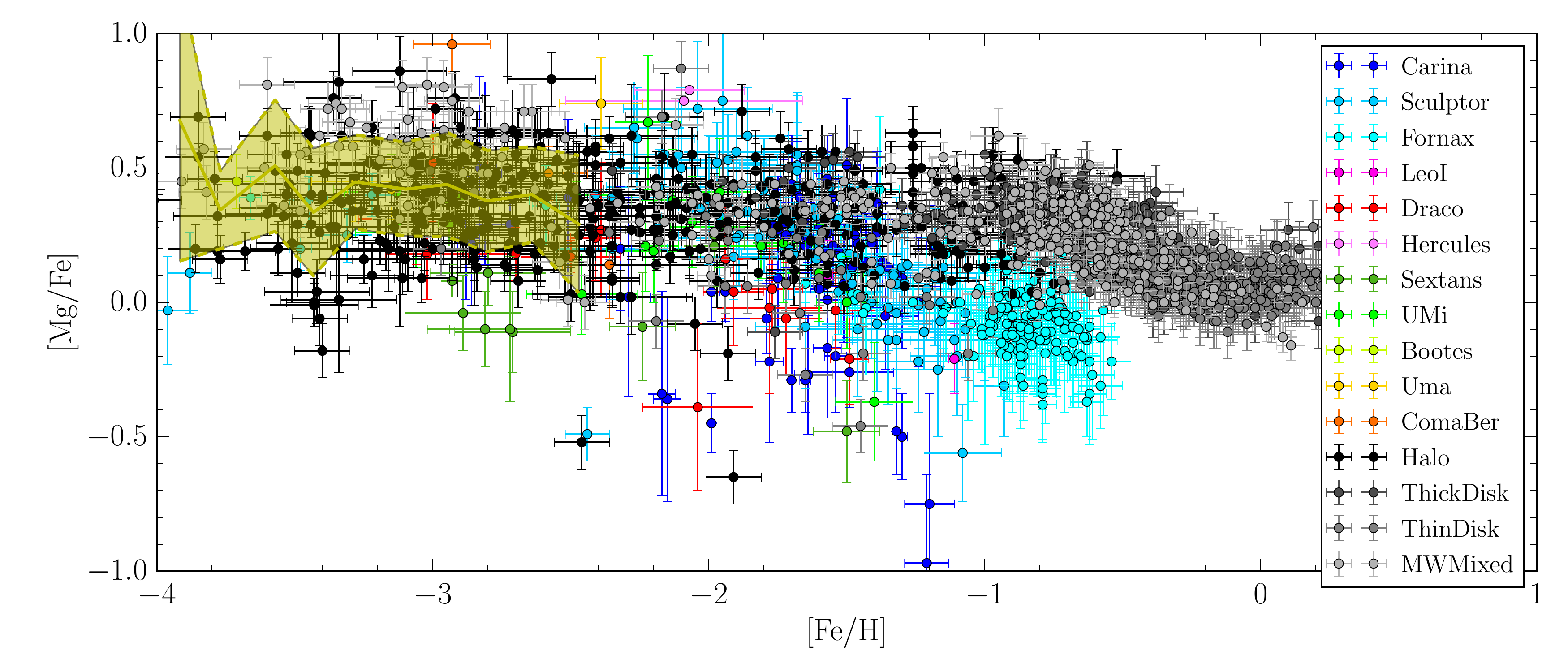}}
\caption{ [Mg/Fe] as a function of [Fe/H] obtained from high-resolution spectroscopy  of individual stars in the Milky Way or in Local Group dSphs. 
The yellow shaded region shows the $1-\sigma$ dispersion around the mean [Mg/Fe] for stars with a metallicity below $-2.5$. 
The data are obtained from high spectroscopy abundances determination:
Fornax \citep{shetrone2003,tafelmeyer2010,letarte2010},
Sculptor \citep{shetrone2003,tafelmeyer2010,starkenburg2013,jablonka2015,hill2015},
Sextans \citep{shetrone2001,aoki2009,tafelmeyer2010},
Carina \citep{shetrone2003,venn2012,lemasle2012},
ComaBer \citep{frebel2010b},                                                                 
Bootes \citep{norris2010},                                                                   
LeoI \citep{shetrone2003},
Hercules \citep{koch2008},
Uma \citep{frebel2010b},
UMi \citep{shetrone2001,cohen2010},
Draco \citep{shetrone2001,fulbright2004,cohen2009},
and MilkyWay \citep{gratton03,cayrel2004,venn2004,honda2004,gehren06,reddy06,andrievsky2010,cohen2013,aoki2014}.
All values have been scaled to the solar abundances of \citet{asplund2009}. }
  \label{fig:EMPS}
\end{figure*}

  As our aim is to properly reproduce the abundance ratio scatter seen in observations, we start by presenting observational facts.
  Figure~\ref{fig:EMPS} shows the compilation of 25 different samples in the Milky Way and its satellites.
  The relation between $[\rm{Mg}/\rm{Fe}]$ and $[\rm{Fe}/\rm{H}]$ arises from more
  than 1700 individual stars with abundances derived from high-resolution
  spectroscopy ($R~\ge 20,000$). All values have been scaled to the solar
  abundances of \citet{asplund2009}.  Three major conclusions can be drawn: (i)
  Stars with $[\rm{Fe}/\rm{H}]<-2.5$ form a plateau around
  $[\rm{Mg}/\rm{Fe}]\sim0.4$ with an error weighted standard deviation of about
  $0.2-0.3\,\rm{dex}$. This is illustrated by the yellow shaded
  region of Figure~\ref{fig:EMPS} corresponding to the 1-$\sigma$ deviation around
  the mean $[\rm{Mg}/\rm{Fe}]$ value, computed for all stars with a metallicty
  lower than $-2.5$.
  (ii) Very few metal-poor stars ($[\rm{Fe}/\rm{H}]<-2.5$) are found at subsolar [Mg/Fe]
  values.  (iii) With the exception of the peculiar cases of ET0381
  \citep{jablonka2015} and SDSSJ0018-0939 \citep{aoki2014}, none of these subsolar cases are found
  at [Mg/Fe] below $-0.5$.  
%\citep{honda2004,cayrel2004,cohen2004, spite2005,aoki2005,cohen2006,  spite2006,aoki2007, lai2008, yong2013,cohen2013,ishigaki2013}
%This is also the case of the dSphs
%\citep{shetrone2003,fulbright2004,koch2008,aoki2009,cohen2009,cohen2010,norris2010,tafelmeyer2010,frebel2010b,venn2012,lemasle2012}
%and UFDs \citep{ishigaki2014,vargas2013}.
This three major features guide our discussion and choices in the following.

%

%%%%%%%%%%%%%%%%%%%%%%%%%%%%%%%%%%%%%%%%%%%%%%%%%%%%%%%%%%%%%%%%%%%%%%%%%%%%%%%%%%%%%%%%%%%%%%%%%%%%%%%%%%%

\section{The code \texttt{GEAR} and its improvements}\label{code}

%%%%%%%%%%%%%%%%%%%%%%%%%%%%%%%%%%%%%%%%%%%%%%%%%%%%%%%%%%%%%%%%%%%%%%%%%%%%%%%%%%%%%%%%%%%%%%%%%%%%%%%%%%%

  Our tests and analyses are conducted with the code \texttt{GEAR}, which was
  developed by \citet{revaz2012}. It is a fully parallel chemo-dynamical Tree/SPH
  code based on \texttt{Gadget-2} \citep{springel2005}. The code \texttt{GEAR} has been
successfully used in the context of dwarf spheroidal galaxies, ensuring good
performance, numerical convergence, conservation of the total energy budget and
reproduction of the main observable properties of dSphs. In addition to the
hydrodynamics of the gas, \texttt{GEAR} includes the complex treatment of baryonic
physics, namely gas cooling, star formation, chemical evolution, and Type Ia (SNeIa) and
Type II  supernova feedback. Numerical codes are constantly improved and we list and
comment hereafter the recent changes compared to the initial version, excluding
the initial mass function, and mixing schemes which we discuss in more depth in
Sections \ref{imf} and \ref{feedback}.

%%%%%%%%%%%%%%%%%%%%%%%%%%%%%%%%%%%%%%%%%%%%%%%%%%%%%%%%%%%%%%%%
\subsection{Pressure-entropy formulation of SPH}
%%%%%%%%%%%%%%%%%%%%%%%%%%%%%%%%%%%%%%%%%%%%%%%%%%%%%%%%%%%%%%%%

We have upgraded the SPH formulation of \texttt{GEAR} using the pressure-entropy formulation described
in \citet{hopkins2013}. This new formulation alleviates difficulties traditional methods have had in the treatment of fluid mixing instabilities. 
This is particularly useful in the context of galaxy formation and evolution to correctly treat the 
dynamics of the ISM when, for example, the galaxy is subject to ram pressure stripping \citep{nichols2015}.

%%%%%%%%%%%%%%%%%%%%%%%%%%%%%%%%%%%%%%%%%%%%%%%%%%%%%%%%%%%%%%%%
\subsection{Individual and adaptive timesteps}\label{durier-th}
%%%%%%%%%%%%%%%%%%%%%%%%%%%%%%%%%%%%%%%%%%%%%%%%%%%%%%%%%%%%%%%%

The individual and adaptive timesteps scheme now precisely follows the algorithm proposed by \cite{durier2012}, which extends the timestep limiter of \cite{saitoh2009}. The algorithm can be summarised as:

\begin{itemize}
 \item \textit{Timestep limiter}: Each particle ensures that its timestep is shorter than or equal to a multiple (here taken as four) of any neighbouring particle.
 \item \textit{Timestep update}: An inactive particle (one whose timestep does not coincide with the current one) becomes instantaneously active if it is touched by the feedback energy of any other particle.
 \item \textit{Timestep criterion}: The default acceleration timestep criterion of \texttt{Gadget-2} involving the ratio between the gravitational 
softening and the acceleration is supplemented by a restriction including the SPH smoothing length (see Eq.~B3 in \citet{durier2012}). 
\end{itemize}

These new features are necessary to correctly reproduce the blast waves of a
single supernova explosion within high-resolution simulations (see
Section~\ref{sedov_taylor_solution}).

%%%%%%%%%%%%%%%%%%%%%%%%%%%%%%%%%%%%%%%%%%%%%%%%%%%%%%%%%%%%%%%%
\subsection{Artificial viscosity}\label{viscosity}
%%%%%%%%%%%%%%%%%%%%%%%%%%%%%%%%%%%%%%%%%%%%%%%%%%%%%%%%%%%%%%%%

The artificial viscosity is an important ingredient in SPH methods designed to
correctly capture shocks. Instead of using the default \texttt{Gadget-2} artificial viscosity 
based on \citet{monaghan1997},
we now use the original formulation of \citet{monaghan83} with the Balsara switch $f_{ij}$ \citep{balsara1995}
which alleviates the spurious transport of angular momentum in the presence of shear flows.
The viscosity between two particles $i$ and $j$, $\Pi_{ij}$  is therefore written as

\begin{equation}
	\Pi_{ij} =  -\frac{  (\alpha c_{ij}  + \beta\, \,\mu_{ij})\mu_{ij} }{\rho_{ij}} \,f_{ij},
	\label{artificial_viscosity2}
\end{equation}
where  $\alpha$ and $\beta$ are two dimensionless parameters determining the viscosity strength,
$c_{ij}$ their averaged sound speed and $\mu_{ij}$ is given by
\begin{equation}
	\mu_{ij} = \frac{\vec{v}_{ij}\cdot\vec{r}_{ij}  \,h_{ij}}{{|\vec{r}_{ij}|}^2 + \epsilon\,h_{ij}^2}, 
	\label{muij}
\end{equation}
where $\vec{r}_{ij}$ and $\vec{v}_{ij}$ are the differences in position and
velocity of the two particles, respectively, $h_{ij}$ is the mean of their SPH
smoothing lengths, and $\epsilon=0.01$ is a small parameter to avoid numerical
divergence when the particles are close.  This formulation is similar to
that implemented in \texttt{Gadget-2} if we impose the parameter $\beta = 3/2
\,\alpha$ and replace $\mu_{ij}$ by
$\omega_{ij}={\vec{v}_{ij}\cdot\vec{r}_{ij}}/{|\vec{r}_{ij}|}$ as discussed in
\citep{springel2005}. We show in Section~\ref{sedov_taylor_solution} that
this modification improves the reproduction of the blast waves in Sedov-Taylor
experiments.

To prevent spurious dissipation even far away from shocks, \citet{morris1997} and
later \citet{rosswog2000} introduced particle-based time dependent viscosity
coefficients $\alpha_i(t)$.  The value of $\alpha_i(t)$ depends on a decay timescale 
and a source term, making it maximal in the presence of a shock and minimal
away from it.
This is further improved with the approach proposed by \citet{cullen2010}, which avoids some problems related to the previous
methods, such as the damping of sound waves or the delay between the peak of the $\alpha_i(t)$ coefficients and the shock front.
However, this last improvement occurs at unacceptable increase of CPU time for our purposes.

In the present version of \texttt{GEAR}, we use individual time dependent
viscosity coefficients\footnote{For the usual symmetry necessary to ensure the
  conservation of integrals in SPH, the mean value $\alpha_{ij}=\frac12
  (\alpha_i+\alpha_j)$ is taken in the viscosity equations.} $\alpha_i(t)$ as
proposed by \citet{rosswog2000} and fix $\beta_i(t) = 2\,\alpha_i(t)$.

\section{Initial mass function sampling}\label{imf}

%%%%%%%%%%%%%%%%%%%%%%%%%%%%%%%%%%%%%%%%%%%%%%%%%%%%%%%%%%%%%%%%%%%%%%%%%%%%%%%%%%%%%%%%%%%%%%%%%%%%%%%%%%%

In most of the chemo-dynamical codes, each stellar particle is considered a single
stellar population (SSP) and its stellar mass distribution at birth follows the
shape of a given IMF.  As time passes, the energy and elements released by dying
stars must be spread over the ISM.  There are various ways of modelling this.

The simplest approach, which is still widely used in N-body simulations \citep[see
  for example][]{rosdahl2015}, is to calculate the total energy and metals 
that are released over time by a full stellar particle, given its mass and IMF.
The energy and metals are then ejected in the ISM at once.  At the opposite end,
pure chemical models \citep[e.g.,][]{matteucci2009} precisely compute the
equations of chemical evolution \citep{tinsley80}.  In this case ejecta are
gradually released over a longer period of time, the longest of which corresponds
to the lifetime of the least massive star that ends its life as a supernova (or
AGB when winds are considered).

In \texttt{GEAR}, each newly formed stellar particle is treated along the pure
chemical evolution model approach. Energy and metals released by the SNeII and
SNeIa are spread over the nearest neighbouring particles (see
Section~\ref{feedback} for more details).  This procedure requires accurate computation of, for each stellar particle $i$ and at every timestep, the number of exploding SNeIa ($N_{\rm{SNIa},i}(t)$) and SNeII
($N_{\rm{SNII},i}(t)$). These numbers are directly dependent on the choice of IMF
and its implementation.

Hereafter, we explore three different techniques to numerically sample the IMF.
We study their impacts on the chemical evolution of a galaxy, in particular on the
scatter in abundance ratios which are discussed in Section~\ref{dsph} and
\ref{spiral_galaxies}.

%In \texttt{GEAR}, each stellar particle formed is considered as a single stellar population (SSP)
%and treated as a semi-analytical model, following the  Tinsley equations \citep{tinsley80}. At every timestep, the amount of energy  and 
%elements released by SNII and SNIa are spread over the nearest neighbouring particles using the standard spline kernel function as a weight.

%This procedure request to accurately compute for each stellar particle and at every timestep following its creation, the number of exploding supernovae
%issued by the latter. This number is directly dependent on the IMF assumed and its modelisation.

%We have explored hereafter three different techniques of numerically sampling the IMF and studied its impact on the
%chemical evolution of a galaxy.  In particular its impact on the scatter of abundances in realistic galactic systems
%will be discussed in Section~\ref{simulations}.

\subsection{Continuous IMF sampling (CIMFS)}\label{continuous_sampling}

The continuous IMF sampling (CIMFS) method \citep[e.g.,][]{kobayashi2000,
revaz2012} involves discretising the Tinsley's equations to obtain an analytic
form of the number of Type II and Type Ia supernovae, $N_{\rm{SNIa},i}(t,\Delta
t)$ and $N_{\rm{SNII},i}(t,\Delta t)$ \citep[][]{poirier2004, revaz2009}, which
explode within a timestep $\Delta t$.  
In this formulation, both values are `real' numbers that depend continuously on 
$\Delta t$ as well as on the stellar particle mass considered, the smaller 
mass producing less supernovae.
This method works well for large
stellar particle masses. Though with the advent of high-resolution simulations,
it can lead to odd situations where $N_{\rm{SNII},i}$ or $N_{\rm{SNIa},i}$ are
much smaller than $1$, i.e. fractions of a supernova explode during a timestep.
This is equivalent to the dilution of a single supernova energy and elements
over a large time interval. Precisely, the time interval over which one supernova
fully explodes is the one for which $N_{\rm{SNIa},i}(t,\Delta t)=1$ or
$N_{\rm{SNII},i}(t,\Delta t)=1$. Diluting these explosions over time
has an obvious impact on the ISM both in terms of dynamics and chemical
enrichment.

For timesteps imposed by the Courant-Friedrichs-Levy condition, the situation
described above comes to a head for Type Ia SNe at medium resolution, where the
mass of a stellar particle becomes smaller than $10^5\,\rm{M_{\odot}}$.  
As an example, for a gas density of $0.1\,\rm{atom/cm^3}$ and a temperature of $10^6\,\rm{K}$, 
the timesteps are constrained to be of about $0.1\,\rm{Myr}$. Assuming a Kroupa IMF 
\citep{kroupa2001} and the \citet{kobayashi2000} SNIa model, the typical number of 
type Ia supernovae exploding during this timestep is $0.002$. This corresponds to 
a dilution of the explosion over $500\,\rm{Myr}$.
This problem is worse for smaller masses or timesteps. Because Type II SNe explode at a
higher rate, they are less prone to this problem. However, for a stellar mass
resolution of $10^3\,\rm{M_{\odot}}$, the dilution of the Type II supernova ejecta
can reach $50\,\rm{Myr}$, a period larger than the longest SNII lifetime.

%These results are of course directly dependent on the sampling of the IMF.  For
%high stellar particle mass the IMF can be considered as continuous, with many
%stars found in each mass bin\footnote{A mass bin $[m_1,m_2]$ is defined by using a
%  time step $\Delta_t$ fixed by the Courant-Friedrichs-Levy condition with an
%  extra requirement of being at least a fraction of the shortest stellar lifetime,
%  i.e., the lifetime of the most massive star considered.  Time interval is then
%  transformed into a mass bin using a mass-age relation.} ensuring a large number
%of exploding stars at every time step.  This however no longer holds true for
%smaller stellar particles.

%We describe two alternative solutions where the IMF is considered truly discrete.
%These methods do not induce any overhead in memory or CPU needs compared to the continuous approach.   

\subsection{Random discrete IMF sampling (RIMFS)}\label{random_sampling}

In principle, one could randomly sample the IMF at particle creation to determine
when and which stars would explode as supernovae.  However, such an approach would
require unpractical amounts of memory for any galaxy.  An alternative is to follow
a stochastic approach that reproduces the discretisation of the IMF without the
requirement of storing information for each star.

At every timestep, one calculates, for each stellar particle $i$, the number
$N_{\rm{SNx},i}(t,\Delta t)$ of potentially exploding supernovae (x standing for
Ia and II).  The integer part of $N_{\rm{SNx},i}(t,\Delta t)$ is assumed to
explode as supernova.  The remaining fractional value is compared to a random
variable $\chi$, taken from the uniform distribution over $[0,1]$.  If
$\chi<N_{\rm{SNx},i}^{\rm{frac}}(t,\Delta t)$, the final number of dying stars
is $\floor{N_{\rm{SNx},i}(t,\Delta t)}+1$ (where $\floor{\cdot}$ corresponds to the 
largest previous integer), otherwise it is
$\floor{N_{\rm{SNx},i}(t,\Delta t)}$.  In the following, we refer to this
method as the random IMF sampling (RIMFS) method.

The drawback of this technique is that when the stellar particle mass is small,
typically smaller than $10^4\,\rm{M_\odot}$, the number of supernovae generated
from the IMF is also small, resulting in Poissonian noise.  This
is particularly pronounced at the high-mass end of the IMF as massive stars are
comparatively rare.  This is illustrated in the top panel of
Figure~\ref{fig:NSNvsTime}, which shows the cumulative number of SNII and the number of exploding
SNII per timestep issued from a $2048\,\rm{M_{\odot}}$ stellar particle.  As stellar masses
are directly related to the chemistry via metal ejection, an artificial scatter in
masses induces a scatter in abundance ratios.

\subsection{Optimal discrete IMF sampling (OIMFS)}\label{optimal_sampling}

The second choice for the discretisation of the IMF is the so-called \text{optimal
  IMF sampling} (OIMFS), described by \cite{kroupa2013} and originally based on
the work of \citet{weidner2004}.  It aims to decrease the noise introduced 
by a random discrete sampling of the IMF, i.e. it avoids  gaps in the initial
stellar mass distribution.

Two important masses are introduced. First, $m_{\rm{max}}^\star$, the \textit{absolute upper} mass 
limit for stars above which no stars may form. It defines the absolute maximum mass of the IMF. 
The second mass is $m_{\rm{max}}$, the maximal star mass of the sampled IMF, i.e.
the maximal star mass among all stars constituting an SSP.
The OIMFS formalism imposes that there is only one star of mass $m_{\rm{max}}$, i.e.
\begin{equation}
	\int_{m_{\rm{max}}}^{ m_{\rm{max}}^\star  }\frac{\Phi(m)}{m}\,dm = 1,
\end{equation}
where $\Phi(m)/m\,dm$ is the number of stars  in the mass interval $[m,m+dm]$.
The other lower stellar masses constituting the discrete IMF sampling are determined iteratively
by ensuring the interval between $[m_{i+1},m_{i}]$ contains only one star with a mass
we set to be $m_{i+1}$.
The total mass of an SSP for a given IMF and a maximum stellar mass $m_{\rm{max}}$ is therefore
\begin{equation}
M_{\rm{SSP}}(m_{\rm{max}}) = \int_{{m_{L}}}^{m_{\rm{max}}} \Phi(m) \,dm + m_{\rm{max}} ,
\label{eq:mmax}
\end{equation}
where $m_{L}$ is the minimum stellar mass.

\citet{pflamm_altenburg2006} developed an algorithm to sample an IMF following this scheme, releasing a set 
of C routines\footnote{\texttt{http://www.astro.uni-bonn.de/uploads/media/optimal\_sampling.tar.gz}}.
This algorithm has been implemented in \texttt{GEAR} fixing the parameter
$m_{\rm{max}}^\star$ to $50$ M$_\odot$, which is the upper mass limit of our SNe nucleosynthesis tables.
The mass $M_{\rm{SSP}}$ is given by the stellar particle mass and is directly related to the resolution of the simulation.
%The stellar IMF mass should not be a numerical parameter but should rather be sampled from a empirical distribution. 
%Figure~\ref{fig:IMFro} illustrates the improvement of the OIMFS method over the RIMFS one, in its ability to reduce the noise present 
%at high stellar mass.
%
\begin{figure}
	\centering
	\leavevmode   
	\subfigure[RIMFS]{\resizebox{1.\hsize}{!}{\includegraphics[angle=0]{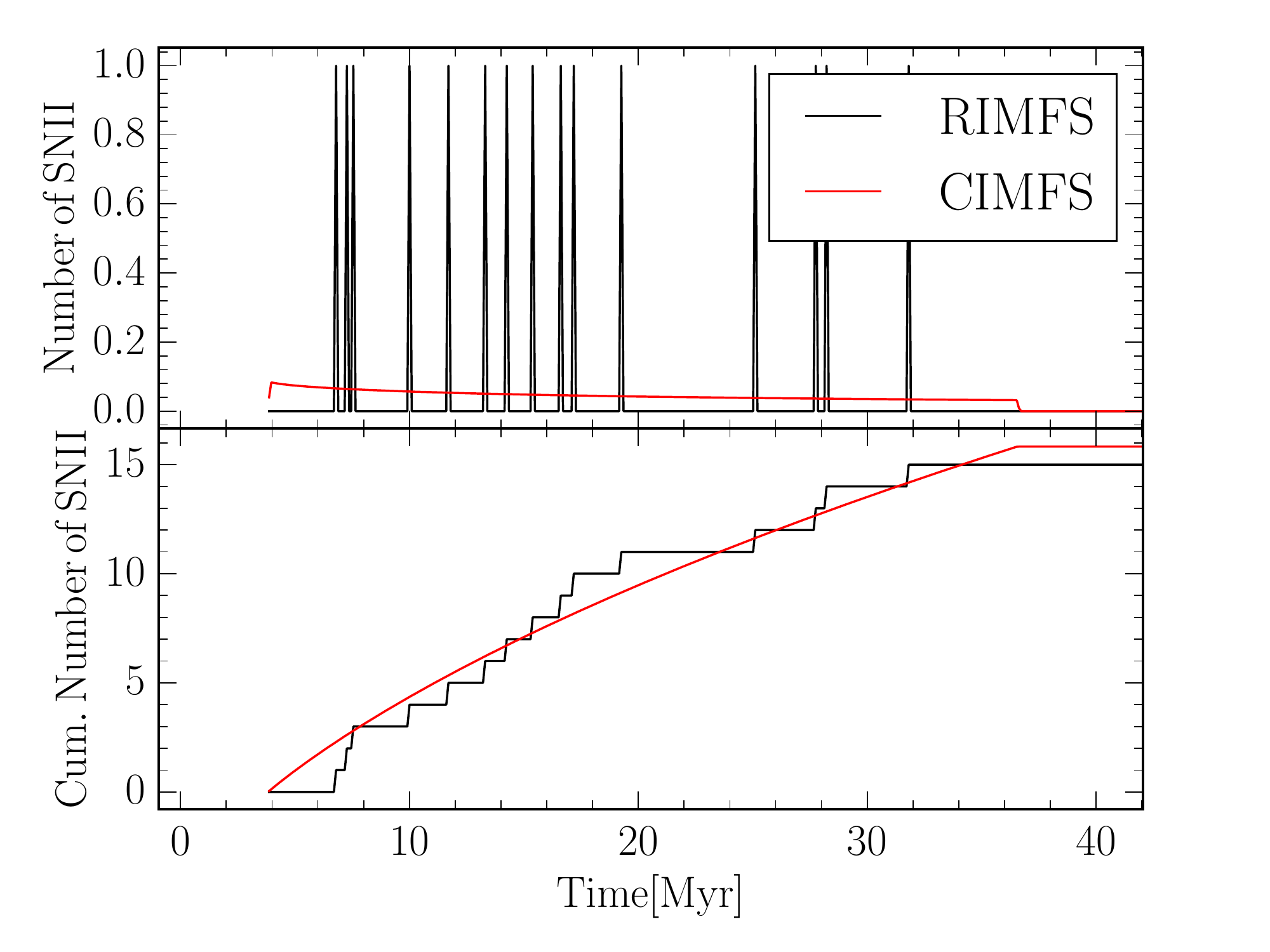}}}
	\subfigure[OIMFS]{\resizebox{1.\hsize}{!}{\includegraphics[angle=0]{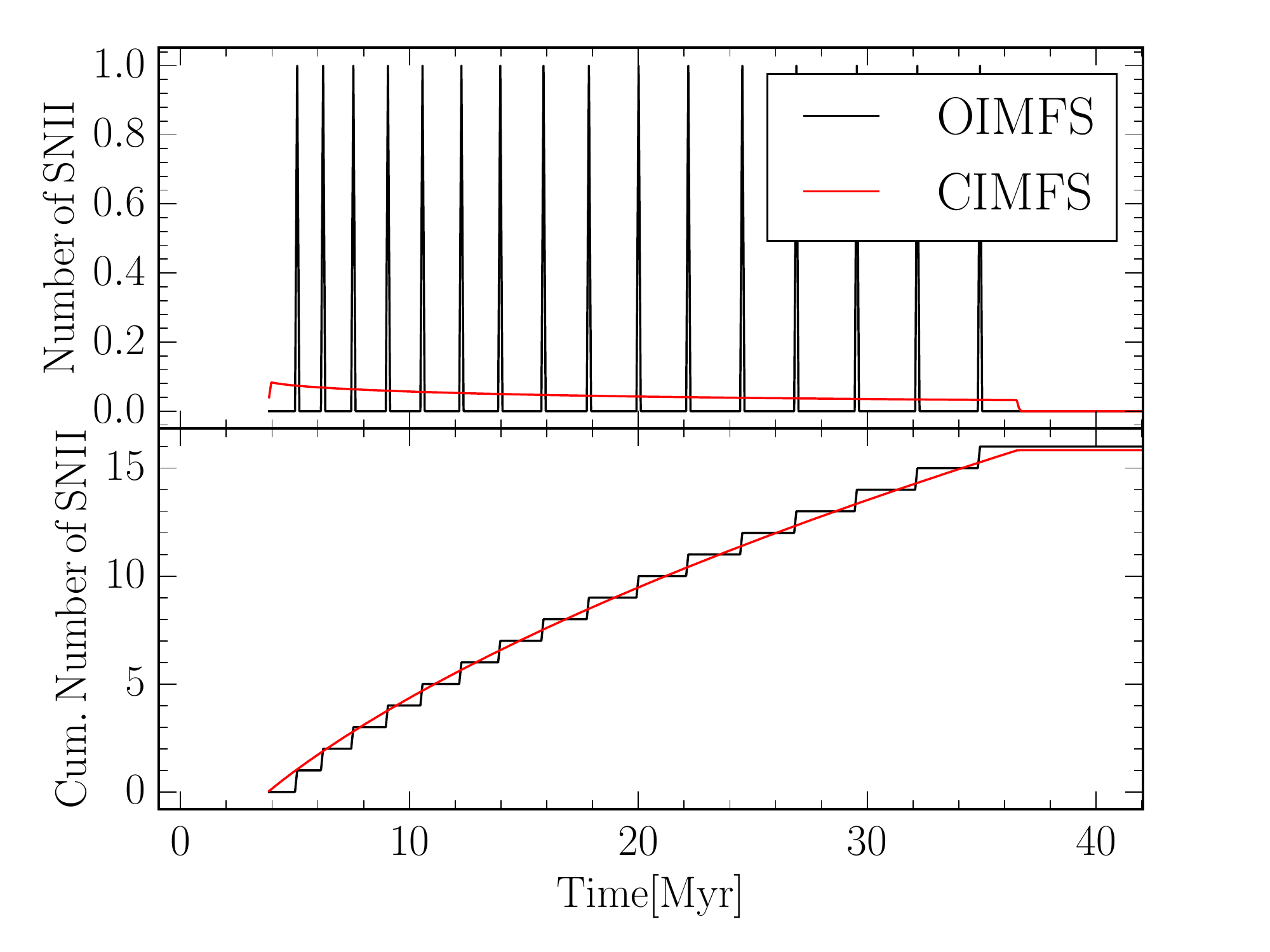}}}   
	\caption{Number of SNII explosions, upper panel (a); cumulative number of SNII, lower panel (b), both as a function of time for a $2048\,\rm{M_{\odot}}$ stellar particle. The red lines correspond to the analytical IMF,
		(CIMFS), while the black lines correspond either to the RIMFS (a) or the OIMFS (b).}
	\label{fig:NSNvsTime}
\end{figure}
Figure~\ref{fig:NSNvsTime} illustrates the reduction of noise in the IMF obtained with the OIMFS (b) compared to the RIMFS (a).
The number of exploding  SNeII per unit time is shown for a $2048\,\rm{M_{\odot}}$ stellar particle.

In the RIMFS approach, the individual stellar masses are uncorrelated, allowing multiple supernovae within short periods of
time or a long gap between two explosions. This results in a momentary change in the cumulative number of SNeII compared to the
CIMFS scheme.  With the OIMFS scheme (panel b), these deviations are erased, and
the fit to the cumulative number of SNeII in the continuous IMF sampling is clearly
improved.

\begin{figure}
	\centering
	\leavevmode   
	\includegraphics[width=9cm]{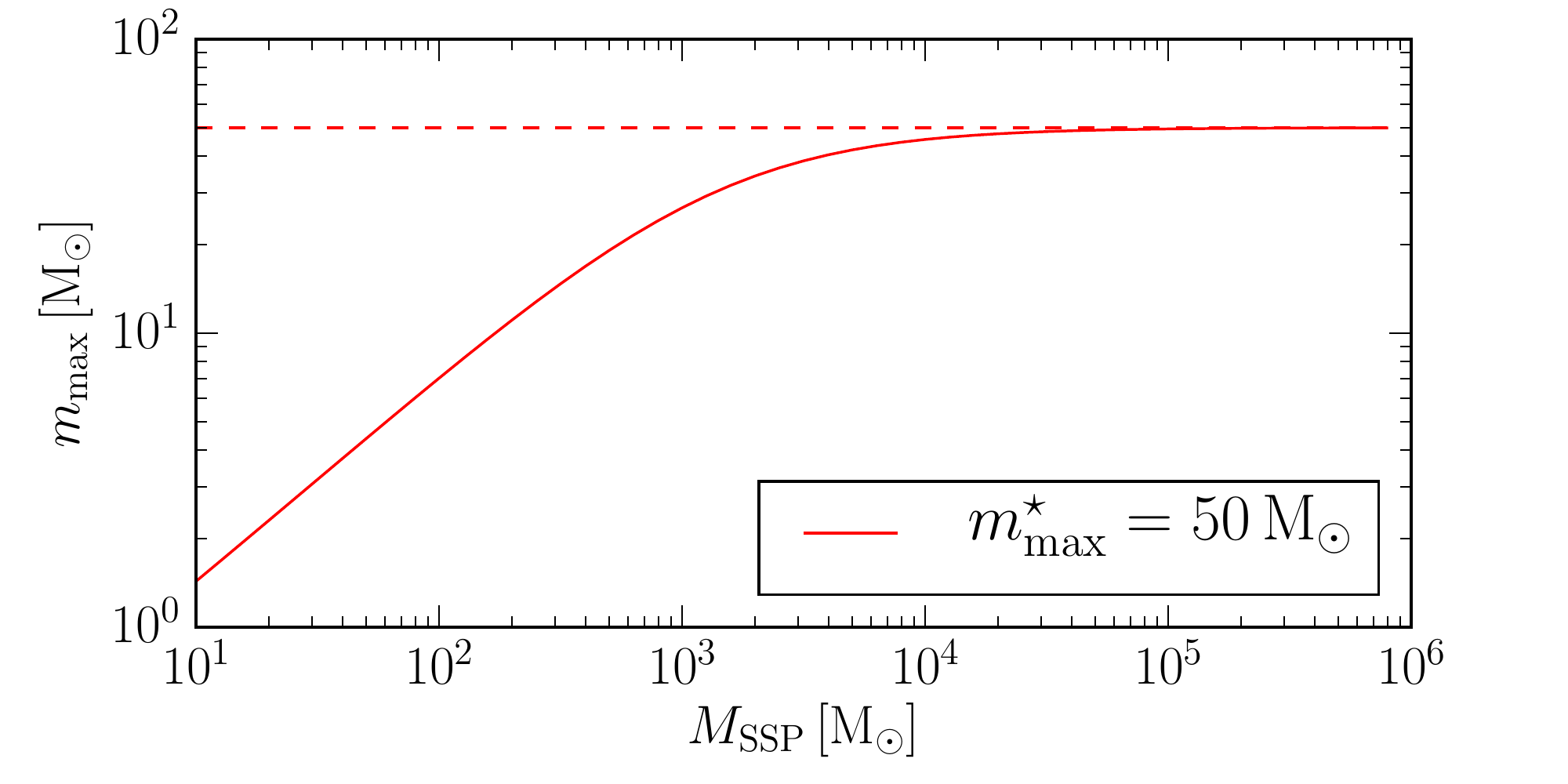}
	\caption{Maximal stellar mass $m_{\rm{max}}$ that a SSP of mass $M_{\rm{SSP}}$ may contain in the OIMFS method
		when the \textit{absolute upper} mass limit $m_{\rm{max}}^\star$ is set to $50\,\rm{M_\odot}$ (dashed line).}
	\label{fig:MmaxvsMcl}
\end{figure}

The OIMFS scheme is however limited in its application by the direct relation
given by Eq.~\ref{eq:mmax}
between the most massive star in the SSP $m_{\rm{max}}$ as well as the SSP mass
$M_{\rm{SSP}}$. This relation shown in Figure~\ref{fig:MmaxvsMcl}.  
Clearly, if the resolution is too high, for
example in the extreme case where $M_{\rm{SSP}}=100\,\rm{M_{\odot}}$, the maximal
stellar mass considered is only $7\,\rm{M_{\odot}}$, lower than the minimal
mass of a supernova.  A more reasonable choice of stellar particle mass is thus
around $M_{\rm{SSP}}=10^4\,\rm{M_{\odot}}$. Further discussions are
conducted in Section~\ref{effect_of_IMFS}

% \begin{figure}[h]
%   \centering
%   \leavevmode   
%    \includegraphics[width=9cm]{./images/IMFro.pdf}
%   \caption{Comparison of the IMF high mass end between the random and optimal sampling.
%   Here, the total mass contained in the IMF is $10^5\,\rm{M_\odot}$.}
%   \label{fig:IMFro}
% \end{figure}

\subsection{An implicit limit in the mass resolution?}\label{implicit_limit}

From the above description, one sees that each method presents an implicit limitation in mass
resolution: the CIMFS is no longer reliable below a mass resolution of about
$10^5\,\rm{M_{\odot}}$.  For higher resolutions, the ability to discretise the IMF
is lost and energy and metals are diluted over unrealistic time intervals.  The
OIMFS breaks down below $10^4\,\rm{M_{\odot}}$ owing to its intrinsic
formulation that sets up a maximal star mass lower than the maximal IMF mass.
The RIMFS induces noise when the stellar particle mass resolution is
below about $10^4\,\rm{M_{\odot}}$.

\begin{figure}
  \resizebox{\hsize}{!}{\includegraphics[angle=0]{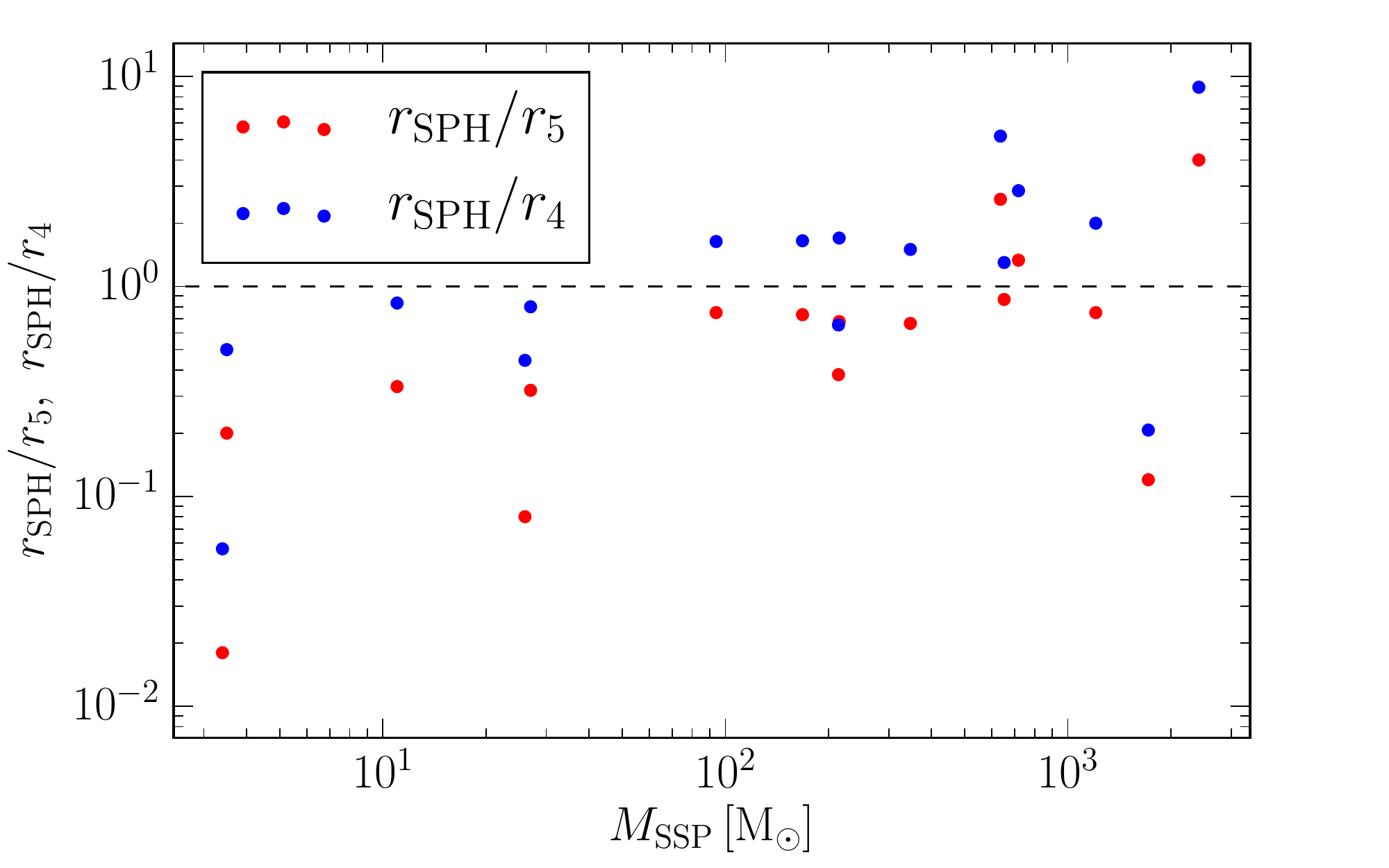}}
  \caption{Ratios $r_{\rm{SPH}}/r_5$ and $r_{\rm{SPH}}/r_4$ as a function of the stellar mass.  
  Below the dashed curve, the assumption that several particles may adequately sample an IMF begins to break down.
  }
  \label{fig:RsphR5R4}
\end{figure}

One can argue that a stellar particle does not have to sample a full IMF by
itself, but instead a complete IMF should result from different contributions
of several stellar particles.  In this sense, the RIMFS method correctly
samples the IMF as long as enough mass/particles are considered and are close to
each other.  However, as gravity quickly decorrelates stellar particles in numerical simulations, this assumption is only
correct as long as $10^4\,\rm{M_{\odot}}$ of stellar particles occupy a volume
corresponding to the zone where one particle injects energy and elements.
This corresponds to the volume defined by the SPH radius
($r_{\rm{SPH}}=h_i$) of the particle.  Computing the ratio between $r_{\rm{SPH}}$
and the size $r_4$ of a region containing $10^4\,\rm{M_{\odot}}$ is not trivial.
To accomplish this, we have extracted those values directly from simulations taken
from \citet{revaz2012}, where a large range of mass resolutions were explored.
$r_{\rm{SPH}}$ is taken as the minimum SPH radius among the stellar particles and
$r_4$ is computed in the central regions of the galaxies.  The ratio
$r_{\rm{SPH}}/r_4$ is displayed on Figure~\ref{fig:RsphR5R4} as well as the ratio
$r_{\rm{SPH}}/r_5$ where $r_5$ similarly defines a region containing
$10^5\,\rm{M_{\odot}}$.
Despite the scatter, it appears that bellow a stellar mass of $10^3\,\rm{M_{\odot}}$,
the SPH volume of a stellar particle contains fewer stars than that needed to correctly sample
a full IMF.
While this allow us to extend the resolution over one order of magnitude in mass when considering the RIMFS method,
we still face a mass resolution limit.
%Moreover, in any case, it is obvious that the RIMFS is limited when the mass 
%of the stellar particles reaches the mass of the most massive stars considered.

We further discuss the impact of this mass limit onto the
chemical evolution of galactic systems in Sect.~\ref{dsph}.

%%%%%%%%%%%%%%%%%%%%%%%%%%%%%%%%%%%%%%%%%%%%%%%%%%%%%%%%%%%%%%%%%%%%%%%%%%%%%%%%%%%%%%%%%%%%%%%%%%%%%%%%%%%

\section{Spreading of elements}\label{feedback}

%%%%%%%%%%%%%%%%%%%%%%%%%%%%%%%%%%%%%%%%%%%%%%%%%%%%%%%%%%%%%%%%%%%%%%%%%%%%%%%%%%%%%%%%%%%%%%%%%%%%%%%%%%%

For a given stellar particle, when a supernova explodes
both the energy and the synthesised elements must be ejected into the surrounding gas.
In practice, one wants to 
(i) choose $N_{\rm{ngb}}$ neighbouring gas particles $j$ around a stellar particle $i$, and
(ii) send to each of them a fraction $\alpha_{ij}$ of the total elements released by the particle $i$.
The fraction $\alpha_{ij}$ can be written in a general form as
\begin{equation}
     \alpha_{ij}  = \frac{X_{ij}}{\sum_{k=1}^{N_{\rm{ngb}}} X_{ik}}.
\end{equation}
Possible values for $X_{ij}$ will be discussed below.

\subsection{Setting the proper weighting}

The most natural way of setting $\alpha_{ij}$ is to use the SPH formulation
\begin{equation}
       X_{ij} = \frac{m_j W(r_{ij}, h_i)}{\rho_j},
       \label{aij1}
\end{equation}
where, in addition to the kernel function $W$, which depends on $h_i$ (the SPH smoothing length), and $r_{ij}$
(the distance between particles $i$ and $j$), 
the amount of each element distributed to the particle $j$ is weighted according to its volume $V_j=m_j/\rho_j$.
A simpler alternative is to weight the kernel according to the particles mass
%An alternative slightly simpler is to use the mass instead of the volume as a weight
%
\begin{equation}
       X_{ij} = m_j W(r_{ij}, h_i).
       \label{aij2}
\end{equation}
In both cases, the final metallicity inevitably traces the shape of the kernel.
As shown in Section~\ref{isolated}, using a simple isolated supernova explosion, this
radial dependency of the ejecta generates an artificial abundance scatter in the gas.
Its impact on the final stellar abundances is discussed in Section~\ref{dsph} in the context
of dSph galaxies.

By default, the kernel function is a cubic spline \citep{monaghan85}.
A simpler alternative consists in using a step function
\begin{equation}
  W(r_{ij}, h_i)= \Gamma_{ij} = \left\{
  \begin{array}{l l}
  1 & \quad \text{if $r_{ij}\le h_i$,}\\
  0 & \quad \text{otherwise,}
  \end{array}    
  \right.
\end{equation}
assuming that each particle within the SPH radius receive the feedback independent of its distance to the source. 
Under this Eq.~\ref{aij1} and \ref{aij2} then become
\begin{equation}
       X_{ij} = \frac{m_j}{\rho_j}  \quad \textrm{and}\quad    X_{ij} = m_j, \quad \textrm{respectively.}
\end{equation}
Based on different simulations, we observe no statistically distinguishable difference
in terms of star formation or final chemical properties, when using either the particle volume or the particle mass-weighting recipe.
This agrees with the results from \citet{tornatore2007} obtained in the context of galaxy clusters, where only a few tenths of dex
difference in the final [Fe/H] of the intra cluster medium exists between the volume or mass-weighting schemes.
Hereafter, we use the mass-weighting.
However, the choice of the kernel is important in this case, as illustrated both by simple supernovae explosion
experiments (Section~\ref{isolated}) and more complex dSphs simulations (Section~\ref{dsph}).

\subsection{Choosing neighbouring particles}
\label{choosing_neighbouring_particles}

A natural way of choosing the number of neighbours affected by a supernova is to follow the SPH weighting scheme. This is our default
choice where we have set $N_{\rm{ngb}}=50$. However, in the rigorous Lagrangian SPH formulation, such as that used
in \texttt{Gadget-2}, the number of particles is defined as \citep{springel2002,hopkins2013}
\begin{equation}
       \bar{N}_{\rm{ngb}_{i}} = \frac{4}{3}\pi h_i^3 \sum_{j=1}^{N_{\rm{ngb}_{i}}}  W(r_{ij}, h_i)\quad \textrm{or equivalently}\quad  \bar{M}_i = \frac{4}{3}\pi h_i^3 \rho_i,
       \label{nngb}
\end{equation}
ensuring that a fixed mass is contained within a given volume. 
Here, $N_{\mathrm{ngb},i}$ is the effective number of neighbouring particles (an integer number) and is not strictly equal to $\bar{N}_{\mathrm{ngb},i}$
(a real number).
It is important to emphasise that $h_i$ as well as $N_{\mathrm{ngb},i}$ are affected by the mass distribution around the particle
via their kernel dependency. This is particularly important if the mass is distributed in a ring-like structure, like it is, for example, after 
a Sedov-Taylor explosion (see Section~\ref{sedov_taylor_solution}). In this case, the effective number of neighbouring particles may be much larger than
expected.
For larger values of $N_{\mathrm{ngb},i}$, each neighbouring gas particle
receives a lower mass of elements and induces a bias.

An alternative to this scheme is to fix the size of the neighbouring region instead of setting a constant number of particles,
where the ejecta is released into a given volume instead of a given mass.
The radius of the volume can be fixed as in \citet{kobayashi2011} or set according to the propagation of the supernova blast wave which depends on 
the local properties of the gas. In the latter case, 
we have used the blast radius $R_{\rm{E}}$ 
\citep[see Eq.~(9) of][]{stinson2006} computed by \citet{chevalier1974,mckee1977},
corresponding to the size of a supernova bubble has reached when its internal pressure drops to that of the ambient interstellar pressure.

An advantage of this method is that it is independent of the resolution as
the amount of mass affected by ejecta is approximatively constant. The drawback is a drastic increase in the CPU time, as, according to our tests, the number of 
neighbouring particles is found to be on average larger.

We discuss the effect of these methods further in Sections~\ref{isolated} and \ref{dsph}.

%%%%%%%%%%%%%%%%%%%%%%%%%%%%%%%%%%%%%%%%%%%%%%%%%%%%%%%%%%%%%%%%%%%%%%%%%%%%%%%%%%%%%%%%%%%%%%%%%%%%%%%%%%%

\section{Mixing of metals}\label{diffusion}

%%%%%%%%%%%%%%%%%%%%%%%%%%%%%%%%%%%%%%%%%%%%%%%%%%%%%%%%%%%%%%%%%%%%%%%%%%%%%%%%%%%%%%%%%%%%%%%%%%%%%%%%%%%
In common SPH methods, star formation feedback transfers metals amongst the nearest neighbours where the metals remain for all time.
Such a scheme ignores the natural mixing that occurs as a reslut of small-scale turbulence or other unresolved processes. In most cases the metal content of particles is assumed to be constant until enriched by the next supernova.
It also results in an artificial scatter as particles at different distances from the source receive varying amount of metals. 
Furthermore, as the number of neighbours is typically resolution dependent, this scheme results in a strong resolution dependence with the scatter becoming worse at higher resolutions.

These problems have been extensively discussed in the literature \citep{carraro1998,okamoto2005,tornatore2007,martinez-serrano2008,greif2009,shen2010}, and two solutions have emerged:
(1) the introduction of a diffusion equation into the SPH scheme and (2) considering the metallicity to be a smoothly varying function.

\subsection{Metal diffusion}\label{metal_diffusion}

Including a classical diffusion equation in SPH has been proposed by different authors \citep[e.g.][]{carraro1998,martinez-serrano2008,greif2009,wiersma2009,shen2010}.
Hereafter, we use the approximate solution of \citet{greif2009}. More precisely, we implemented its Eq.~5. which predicts the time
evolution of any scalar quantity associated with a particle. This equation involves a diffusion coefficient $D_i$,  which  reads
\begin{equation}
        D_i = 2\,d\, \rho_i\, h_i \, \tilde v_i,
\end{equation}
where $h_i$ is the SPH smoothing length of the particle $i$, 
$\rho_i$ its density,
$\tilde v_i$ is an estimation of the RMS velocity at the position $i$ \citep{klessen2003}, and $d$ is a free parameter used to calibrate the diffusion. 
Diverging slightly from \citet{greif2009} we use the following estimation of $\tilde v_i$ described as
\begin{equation}
        \tilde v_i^2 = \frac{1}{\rho_i} \sum_{j=1}^{N_{ngb}} m_j\, W(r_{ij}, h_i)\, \left|\vec{v}_i-\vec{v}_j \right|^2,
\end{equation}
where $\vec{v}_i$ is the velocity of particle $i$ and $W(r,h)$ is the SPH kernel function.
Different values of the diffusion parameter $d$ are discussed in 
the context of isolated SN explosions (Section~\ref{isolated}) and dSph modellings (Section~\ref{dsph}).

% The diffusion equation for a scalar quantity $c(\vec{x},t)$ is thus given by:
% \begin{equation}
% 	\frac{\partial}{\partial t}c(\vec{x},t) = D \nabla^2 c(\vec{x},t),
% \end{equation}
% where $D$ is a diffusion parameter.
% In the SPH formulation this equation is discretised as: 
% \begin{equation}
% 	\frac{\rm{d}}{\rm{d} t}c_i = \sum_{j=1}^{N_{ngb}} K_{ij}  \left( c_i - c_j \right),
%   \label{diff-equation}
% \end{equation}
% where
% \begin{equation}
% 	K_{ij} = \frac{m_j}{\rho_i \rho_j}\frac{4\,D_i\,D_j}{D_i+D_j}\frac{\vec{r}_{ij} \cdot \vec{\nabla}_i   W(r_{ij}, h_i)  }{ r_{ij}^2 }.
% \end{equation}
% A diffusion parameter $D_i$ is associated to each particle $i$ and reads
% \begin{equation}
% 	D_i = 2\,d\, \rho_i\, h_i \, \tilde v_i,
% \end{equation}
% where $h_i$ is the SPH smooth length of the particle $i$, 
% $\tilde v_i$ an estimation of the RMS velocity at the position $i$ \citep{klessen2003} and $d$ is a free parameter to calibrate the diffusion. 
% Slightly diverging from \citet{greif2009} we use the following estimation of $\tilde v_i$:
% \begin{equation}
% 	\tilde v_i^2 = \frac{1}{\rho_i} \sum_{j=1}^{N_{ngb}} m_j\, W(r_{ij}, h_i)\, \left|\vec{v}_i-\vec{v}_j \right|^2
% \end{equation}
% which fits better with the SPH scheme.
% We exactly follow the recipe of \citet{greif2009} for integrating the diffusion differential equation \ref{diff-equation}.

\subsection{Smooth metallicity}\label{smooth_metallicity}

In particle based simulations, any elemental abundance can be computed as the ratio between the mass $m_i^{\rm{X}}$ of the element $X$ and the total hydrogen mass $m_i^{\rm{H}}$ for each particle $i$, namely,
\begin{align}
	%[\rm{X}/\rm{H}]_i = \log_{10}\left[\left(\frac{m_i^{{\rm{X}}}}{m_i^{\rm{H}}}\right)/\left(\frac{m_i^{\rm{X}}}{m_i^{\rm{H}}}\right)_\odot\right],
	[\rm{X}/\rm{H}]_i = \log_{10}\left[  \frac{  \left({m_i^{{\rm{X}}}}/{m_i^{\rm{H}}}\right)   }{   \left({m_i^{\rm{X}}}/{m_i^{\rm{H}}}\right)_\odot  } \right],
	\label{XHstandard}
\end{align}
where $(m_i^{\rm{X}} / m_i^{\rm{H}})_\odot$ is the solar abundance ratio of an element $X$.
%As previously stated, in the absence of metal diffusion, the abundances of a gas particle can only be modified by directely adding new elements to this particle via supernovae ejections.
%

With an SPH scheme however, extensive physical quantities linked to one particle are defined through a convolution running over the nearest neighbours. 
Following \citet{wiersma2009}
\footnote{A similar technique has been used by \citet{okamoto2005} and \citet{tornatore2007} to reduce the noise in the cooling function by preventing close particles from having different metallicites and thus different cooling rates.}, 
Eq.~\ref{XHstandard} may be reformulated with the density $\rho_i^{\rm{X}}$ and $\rho_i^{\rm{H}}$ to become
\begin{align}
	%[\rm{X}/\rm{H}]_i^s = \log_{10}\left[\left(\frac{\rho_i^{{\rm{X}}}}{\rho_i^{\rm{H}}}\right)/\left(\frac{m_i^{\rm{X}}}{m_i^{\rm{H}}}\right)_\odot\right],
	[\rm{X}/\rm{H}]_i^s = \log_{10}\left[  \frac{  \left({\rho_i^{{\rm{X}}}}/{\rho_i^{\rm{H}}}\right)   }{   \left({m_i^{\rm{X}}}/{m_i^{\rm{H}}}\right)_\odot  } \right],
	\label{XHsmoothed}
\end{align}
with $\rho_i^{\rm{X}}$  defined by
\begin{align}
	\rho_i^{\rm{X}} = \sum_{j=1}^{N_{ngb}} \rho_j^{\rm{X}}  \frac{m_j}{\rho_j}\, W(r_{ij}, h_i) = \sum_{j=1}^{N_{ngb}} m_j^{\rm{X}}\, W(r_{ij}, h_i).
        \label{rhoiX}
\end{align}

This ratio $[\rm{X}/\rm{H}]_i^s$ is referred to as the \textit{smooth metallicity}  because it results from the weighted contribution of neighbouring particles.
This scheme smooths the metallicity gradient between close particles, as would be expected by a turbulent ISM. 
%
%A similar method has already be used to compute the metallicity dependent radiative cooling acting on single particle \citet{okamoto2005,tornatore2007}. 
%This avoids close particles to have very different metallicities and thus too different cooling rates.
%It is also used to assign a metallicity to new stellar particles created during the star formation process.
%It is important to recall that 
Owing to its definition, this method naturally takes the resolution of the simulation into account.

\subsection{Fundamental differences between the two schemes}

Contrary to the diffusion scheme, the smooth metallicity scheme does not explicitly redistribute
metals among particles. Its effect is local, extending the
enriched region according to the size of the SPH smoothing
length, instantaneous, as it is performed at every timestep; and
static, as the enriched region does not change if particles are
not moving sensitively. In contrast, the diffusion redistributes
metals among particles and continuously extends the enriched
region with time as long as the velocity dispersion is non-zero,
which is always the case in practice. In this sense, it is a
dynamical process.

%The aim of the the smooth metallicity scheme is to improve
%the estimation of the metallicity of a gas particle by taking into
%account the SPH neighbours metallicity, without redistributing
%metals among particles. Its effect is then local, extending the
%enriched region according to the size of the SPH smoothing
%length, instantaneous, as it is performed at every time step, and
%static, as the enriched region will not change if particles are
%not moving sensitively. Contrastingly, the diffusion redistributes
%metals among particles and continuously extends the enriched
%region with time as long as the velocity dispersion is non zero,
%which is always the case in practice. In this sense, it is a
%dynamical process.

%%%%%%%%%%%%%%%%%%%%%%%%%%%%%%%%%%%%%%%%%%%%%%%%%%%%%%%%%%%%%%%%%%%%%%%%%%%%%%%%%%%%%%%%%%%%%%%%%%%%%%%%%%%

%\section{Simulations}\label{simulations}

%%%%%%%%%%%%%%%%%%%%%%%%%%%%%%%%%%%%%%%%%%%%%%%%%%%%%%%%%%%%%%%%%%%%%%%%%%%%%%%%%%%%%%%%%%%%%%%%%%%%%%%%%%%

%%%%%%%%%%%%%%%%%%%%%%%%%%%%%%%%%%%%%%%%%%%%%%%%%%%%%%%%%%%%%%%%%%%%%%%%%%%%%%%%%%%%%%%%%%%%%%%%%%%%%%%%%%%

\section{Isolated supernovae explosions}\label{isolated}

%%%%%%%%%%%%%%%%%%%%%%%%%%%%%%%%%%%%%%%%%%%%%%%%%%%%%%%%%%%%%%%%%%%%%%%%%%%%%%%%%%%%%%%%%%%%%%%%%%%%%%%%%%%

Simple tests provide the first steps to understanding the effect of feedback and elemental dispersion inside numerical simulations.
Towards this purpose we perform a series of simulations of isolated supernovae explosions inside a homogeonous gaseous medium.

We look in particular at the impact of the resolution, the adaptive timesteps, the artificial viscosity, and the method of the metal deposition, including the smooth metallicity scheme and/or the metal diffusion. 
We also take care to ensure the conservation of energy.

\subsection{Initial conditions}

In these scenarios a single supernova is detonated within a periodic box of side $1\,\rm{kpc}$ filled with a pristine gas of uniform density. 
The initial density is set to $0.1\,\rm{atom/cm^3}$ and the temperature to $T=10^4\,\rm{K}$, which corresponds roughly to the ISM of the first supernovae explosions in our dSph models \citep{revaz2012}. The metallicity is set to a primordial abundance of (by mass) 76\%  hydrogen and 24\%  helium. 
In order to avoid undesired noise, the gas particles in the box are first relaxed to form a glass-like structure. 

To examine the impact of the resolution, these simulations were run with five different number of particles between $N_{\rm{gas}}=16^3$ and $256^3$
(from $600\,\rm{M_{\odot}}$ down to $0.15\,\rm{M_{\odot}}$, in terms of mass resolution).
The smallest resolution is of order the dSph simulations of \citet{revaz2012} and the highest is for convergence tests alone. 

For each resolution, we explored the impact of the artificial viscosity improvement (Section~\ref{viscosity}), the adaptive timesteps improvement (Section~\ref{durier-th}), and the diffusion or smooth metallicity schemes (Section~\ref{diffusion}).  
The parameters used for each model are summarised in Table~\ref{tab:SN_res}.
\begin{table}[htpb]
	\centering
	\begin{tabular}{|c|c|c|c|c|}%c|}
		\hline
		Name 	& Art.visc.	 	& Adap. timesteps 	& Diff. 	&	Smooth 	\\%	& cooling \\ 
		        & improved		& improved			& coeff $d$	&	metallicity\\%	&	  \\
		\hline
		\hline
        a       & yes           & yes               & -             &       no     \\%     & yes     \\
		b 	    & no 			& yes				& - 	 	    &	no	\\%	& no	  \\
		c 	    & yes 			& no				& - 		    &	no	\\%	& no	  \\
        d       & yes           & yes               & -             &       yes      \\%     & no      \\		
		e 	    & yes 			& yes				& 0.001 	    &	no	\\%	& no	  \\
        f       & yes           & yes               & 0.003         &       no      \\%     & no      \\
        g       & yes           & yes               & 0.0001        &       no      \\%     & no      \\
        h       & yes           & yes               & 0.0003        &       no      \\%     & no      \\
		\hline
	\end{tabular}
	\caption{Details of parameters used for a single supernova explosion. Each model has been run with the five resolutions considered, $N=16^3, 32^3,64^3,128^3,256^3$, corresponding to a gas particle mass of $600,75,9,1.2,0.15\,\rm{M_{\odot}}$ respectively.}
	\label{tab:SN_res}
\end{table}

\subsection{Sedov-Taylor solution}
\label{sedov_taylor_solution}

\begin{figure*}
  \centering
  \leavevmode
  \includegraphics[width=18cm]{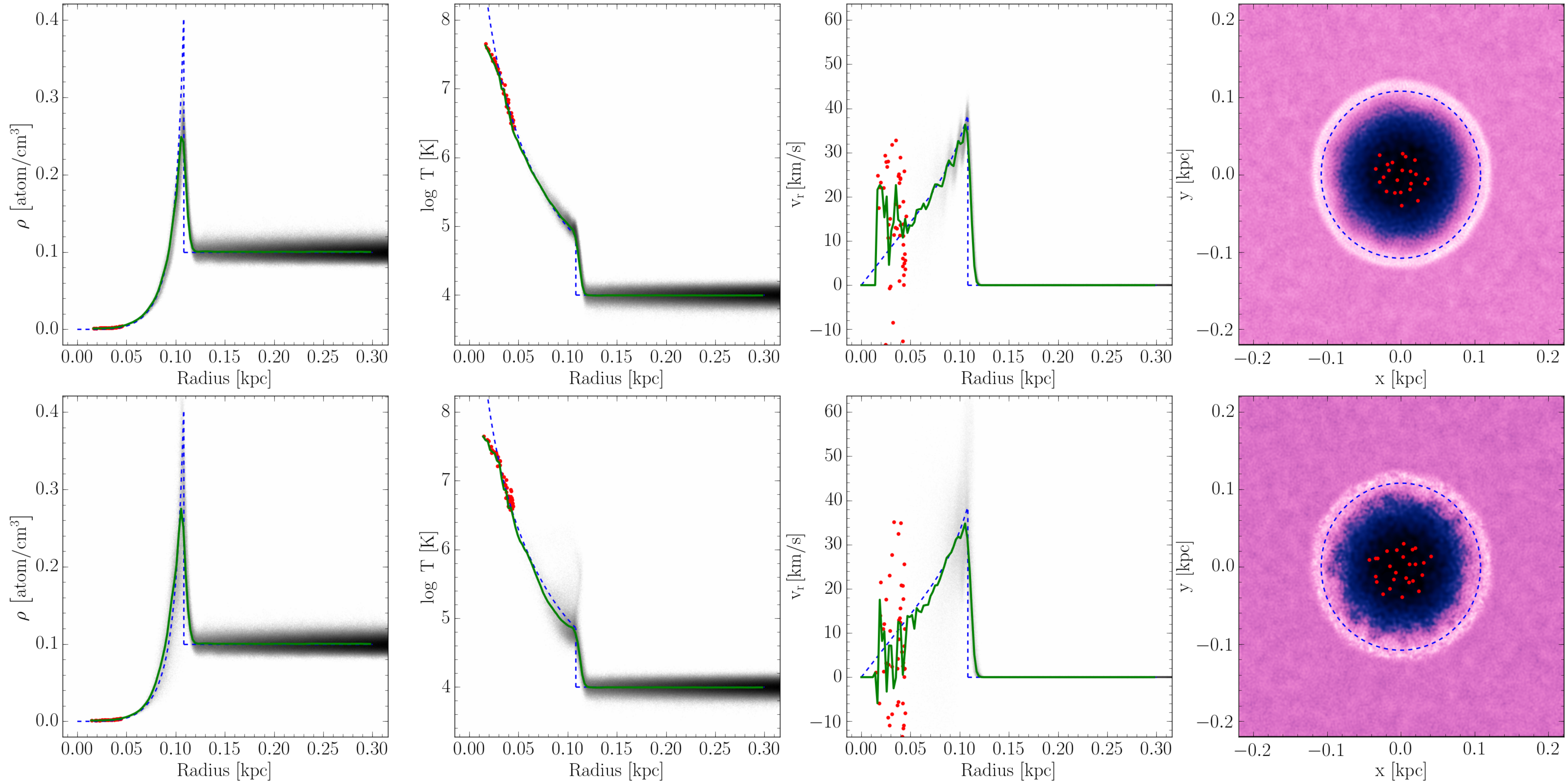}
  \caption{Density, temperature radial velocity, and surface density for two models with $N=256^3$ particles for different artificial viscosities.
  Top: the classical formulation \citep{monaghan83} (model ``a''). Bottom: the formulation proposed by \cite{monaghan1997} and
  used as a default in the \texttt{Gadget-2} code (model ``b''). In each panel, the blue dashed curve corresponds to the analytical Sedov-Taylor solution
  while the green dashed curve is the average particle value.
  The particles indicated in red were directly assigned feedback energy.   
  Because of the lower efficiency of the standard \texttt{Gadget-2} artificial viscosity (bottom), some gas particles can 
  penetrate into the ISM through the supernova shell. This is visible in anisotropies in the density behind the shock front.
  }
  \label{fig:sn256}
\end{figure*}

We confirm that in our highest resolution simulation (top panel of Figure~\ref{fig:sn256}) the \cite{monaghan1997} artificial viscosity (see Section \ref{viscosity}) formulation in combination with the adaptive timestep improvement 
proposed by \cite{durier2012} (see Section \ref{durier-th}), reproduces the Sedov-Taylor solution.
During the supernovas expansion, a non-negligible amount of particles remain at low densities inside the expanding bubble. 
The shell of this bubble is sufficiently thin and dense, agreeing with the analytic density profile. This is also
the case for other variables such as the temperature and radial velocity. The total energy is conserved at a $0.3\%$ level. 
%This demonstrates the ability of the code to reproduce the Sedov-Taylor solution at high resolutions. 

At the lowest resolution, where gas particles have a mass of $500\,\,M_\odot$ or higher, typically used for galaxy formation simulations, the blastwave is not well recovered. 
As the SPH kernel has approximately a fixed number of particles (Eq.~\ref{nngb}), 
the total mass inside the kernel rises at low resolution.
This naturally results in the dilution of the feedback energy per unit mass and consequently the increase of internal energy/temperature per particle is weakened \citep{dallavecchia2012}.
The impact of this can be up to two dex between the $N=16^3$ and the $N=256^3$ models.
However the transfer of the internal energy to kinetic energy only depends weakly on the resolution (at a 2\% level).
As a consequence, the net impact outside the kernel remains similar, independent of the resolution.

When radiative cooling is considered in these Sedov-Taylor experiments, the resolution dependence of temperature and density becomes important since the cooling function is directly related to these quantities.
At low resolution, the central particles do not reach the low density and high temperature regime where the cooling is significantly reduced \citep{dallavecchia2012}.
Following \citet[][and as in \citet{revaz2012}]{stinson2006}
we use an adiabatic time $t_{\rm{ad}}=5\,\rm{Myr}$, during which the cooling of any particle that received energy from the supernova is switched off.
This fix avoids an instantaneous cooling of the hot gas and improves the convergence.

\subsubsection{Adaptive timesteps}

Without the improved timestep scheme (model ``c'') proposed by \cite{durier2012},
gas particles in the ISM surrounding the explosion fail to be activated early on and arbitrarily large errors can then occur.

%The right column of Fig~\ref{fig:SNrho} to \ref{fig:SN} shows the dramatic impact of neglecting  
%the improved individual timestep scheme (model "c") proposed by \cite{durier2012}}.

% as shown on Fig.~\ref{fig:SNenergy}.
%%However, as previously, for low resolution typical of galaxy formation, we found that this correction is no longer essential.
%%This is true as long as the proper blast wave is not correctly reproduced by a single supernova explosion as it is in low resolution case.
%%In the case of realistic galaxies, several simultaneous explosions are common but this reasoning is still valid as the dynamical contrast will 
%%never be large enough to cause any troubles. 
%%%%In Section xxx, we will show that the effect of this correction in dSph simulations is negligible.

\subsubsection{Artificial viscosity}\label{visco-SN}

Figure~\ref{fig:sn256} compares the two artificial viscosity schemes.
The top panel corresponds to the  viscosity scheme adopted by \texttt{GEAR} (model ``a'') while
the bottom panel represents the standard \texttt{Gadget-2} implementation (model ``b'').
%Similar plots corresponding to lower resolution are displayed in Fig~\ref{fig:SNrho} to \ref{fig:SN}.
At high resolutions and using the standard \texttt{Gadget-2} artificial viscosity, some gas particles can penetrate 
into the ISM through the supernova shell (see the gas surface density) whereas this is not the case in the \citet{monaghan83} formulation. 

The difference between the two methods is explained by the presence of the additional factor $h_{ij}/r_{ij}$ in the definition of $\mu_{ij}$ compared with the standard \texttt{Gadget-2} viscosity (see Eq.~\ref{muij}).
This factor increases as particles come closer and boosts the effect of the bulk viscosity\footnote{At very small pair separation, the divergence is softened by a $\epsilon\,h_{ij}^2$ term at the denominator.}.
At low resolution this difference is negligible, however, as the shock is not strong enough for this to become important.
Finally, we stress that the total energy conservation remains very similar between the two schemes.

\subsection{Metal distribution and the effect of smooth metallicity or diffusion}

\subsubsection{Radial distribution of ejecta}

\begin{figure*}
\centering
\leavevmode
  \subfigure[$N=16^3$]{\includegraphics[width=9cm]{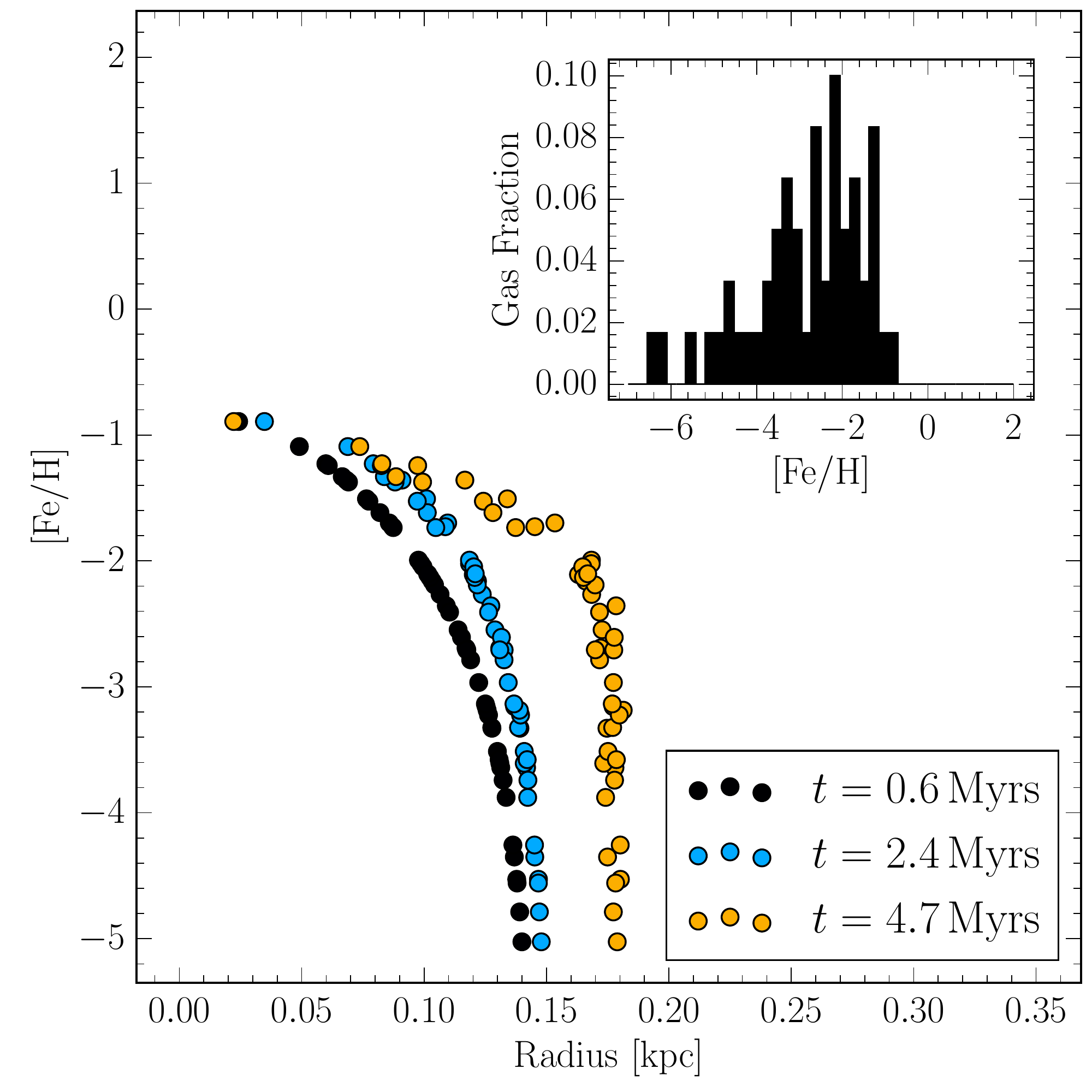}}
  \subfigure[$N=128^3$]{\includegraphics[width=9cm]{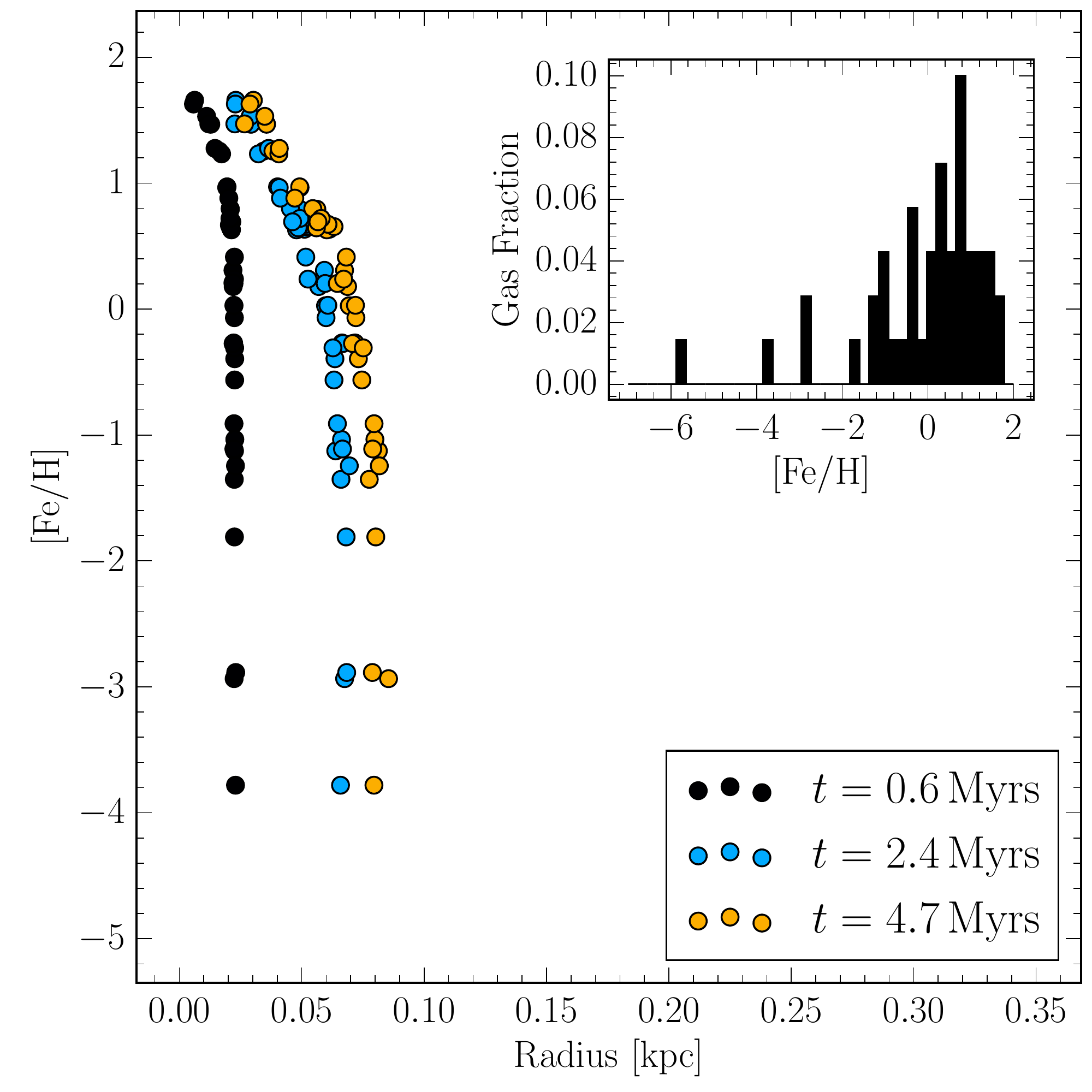}}
  \caption{Effects of the resolution on the evolution of the metallicity profile of particles that received feedback in our fiducial model ``a'',
  with a resolution of $N=16^3$ (a) and $N=128^3$ (b).
  Each curve corresponds to a different times, $0.6$, $2.4$, and $4.7\,\rm{Myr}$ after the supernova explosion.
  The upper right panel shows the iron distribution function which, in this particular experiment, is independent of time.}
  \label{fig:FeTime}
\end{figure*}

The [Fe/H] ratio of gas particles as a function of time is shown in Figure~\ref{fig:FeTime} for model ``a'' with $N=16^3$ and $N=128^3$ particles.
Without the smooth metallicity scheme or a diffusion term, only the particles that are directly touched by feedback contain metals.
Furthermore, the [Fe/H] profile directly traces the kernel at the moment of the supernova explosion
because the metal is distributed according to the SPH kernel,
and higher resolution models distribute metals over a smaller physical volume.
This reduced volume (and consequently reduced mass of gas that receives ejecta) means that higher resolution models end up with much higher metallicities.
%The offset induced in [Fe/H] when changing the stellar mass resolution from $m$ to $m'$ is $\log_{10}(m/m')$
%and is equal to $\log_{10}(512)$ in this particular case.

%It is worth noting that at this stage the radial profile of the ratio of different elements will 
%be constant, as each particle will receive the same fraction of elements according to its distance to the source, and thus 
%the ratio between elements will be conserved, tracing the one of the stellar yields.
%Thus it is only once a second supernova will release its metals that a scatter in elements ratio will be observed.
The width of the distribution however remains unchanged.
Owing to the fact that the energy transfer into a kinetic form is resolution independent, the velocity of 
particles is on average smaller in the low resolutions and the growth of the polluted region is subsequently three to four times slower 
in low-resolution cases.
This reduction in speed helps to reduce the size discrepancy of the final polluted region between different resolution models.

\subsubsection{Effects of smooth metallicity and metal diffusion schemes}

\begin{figure*}
\centering
\leavevmode
  \subfigure[$N=16^3$]{\includegraphics[width=9cm]{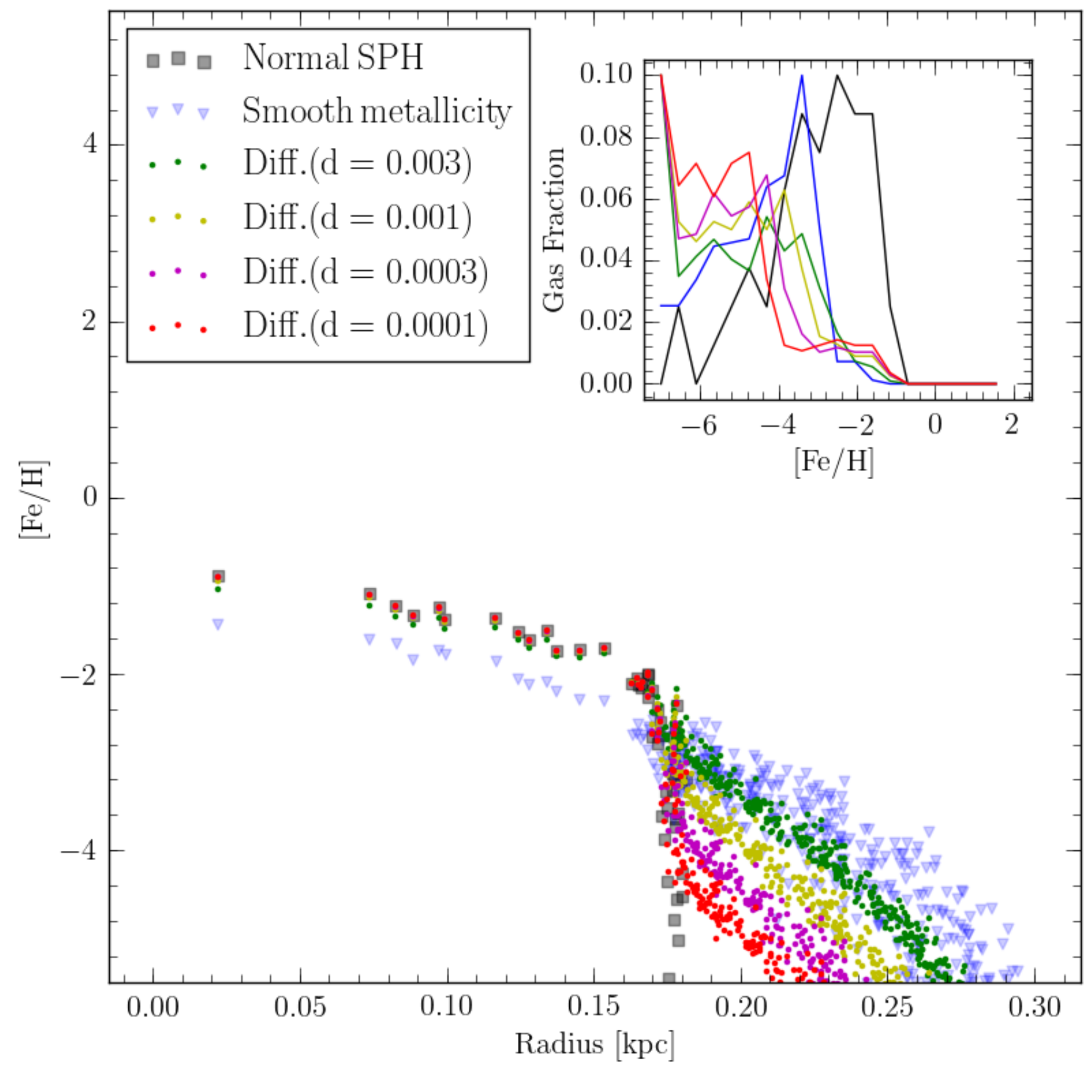}}
  %\subfigure[$N=32^3$]{\includegraphics[width=9cm]{./images/Fe32-100.pdf}}
  %\subfigure[$N=64^3$]{\includegraphics[width=9cm]{./images/Fe64-100.pdf}}
  \subfigure[$N=128^3$]{\includegraphics[width=9cm]{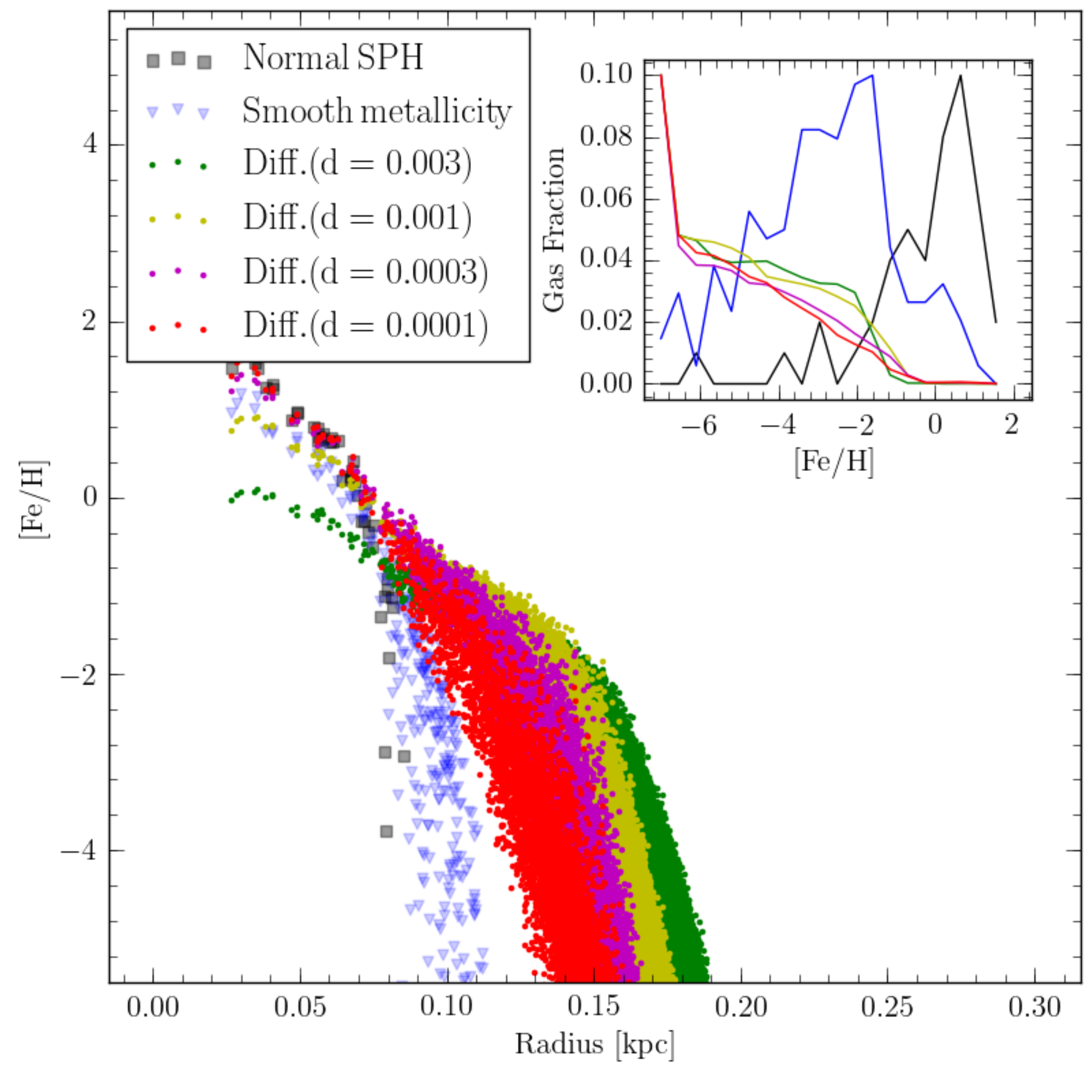}}
  \caption{Metallicity profile after $5\,\rm{Myr}$, for three different schemes: normal SPH (model ``a''), smooth metallicity (model ``d''), and metal diffusion (models ``e'' to ``h'').
  The upper panel shows the metallicity distribution function (independent of time in this particular experiment) for the $N=16^3$ and $N=128^3$ models.}
  \label{fig:FeRadius}
\end{figure*}

In contrast to the standard enrichment schemes, both the smooth metallicity scheme (model ``d'') and metal diffusion scheme (models ``e'' to ``h'') allows particles initially outside the feedback region to contain metals.
Figure~\ref{fig:FeRadius} shows the metallicity $5\,\rm{Myr}$ after a supernova explosion for the standard SPH scheme, the smooth metallicity scheme, and diffusions schemes (covering a diffusion coefficient $d$ of 0.003, 0.001, 0.0003 and, 0.0001).

Both methods reduce the strong metallicity gradient that exists in the standard SPH scheme.
Particles outside the region initially touched by feedback are now able to contain metals produced from the supernova, and the level dependent is on the model (and diffusion coefficient) used.
Consequently, the metallicity within the inner region is reduced in both schemes either because metals diffused outwards (diffusion scheme)
or the presence of low-metallicity neighbours (smooth metallicity) in the kernel of the central particles.
%Consequently the metallicity within the inner region is reduced as metals are conserved in both schemes either explicitly (diffusion scheme) or implicitly (smooth %metallicity).

The major differences between the two methods occurs because of their fundamental nature, i.e. whether they are static and instantaneous or a dynamical process.
At low resolutions ($N=16^3$, $h\cong 0.15\,\rm{kpc}$), the smooth metallicity scheme affects a region approximately twice the size of the region initially enriched; this region is much larger than that affected by diffusion regardless of the diffusion coefficient.
At high resolutions ($N=128^3$, $h\cong 0.035\,\rm{kpc}$), the smooth metallicity scheme only slightly broadens the initial polluted region.
Meanwhile, the diffusion scheme remains weakly affected by the resolution and at high resolutions  becomes the most efficient mechanism of metal mixing.

In Figure~\ref{fig:FeRadius}, the metallicity distribution function (MDF) for each model considered is shown in
the upper right corner.
%The two upper panels of Figure~\ref{fig:FeRadius} display the metallicity distribution function (MDF) for each model considered.
Independent of the resolution, the difference of the two methods is clear. 
With the diffusion scheme, a lot of particles receive a tiny fraction of elements, thus dominating the low-metallicity region of the MDF. 
Hence, the peak of the MDF is washed out.
On the contrary, the smooth metallicity scheme only slightly shifts the MDF towards lower metallicity while
keeping the peak of the distribution at an intermediate metallicity.

We finally point out that the effect of both methods is, by construction, resolution dependent, and the 
mixing is always strongest at the lowest resolution.

% \subsection{Discussion}
% 
% {\color{red}
% Ideas for a first small summary here or on the discussion section:
% \begin{enumerate}
% 	\item Able to reproduce SNr at high resolution
% 	\item We improve the viscosity and the timestep
% 	\item Seems necessary but at normal resolution the effect is negligible
% 	\item We can't increase the resolution due to the chemistry (see REF)
% 	\item We keep this improvement for future amelioration of the code to reach high resolutions
% \end{enumerate}
% }

\subsection{Two supernovae and abundance ratio scatter}
\label{TwoSNs}

In the previous experiments where only a single supernova exploded in a box of pristine gas, the final abundance ratios of all gaseous particles were 
strictly identical since they all received the same ratio of elements.
Studying the scatter in elemental abundances requires simulations with multiple supernovae of different yields.

Here, we study the impact of various processes on the scatter in a dual supernovae scenario.
First, an $\alpha$-rich ($\rm{[Mg/Fe]}\cong 1$) $40\,\rm{M_\odot}$ supernova is exploded and then $5\,\rm{Myr}$ later, a second, $\alpha$-depleted ($\rm{[Mg/Fe]}\cong -0.35$) $15\,\rm{M_\odot}$ supernova explodes, ensuring a large scatter in abundances ratio.

We performed 24 simulations,  each a combination of (i) a mixing scheme: the smooth metallicity, 
a strong diffusion ($d=0.003$), a weak diffusion ($d=0.0001$), or no mixing; and (ii) an ejection scheme: the normal SPH kernel, no kernel dependence ($w_{ij}=\Gamma_{ij}$), a fixed radius ($R_{\rm{ej}}=0.125\,\rm{kpc}$), or a blast radius ($R_{\rm{ej}}=R_{\rm{E}}$), 
as described in Section~\ref{feedback}.
For each of these simulations, we explored two resolutions: $N=16^3$ and $N=64^3$.
In the fixed radius scheme, the value $R_{\rm{ej}}=0.125\,\rm{kpc}$ corresponds to the blast radius $R_{\rm{E}}$ obtained for a medium
with a density of $0.1\,\rm{atom/cm^3}$ and a temperature of $10^4\,\rm{K}$.
As we found only a small difference between the fixed radius and blast radius, we only show results for the former to avoid clutter.

Figure~\ref{fig:twoSNs} shows the [Fe/H] of the gas as a function of the distance to the centre of the explosion, and [Mg/Fe] of the gas as a function of [Fe/H]
for a subset of the most relevant models with a resolution of $N=64^3$.
At the top right of each panel, we also show the final metallicity distribution.
For each simulation, the abundance of the gas at four different times is indicated:
$t_1=0.6\,\rm{Myr}$ (black), just after the first explosion;
$t_2=5.2\,\rm{Myr}$ (blue), just before the second explosion;
$t_3=5.3\,\rm{Myr}$ (green), just after the second explosion;
$t_4=9.4\,\rm{Myr}$ (red), at the end of the run.

%In the next section, we will present different simulations of galaxies, with a stellar masses around  $2048\,\rm{M_\odot}$.
%According to the Kroupa IMF \citep{kroupa2001}, about 15 supernova explode from a $2048\,\rm{M_\odot}$ SSP. 

%\subsubsection{T1.0: normal SPH kernel, no mixing}
\subsubsection{Normal SPH kernel, no mixing}

With the normal SPH kernel and without mixing, $49$ particles are enriched by the first explosion, which is in agreement with the choice of $50$ neighbouring particles.
As a result of the absence of mixing, the metallicity distribution traces the SPH kernel with the central regions more heavily enriched than the other regions.
As each metal species are distributed identically, the touched particles have a constant $[\alpha/{\rm Fe}]$ value, forming the blue horizontal line.
Between the two explosions, the particles enriched by the first supernovae are overpressurised and travel outwards forming the supernovae blast wave.
Interestingly, the second explosion impacts 84 particles.
As evoked in Section~\ref{choosing_neighbouring_particles}, such a large deviation compared to the expected $\pm 50$ particles is a direct consequence 
of the inhomogeneous medium generated by the blastwave.  
In addition to the $49$ particles enriched by the first supernova, $35$ particles comprised of pristine gas receive metals and these all share the same $\alpha$-depleted yields, that of the  $15\,\rm{M_\odot}$ supernova (forming the red horizontal line at the bottom of the [Mg/Fe] vs [Fe/H] plot).
In the central regions of the blast wave, the initially enriched particles are mixed with $\alpha$-depleted materials and end up with a $[{\rm Mg}/{\rm Fe}]$ of around 0.5. The resulting scatter is large due to the presence of extreme $\alpha$-rich and $\alpha$-depleted regions.
As the number of particles touched by the second supernova increases, the ejecta are also more diluted.

%\subsubsection{T1.1: no kernel dependence ($w_{ij}=\Gamma_{ij}$), no mixing}
\subsubsection{Step function kernel ($w_{ij}=\Gamma_{ij}$), no mixing}

In this situation, all particles touched by one explosion receive the same amount of ejecta. 
The radial [Fe/H] distribution is thus perfectly flat until the edge of the kernel.
The second explosion simply shifts the metallicity upwards and all particles are grouped in a small cluster on the [Mg/Fe] vs [Fe/H] plot.
The scatter in abundance ratio is thus extremely low.
As opposed to the previous case, the particles touched by the feedback receive equal amounts of energy and thus travel the same 
length and maintain a similar density.
Consequently, as the density is kept constant and homogeneous at the time of the second explosion, only the same 49 particles are touched by the ejecta of the second supernova.

%\subsubsection{T1.2: fixed radius ($r=0.125\,\rm{kpc}$), no mixing}
\subsubsection{Fixed radius ($r=0.125\,\rm{kpc}$), no mixing}

When the ejection radius is fixed, the bias related to the adaptive SPH radius is staved off. The second supernova only concerns
the particles that are still lying  in $0.125\,\rm{kpc}$ and, thus, because of the kernel, are the most enriched. 
This double enrichment results in the radial profile bump traced by the green line.
As a result of the $\alpha$-depleted nature of the second supernova, these particles see their [Mg/Fe] decreased.
Further away (where particles have low $[{\rm Fe}/{\rm H}]$), the $\alpha$ abundances remain unchanged.
The low [Mg/Fe] floor present in the normal SPH simulation without mixing is completely absent.

%\subsubsection{T2.0, T2.1, T2.2: smooth metallicity}
\subsubsection{Normal SPH kernel, smooth metallicity}

The smoothed metallicity dilutes the ejecta across a larger number of particles, shifting the mean to a lower metallicity.
Interestingly, this mixing scheme does not avoid the large scatter in [Mg/Fe] but only reduces it slightly by averaging some extreme values.

%When removing the radial dependence for the ejecta (T2.1), particles are no longer forming isolated cluster,
%but a tail at low metallicity appears.
%Using a constant radius (T2.2), at higher metallicity, the gas distribution is not really different from the T1.2 situation,
%expect that more low metallicity indirectly polluted by the mixing is present.

%\subsubsection{T3.0, T3.1, T3.2, strong diffusion ($d=0.003$)}
\subsubsection{Normal SPH kernel, diffusion}

The addition of a strong diffusion coefficient ($d=0.003$) allows the transfer of metals from the inner metal-rich regions to the outer metal-poor regions.
When the second supernova explodes, no pristine gas is affected by the $\alpha$-depleted ejecta.
Particles far from the second supernova, which would be pristine if not for diffusion, receive only a small fraction of ejecta relative to that received via diffusion and consequently maintain a [Mg/Fe] ratio around 1.
On the contrary, in the central regions that have lost elements
through diffusion, the impact of the second supernova is stronger and makes the [Mg/Fe] ratio gradually decrease down to around 0. 
%Similarly to T2.1, when using a constant kernel (T3.1), two phases are observed. The particle directly touched by the supernova, sharing similar
%abundances and a long tail of particles enriched by diffusion. The metallicity distribution is thus bimodal.
%The effect of adding  diffusion to the fixed radius scheme (T3.2) is rather weak.
%We can however observe the smoothing of the knee in the [Mg/Fe] vs [Fe/H] between $t_3$ and $t_4$, the direct effect of the diffusion.

When the diffusion coefficient is decreased ($d=0.0001$), the result is intermediate between the normal SPH case and the case with the strong diffusion.
In this case the high-metallicity regions follow the standard case, while the low-metallicity are similar to the strong diffusion case.

%\subsubsection{T4.0, T4.1, T4.2: weak diffusion ($d=0.0001$)}

%
\begin{figure*}  
  \subfigure[normal SPH, no mixing]                 {\resizebox{0.33\hsize}{!}{\includegraphics[angle=0]{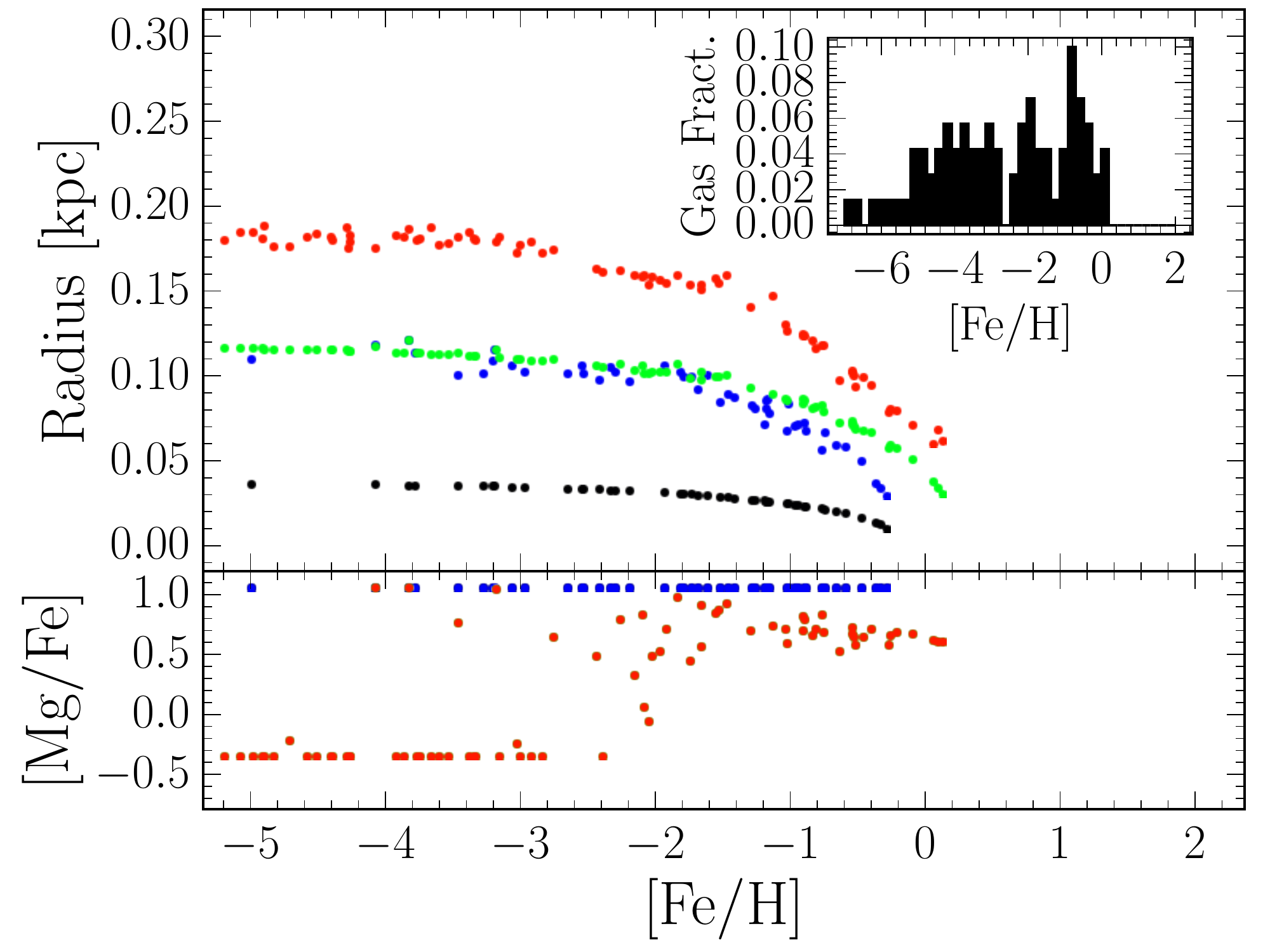}}}
  \subfigure[$w_{ij}=\Gamma_{ij}$, no mixing]              {\resizebox{0.33\hsize}{!}{\includegraphics[angle=0]{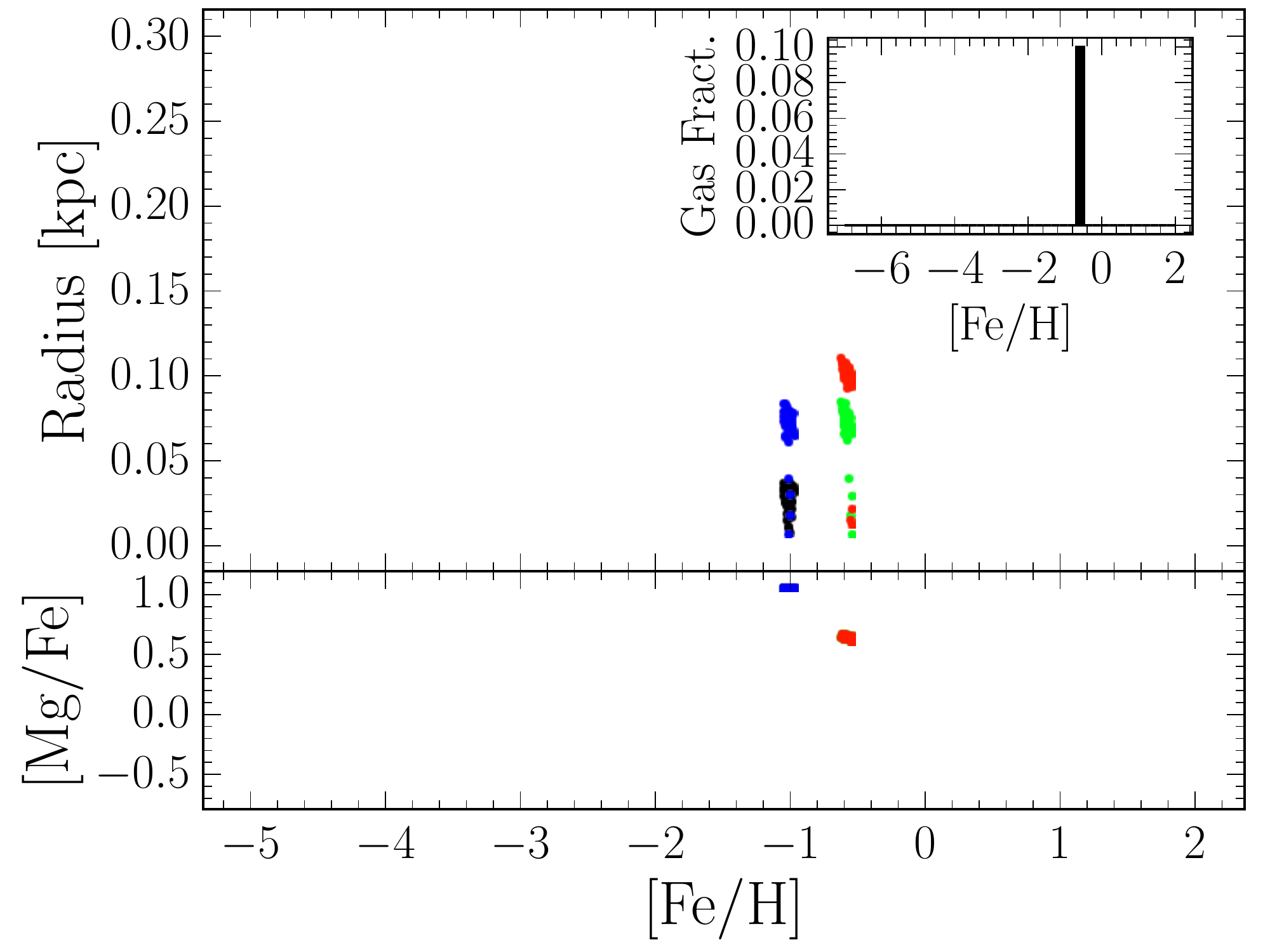}}}
  \subfigure[$R_{\rm{ej}}=0.125\,\rm{kpc}$, no mixing]{\resizebox{0.33\hsize}{!}{\includegraphics[angle=0]{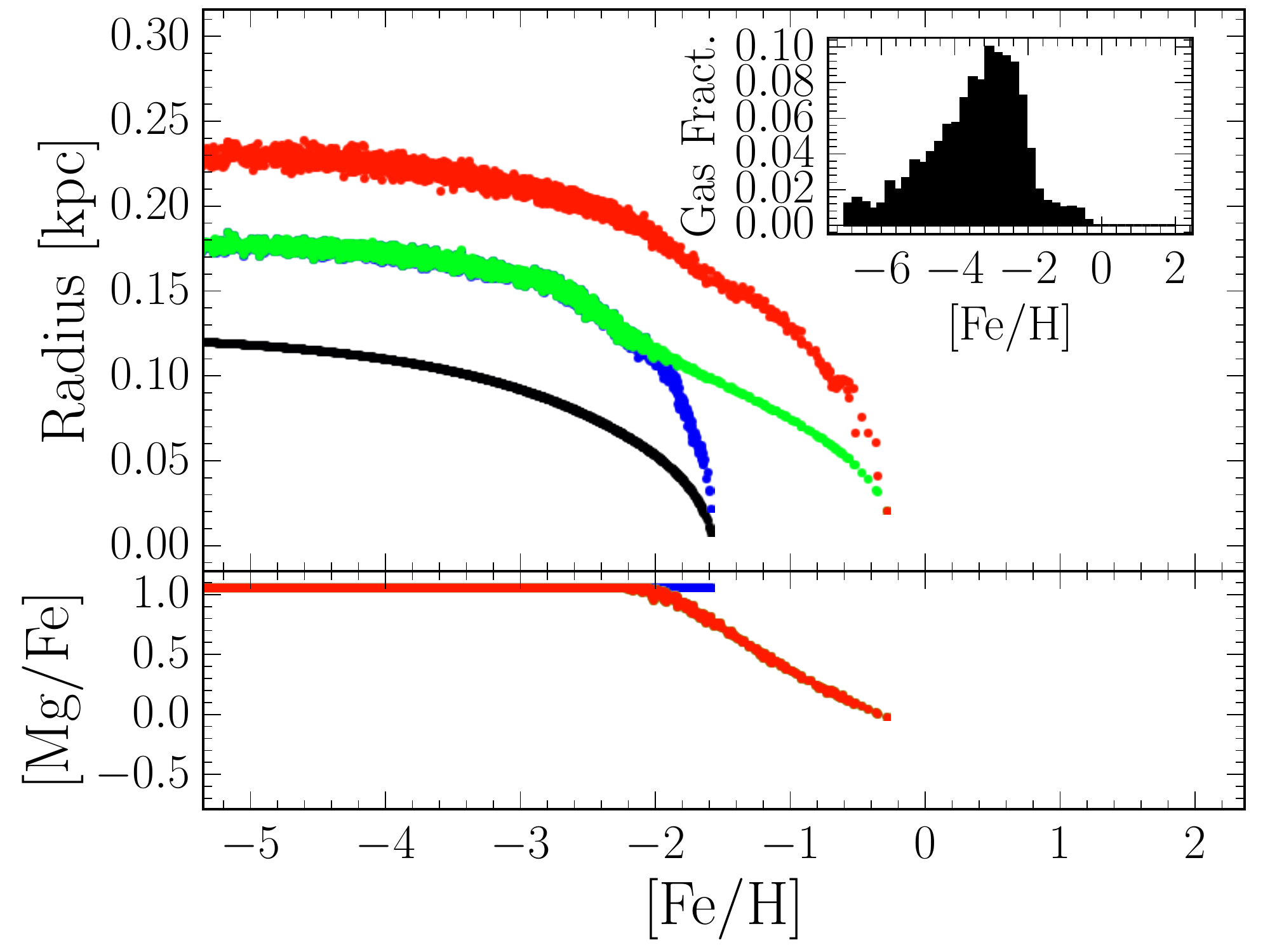}}}
  \subfigure[normal SPH, smooth metal.]{\resizebox{0.33\hsize}{!}{\includegraphics[angle=0]{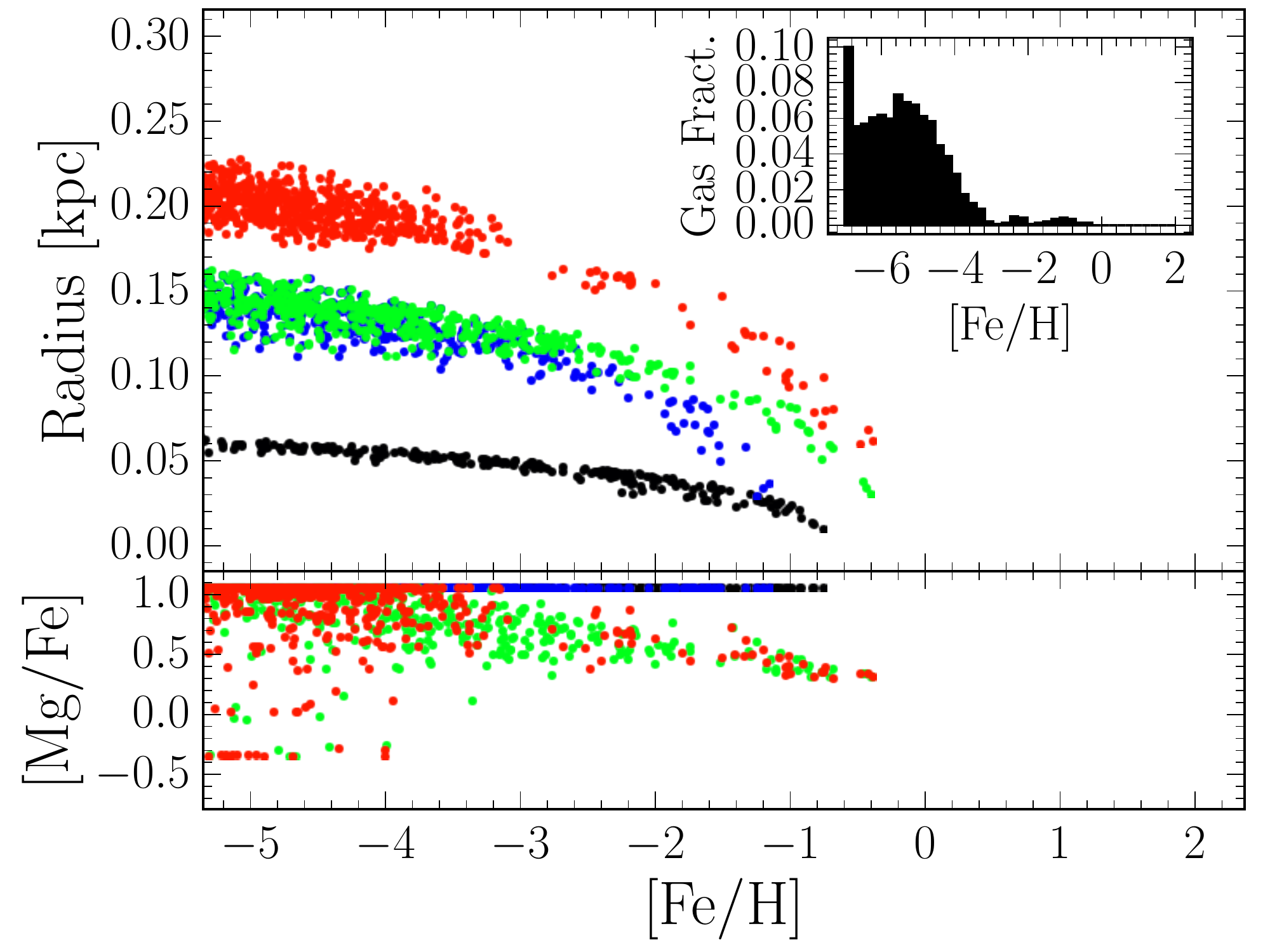}}}
  %\subfigure[$w_{ij}=\Gamma_{ij}$, smooth metal.]{\resizebox{0.33\hsize}{!}{\includegraphics[angle=0]{./graphs/SNs/11.pdf}}}
  %\subfigure[$R_{\rm{ej}}=0.125\,\rm{kpc}$, smooth metal.]{\resizebox{0.33\hsize}{!}{\includegraphics[angle=0]{./graphs/SNs/12.pdf}}}
  \subfigure[normal SPH, diffusion ($d=3\times 10^{-3}$)]{\resizebox{0.33\hsize}{!}{\includegraphics[angle=0]{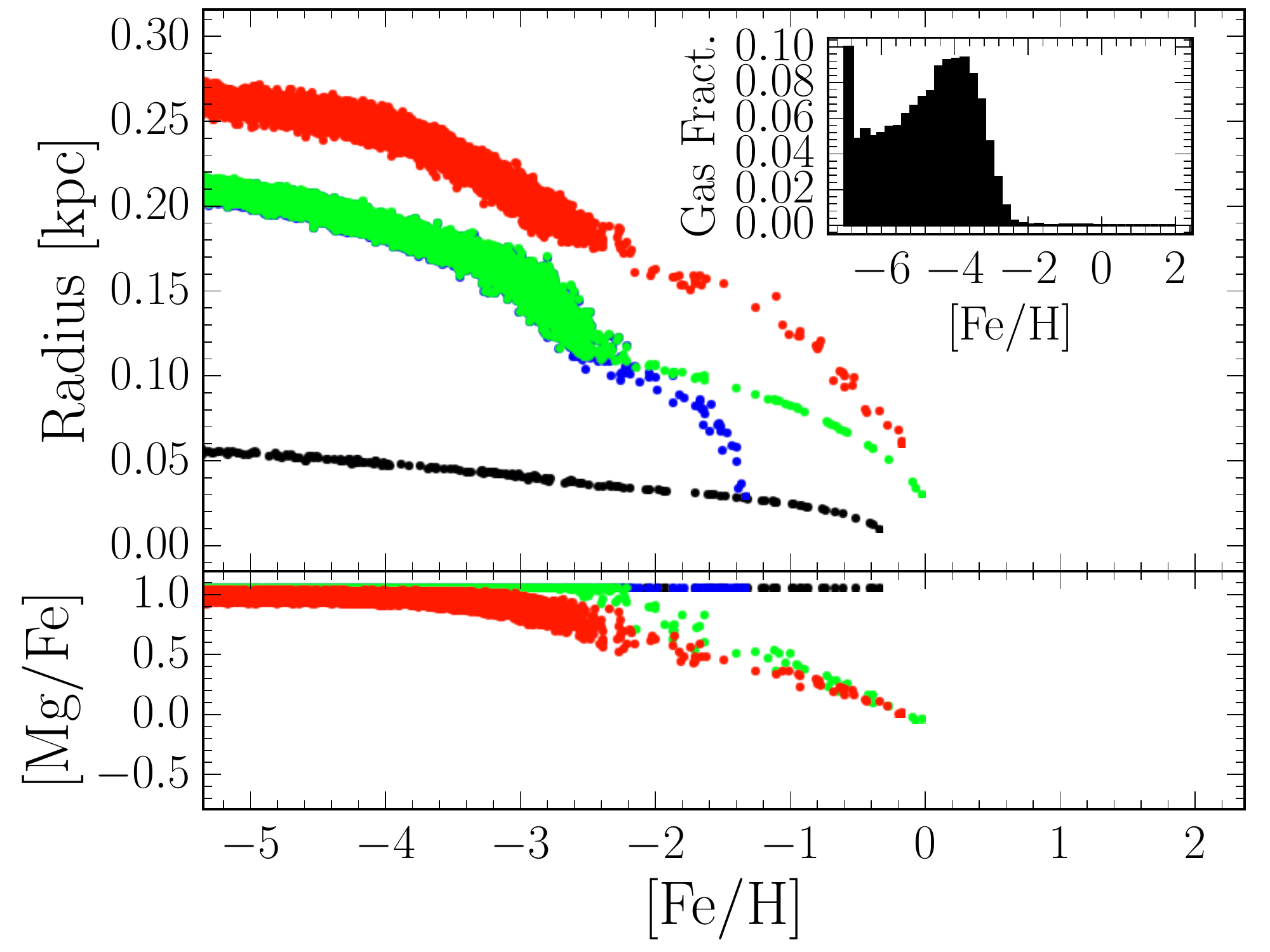}}}
  %\subfigure[$w_{ij}=\Gamma_{ij}$, diffusion ($d=0.003$)]{\resizebox{0.33\hsize}{!}{\includegraphics[angle=0]{./graphs/SNs/21.pdf}}}
  %\subfigure[$R_{\rm{ej}}=0.125\,\rm{kpc}$, diffusion ($d=0.003$)]{\resizebox{0.33\hsize}{!}{\includegraphics[angle=0]{./graphs/SNs/22.pdf}}}
  \subfigure[normal SPH, diffusion ($d=10^{-4}$)]{\resizebox{0.33\hsize}{!}{\includegraphics[angle=0]{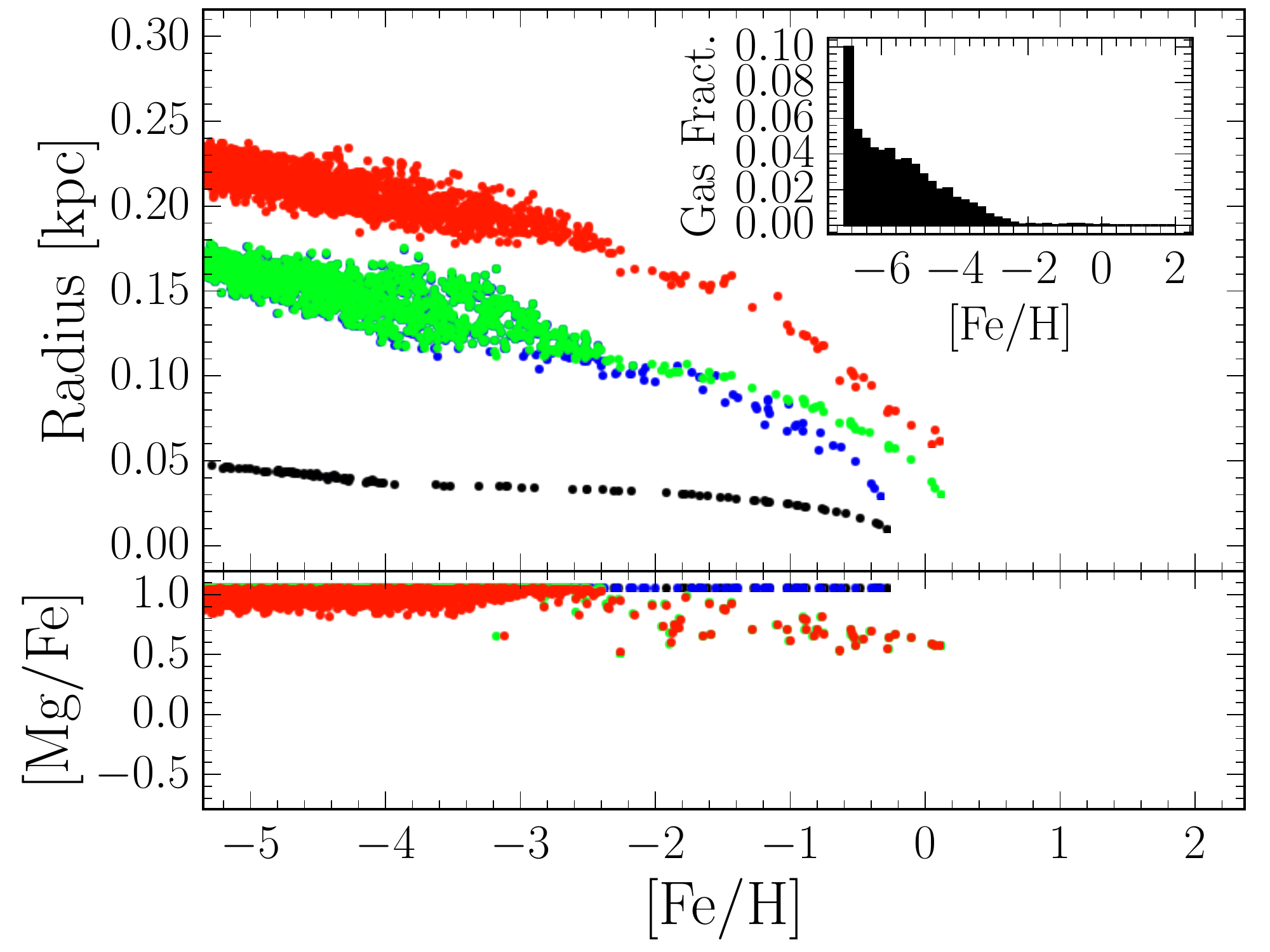}}}
  %\subfigure[$w_{ij}=\Gamma_{ij}$, diffusion ($d=10^{-4}$)]{\resizebox{0.33\hsize}{!}{\includegraphics[angle=0]{./graphs/SNs/31.pdf}}}
  %\subfigure[$R_{\rm{ej}}=0.125\,\rm{kpc}$, diffusion ($d=10^{-4}$)]{\resizebox{0.33\hsize}{!}{\includegraphics[angle=0]{./graphs/SNs/32.pdf}}}  

\caption{Evolution of the particles touched by the supernova feedback is shown for different element ejection schemes (see text). The upper parts of the plots show the
radius as a function of the metallicity. The lower parts indicate the [Mg/Fe] ratio as a function of [Fe/H].
The black dots show particles at 
$t_1$, just after the first explosion;
the blue dots at
$t_2$, just before the second explosion;
the green dots at
$t_3$, just after the second explosion;
and the red dots at
$t_4$, at the end of the run.
The histograms on the upper right indicate the metallicity distribution function at $t_4$.
}
  \label{fig:twoSNs}
\end{figure*}
%

%%%%%%%%%%%%%%%%%%%%%%%%%%%%%%%%%%%%%%%%%%%%%%%%%%%%%%%%%%%%%%%%%%%%%%%%%%%%%%%%%%%%%%%%%%%%%%%%%%%%%%%%%%%

\section{Simulations of dwarf spheroidal galaxies}\label{dsph}

%%%%%%%%%%%%%%%%%%%%%%%%%%%%%%%%%%%%%%%%%%%%%%%%%%%%%%%%%%%%%%%%%%%%%%%%%%%%%%%%%%%%%%%%%%%%%%%%%%%%%%%%%%%

We simulate dwarf spheroidal galaxies (dSph) to study the effect of the numerical schemes on observable properties.
Here, we systematically alter IMF modelisation (Section~\ref{imf}),
how elements are distributed after supernovae explosions (Section~\ref{feedback}), and mixing of metals via the smooth metallicity approach (Section~\ref{smooth_metallicity}) or 
diffusion (Section~\ref{metal_diffusion}).

We focus on star formation rates as well as on stellar abundances, both of which are well constrained
by recent observations (see Section~\ref{introduction}).
As dSphs form nearly all stars in situ with only minimal accretions, they are
more relevant for a direct comparison than more massive galaxies.
Furthermore, as dwarf galaxies
are small systems, their modelling is less CPU intensive, which allows us to explore a 
large parameter space.

\subsection{Initial conditions}

The dSph models are based upon updated versions of the models described in \citep{revaz2012}. The initial 
conditions are based on a two-slope density spherical profile
\begin{equation}
        \rho(r) = \frac{\rho_0}  {\left( \frac{r}{r_s} \right)^a  \left(1+\frac{r}{r_s}\right)^{b-a}}  
\end{equation}
where $a$ is the slope of the inner profile, fixed to $0$ to correspond to a core favoured by observations 
of normal low brightness and dwarf galaxies \citep{blaisouellette01,deblok02,swaters03,gentile04,gentile05,spekkens05,deblok05,deblok08,spano08,walker11,oh11},
$b$ is the slope of the outer profile, fixed to $3$, in agreement with the outer NFW profile \citep{navarro96,navarro97} from $\Lambda$CDM simulations.
$r_s$, the scale radius is set to $2\,\rm{kpc}$.
The models are truncated at a maximum radius $r_\mathrm{max}$.
Initially, the models only contain dark matter and pristine gas both sharing the same profile with a respective mass ratio of $0.15$ in agreement with the standard cosmological baryonic fraction.

The initial equilibrium of the halo is obtained by calculating the velocity dispersions derived from the Jean's equations,
assuming a spherical symmetry.
For the gas, these velocities are transformed in thermal energy by multiplying by 0.5 such that the gas sphere 
smoothly condenses when the simulation is started.

\subsection{Dwarf spheroidal models}

Two different sets of initial conditions have been considered, as much as possible both reproducing the observed properties of the Local Group dSphs Sextans and Fornax, which are two galaxies representative of the variety among dSphs.
%for which accurate spectroscopic data are available \citep{shetrone2001,shetrone2003,aoki2009,tafelmeyer2010,letarte2010}. 
Fornax is expected to be rather massive owning to it high luminosity ($15.5\times 10^6\,\rm{L_\odot}$) and
metal-rich stellar population ($\langle \rm{[Fe/H]}\rangle=-1.07$).
It is characterised by an extended star formation rate \citep{coleman08}.
At the opposite end, Sextans is substantially less luminous ($L= 0.53\times 10^6\,\rm{L_\odot}$) and
characterised by a metal-poor stellar population ($\langle \rm{[Fe/H]}\rangle=-2.02$ ) dominated by old stars \citep{lee03}.
The parameters used for each one of these two models are given in Tab.~\ref{tab:cinit_parameters}.
\begin{table}[tpb]
        \centering
        \begin{tabular}{|c|c|c|c|c|c|}%c|}
                \hline
                Galaxy~Name            & $M_{\rm{tot}}$                & $r_{\rm{max}}$        & $r_s$         & a     & b\\
                                & $[10^8\,\rm{M_{\odot}}]$      & $[\rm{kpc}]$          & $[\rm{kpc}]$  &       &  \\
                \hline
                \hline
                Sextans         & 3.5                           & 9                     & 2.5           & 0     & 3\\
                Fornax          & 8                             & 9.5                   & 2             & 0     & 3\\
                
                \hline
        \end{tabular}
        \caption{Set of parameters for the initial conditions of dSph models like Fornax and Sextans.}
        \label{tab:cinit_parameters}
\end{table}
Each of these two models is run using different resolutions spanning nearly three orders of magnitude in mass.
The resolution is defined by an integer $\mathfrak{r}$. The corresponding stellar mass $m_{\rm{\star,\mathfrak{r}}}$ and spatial resolution
(gravitational softening) $\epsilon_{\mathfrak{r}}$, are computed using the relations
\begin{equation}
        m_{\rm{\star,\mathfrak{r}}} = m_0/2^\mathfrak{r}\qquad \mathrm{and}\qquad
        \epsilon_{\mathfrak{r}}     = \epsilon_0 \cdot 2^{-\mathfrak{r}/3},
\end{equation}
where $m_0=65536\,\rm{M_{\rm{\odot}}}$ and $\epsilon_0=0.24\,\rm{kpc}$
correspond to the poorest resolution considered ($\mathfrak{r}=0$).
With this definition, incrementing the resolution by 3 corresponds to dividing the mass by a factor eight and the spatial resolution by two.
%For both the Fornax and Sextans models, the effect of the IMF sampling schemes have been tested and will be discussed in Section~\ref{random_sampling} for the RIMFS, 
%in Section~\ref{optimal_sampling} for the OIMFS and in Section~\ref{continuous_sampling} for the CIMFS. 
%The smooth metallicity and the metal diffusion scheme will also be investigated in Section~\ref{smooth_metallicity} and \ref{metal_diffusion},
%as well as the element spreading schemes (Section~\ref{feedback}), a step function SPH kernel ($w_{ij}=\Gamma_{ij}$), a constant ejection radius $R_{\rm{ej}}$ and a blast radius %$R_{\rm{E}}$.

The complete set of simulations including their parameters is given in Table~\ref{tab:fiducial_parameters}, representing more than 80 simulations. 
A parameter set to ``SPH'' means that the default SPH setting is used for both $w_{ij}$ and $R_{\rm{ej}}$. 
We also test additional parameters, such as the adiabatic time (Section~\ref{sedov_taylor_solution}) applied to either SNIa or SNII.
We defined a fiducial set of parameters shown on the top of the table to ease the comparison between the models and increase readability.
When the value of a parameter is not given, the one of the fiducial model is used.
For all models, the star formation parameter $c_{\star}$ is fixed to $0.01$ and the supernova efficiency $\epsilon_{\rm{SN}}$ is set to $0.3$.

\begin{table*}[htpb]
        \centering
        \begin{tabular}{|c|c|c|c|c|c|c|c|c|c|}%c|}
                \hline
                Resolution              & $m_{\star}$           & $\epsilon_{\rm{grav}}$    & IMF            & $t_{\rm{ad,SNII}}$	& $t_{\rm{ad,SNIa}}$  & smooth metal. & diff. coeff. $d$ & $R_{\rm{ej}}$			& $w_{ij}$	      \\
                $\mathfrak{r}$          & $[\rm{M_{\odot}}]$    & $[\rm{pc}]$               & sampling       & $[\rm{Myr}]$		& $[\rm{Myr}]$       &               &    		 & $[\rm{kpc}]$	 		&  \\
                \hline\\
                \multicolumn{10}{c}{Fiducial models}\\
                \hline
                9                       & 128                   & 0.030                     & RIMFS   	     & 5 			& 5		      & yes	      & 0		 &	SPH 	& SPH\\
                6                       & 1024                  & 0.060                     & RIMFS  	     & 5 			& 5		      & yes	      & 0		 &	SPH 	& SPH\\
                5                       & 2048                  & 0.076                     & RIMFS  	     & 5 			& 5		      & yes	      & 0		 &	SPH 	& SPH\\
                4                       & 4096                  & 0.096                     & RIMFS  	     & 5 			& 5		      & yes	      & 0		 &	SPH 	& SPH\\
                3                       & 8192                  & 0.120                     & RIMFS  	     & 5 			& 5		      & yes	      & 0		 &	SPH 	& SPH\\
                0                       & 65536                 & 0.241                     & RIMFS  	     & 5 			& 5		      & yes	      & 0		 &	SPH 	& SPH\\    
                \hline\\
                \multicolumn{10}{c}{IMF sampling schemes}\\
                \hline                          
                9                       & 128                   & 0.030                     & OIMFS          &  			&  		      &  	      &  	     &      	& \\
                6                       & 1024                  & 0.060                     & OIMFS          &  		   	&		      & 	      & 		 &	 		& \\
                5                       & 2048                  & 0.076                     & OIMFS          &  		   	&		      & 	      & 		 &	 		& \\
                4                       & 4096                  & 0.096                     & OIMFS          &  		   	&		      & 	      & 		 &	 		& \\
                3                       & 8192                  & 0.120                     & OIMFS          &  		   	&		      & 	      & 		 &	 		& \\
                0                       & 65536                 & 0.241                     & OIMFS          &  		   	&		      & 	      & 		 &	 		& \\
                \hline
                9                       & 128                   & 0.030                     & CIMFS          &  			& 		      &  	      &  	     &      	& \\
                6                       & 1024                  & 0.060                     & CIMFS          &  		    &   		  &		      &		     &	 		& \\
                5                       & 2048                  & 0.076                     & CIMFS          &  		    &   		  &		      &		     &	 		& \\
                4                       & 4096                  & 0.096                     & CIMFS          &  		    &   		  &		      &		     &	 		& \\
                3                       & 8192                  & 0.120                     & CIMFS          &  		    &   		  &		      &		     &	 		& \\
                0                       & 65536                 & 0.241                     & CIMFS          &  		    &   		  &		      &		     &	 		& \\
                \hline
                9                       & 128                   & 0.030                     & CIMFS          &  			& 0		      &  	      &  	     &      	& \\
                6                       & 1024                  & 0.060                     & CIMFS          &  		    & 0  		  &		      &		     &	 		& \\
                5                       & 2048                  & 0.076                     & CIMFS          &  		    & 0  		  &		      &		     &	 		& \\
                4                       & 4096                  & 0.096                     & CIMFS          &  		    & 0  		  &		      &		     &	 		& \\
                3                       & 8192                  & 0.120                     & CIMFS          &  		    & 0  		  &		      &		     &	 		& \\
                0                       & 65536                 & 0.241                     & CIMFS          &  		    & 0  		  &		      &		     &	 		& \\
                \hline\\
                \multicolumn{10}{c}{Elements spreading schemes}\\
                \hline                              
                6                       & 1024                  & 0.060                     &        	     &   			&  		      & no 	      &  		 &	0.125  &  \\             
                6                       & 1024                  & 0.060                     &        	     &   			&  		      & no 	      &  		 &	$R_{\rm{E}}$    &  \\   
                6                       & 1024                  & 0.060                     &        	     &   			&  		      & no 	      &  	     &         & $\Gamma_{ij}$\\                           
                \hline \\
                \multicolumn{10}{c}{Mixing schemes}\\
                \hline                 
                6                       & 1024                  & 0.060                     &                &  			& 		      & no	      &			 & 			& \\
                5                       & 2048                  & 0.076                     &                &  		    &    		  &	no	      &		     &	 		& \\
                4                       & 4096                  & 0.096                     &                &  		    &    		  &	no	      &		     &	 		& \\
				\hline
                6                       & 1024                  & 0.060                     &                &              &             & no        & 0.003    &       	&\\
                5                       & 2048                  & 0.076                     &                &              &             & no        & 0.003    &      	& \\
                4                       & 4096                  & 0.096                     &                &              &             & no        & 0.003    &       	&\\
                \hline
                6                       & 1024                  & 0.060                     &                &              &             & no        & 0.001    &       	&\\
                5                       & 2048                  & 0.076                     &                &              &             & no        & 0.001    &       	&\\
                4                       & 4096                  & 0.096                     &                &              &             & no        & 0.001    &       	&\\
                \hline          
                6                       & 1024                  & 0.060                     &                &              &             & no        & 0.0003   &      	&\\
                5                       & 2048                  & 0.076                     &                &              &             & no        & 0.0003   &       	&\\
                4                       & 4096                  & 0.096                     &                &              &             & no        & 0.0003   &       	&\\
                \hline          
                6                       & 1024                  & 0.060                     &                &              &             & no        & 0.0001   &       	&\\
                5                       & 2048                  & 0.076                     &                &              &             & no        & 0.0001   &       	&\\
                4                       & 4096                  & 0.096                     &                &              &             & no        & 0.0001   &       	&\\
                \hline                          

        \end{tabular}
        \caption{Complete set of dSph simulations including their proper parameters. Each simulation has been run for both for Fornax and Sextans model. When the value
        	of a parameter is not given the value of the fiducial model is used.}
        \label{tab:fiducial_parameters}
\end{table*}

\subsection{The effect of the IMF sampling}\label{effect_of_IMFS}

In Section \ref{imf} we presented different approaches to sample the IMF.
Briefly, these can be summarised as: continuous IMF sampling (CIMFS, see Section~\ref{continuous_sampling}), where each stellar mass contributes in ratio to its proportion of the IMF (including fractional contributions, if this is less than one); random sampling (RIMFS, Section~\ref{random_sampling}), where the IMF is treated as a probability distribution and supernovae are generated stochastically; and ``optimal'' IMF sampling (OIMF, Section~\ref{optimal_sampling}), where the IMF is split into mass bins such that each bin contains one supernova at the lower end of the mass bin.
The choice of IMF sampling method has a direct impact on both the feedback energy and the elements injected as supernovae go off inside the galaxy.
Our aim here is to address the impact of the IMF choice on the global chemical enrichment of a galaxy.
\begin{figure*}
  \subfigure[Fornax : $\mathfrak{r}=4, m_{\star}=4096\,\rm{M_{\odot}}$]{\resizebox{0.33\hsize}{!}{\includegraphics[angle=0]{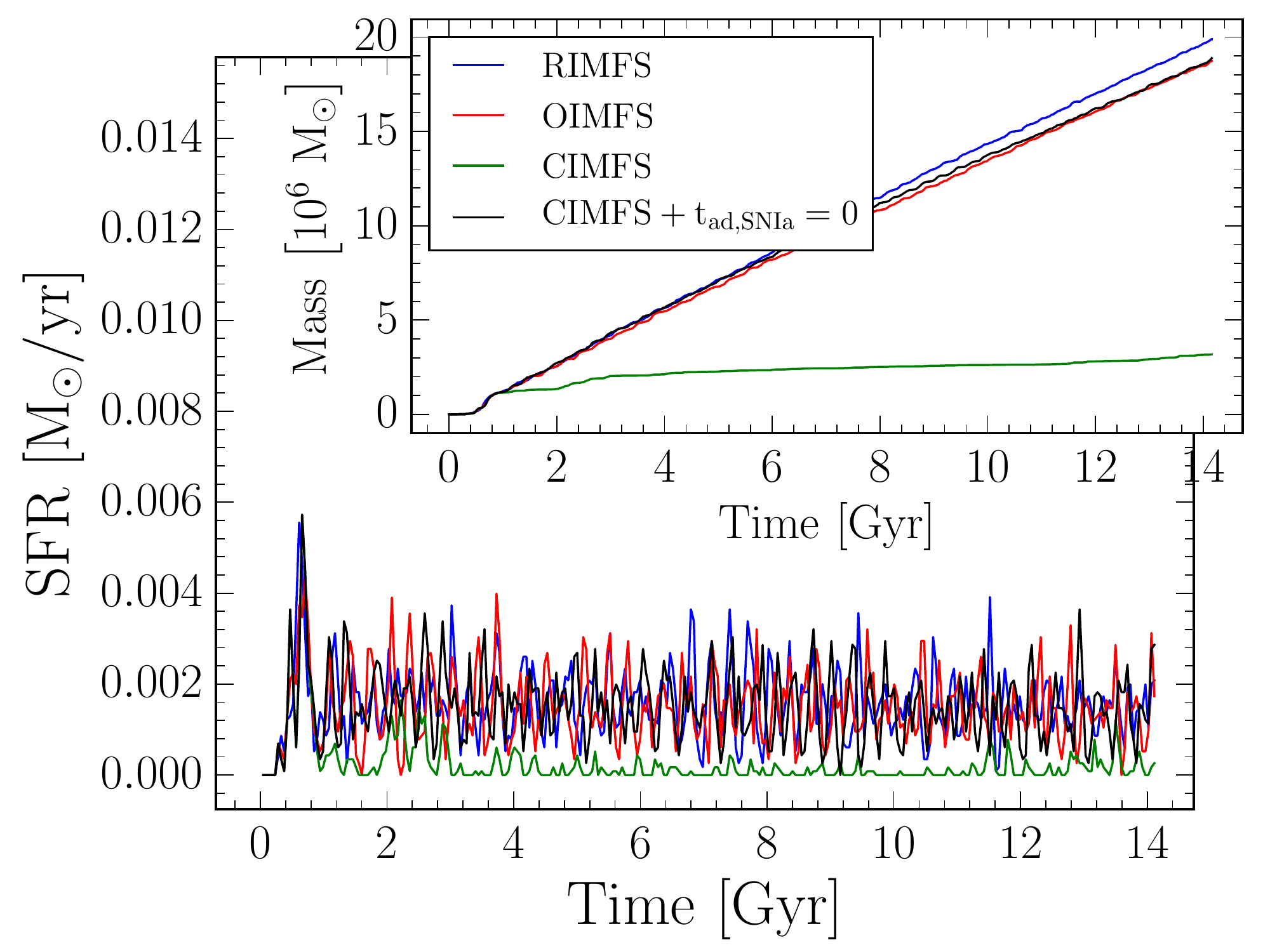}}}
  \subfigure[Fornax : $\mathfrak{r}=5, m_{\star}=2048\,\rm{M_{\odot}}$]{\resizebox{0.33\hsize}{!}{\includegraphics[angle=0]{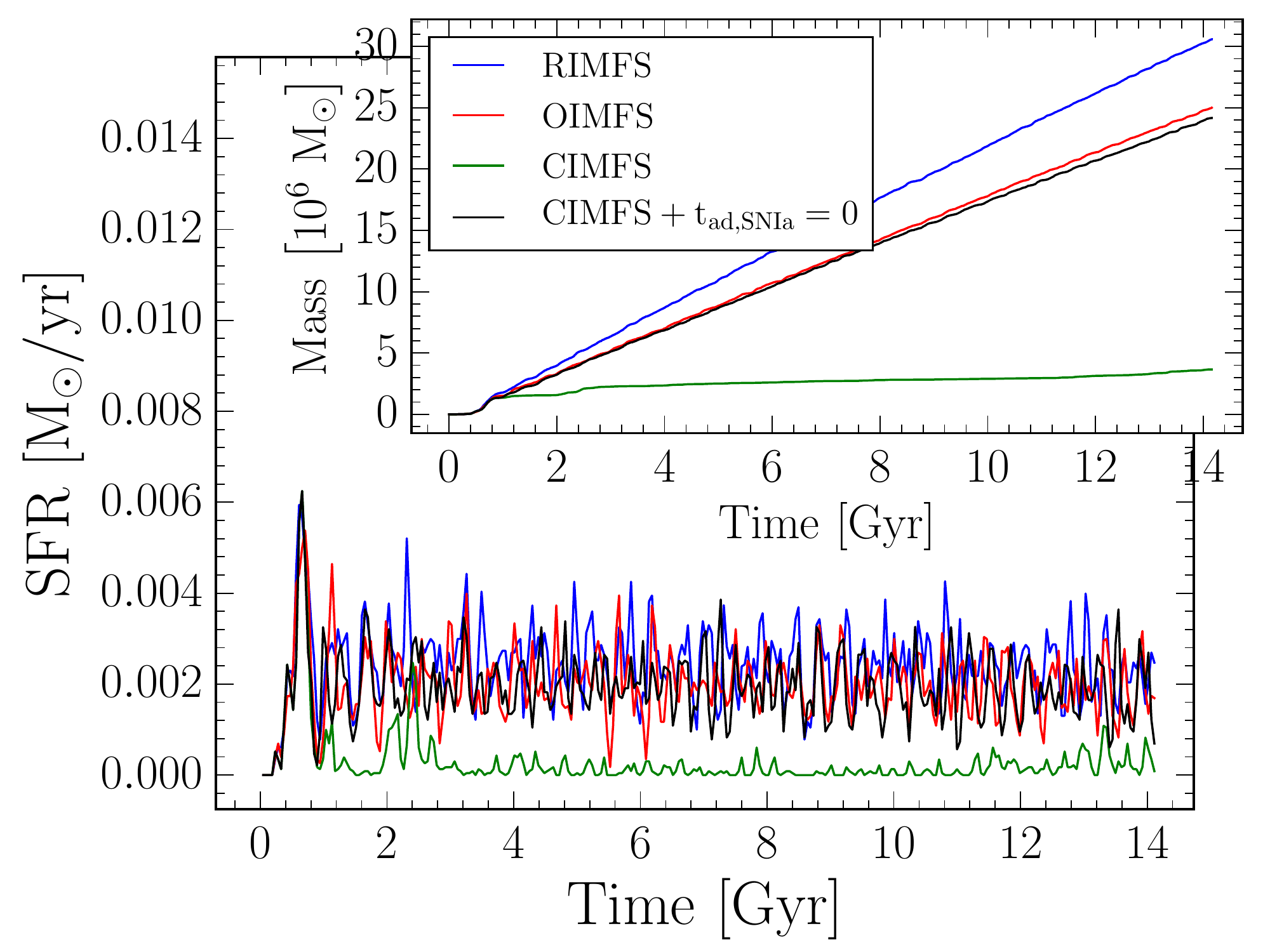}}}
  \subfigure[Fornax : $\mathfrak{r}=6, m_{\star}=1024\,\rm{M_{\odot}}$]{\resizebox{0.33\hsize}{!}{\includegraphics[angle=0]{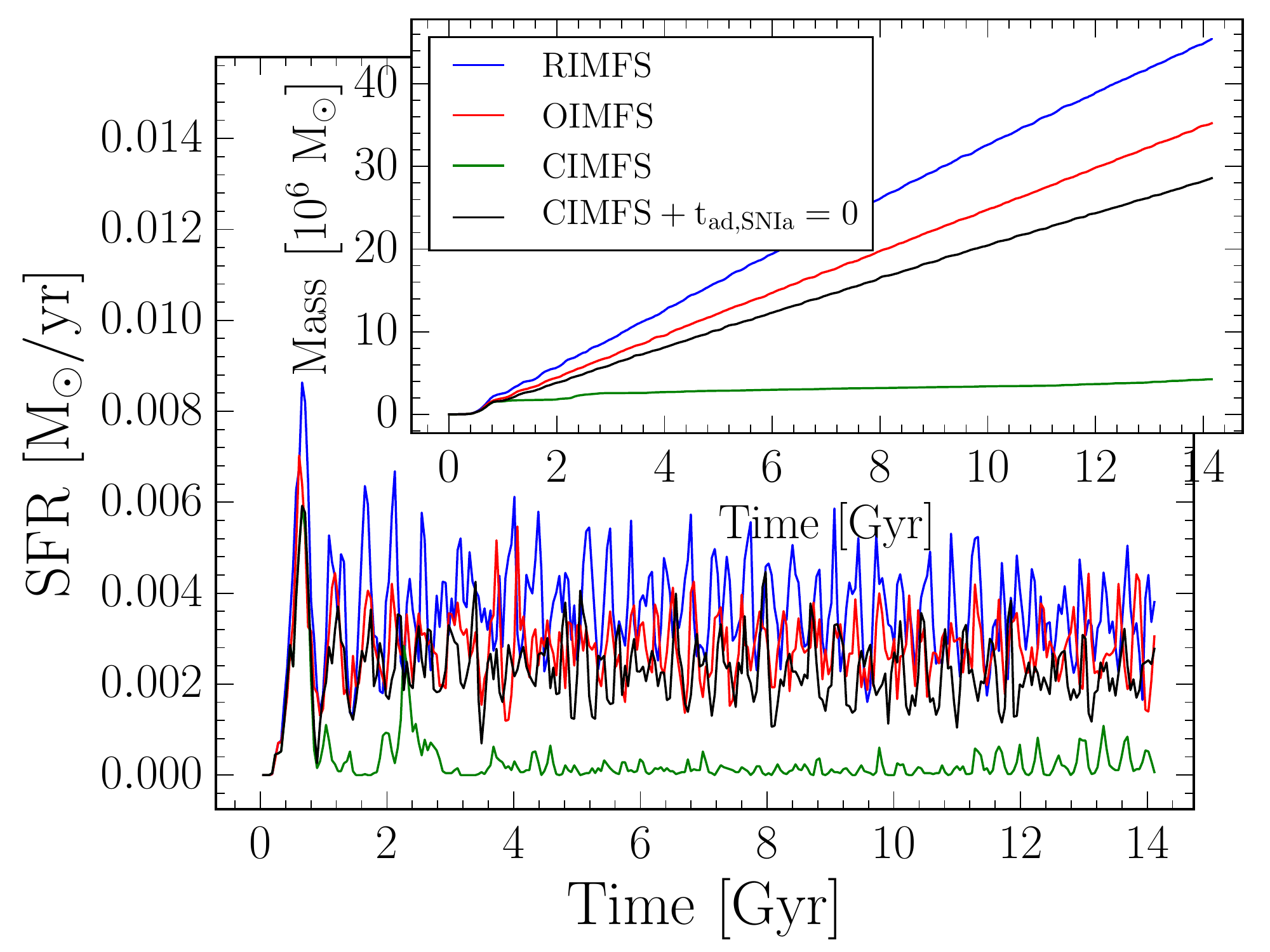}}}

  \subfigure[Sextans : $\mathfrak{r}=4, m_{\star}=4096\,\rm{M_{\odot}}$]{\resizebox{0.33\hsize}{!}{\includegraphics[angle=0]{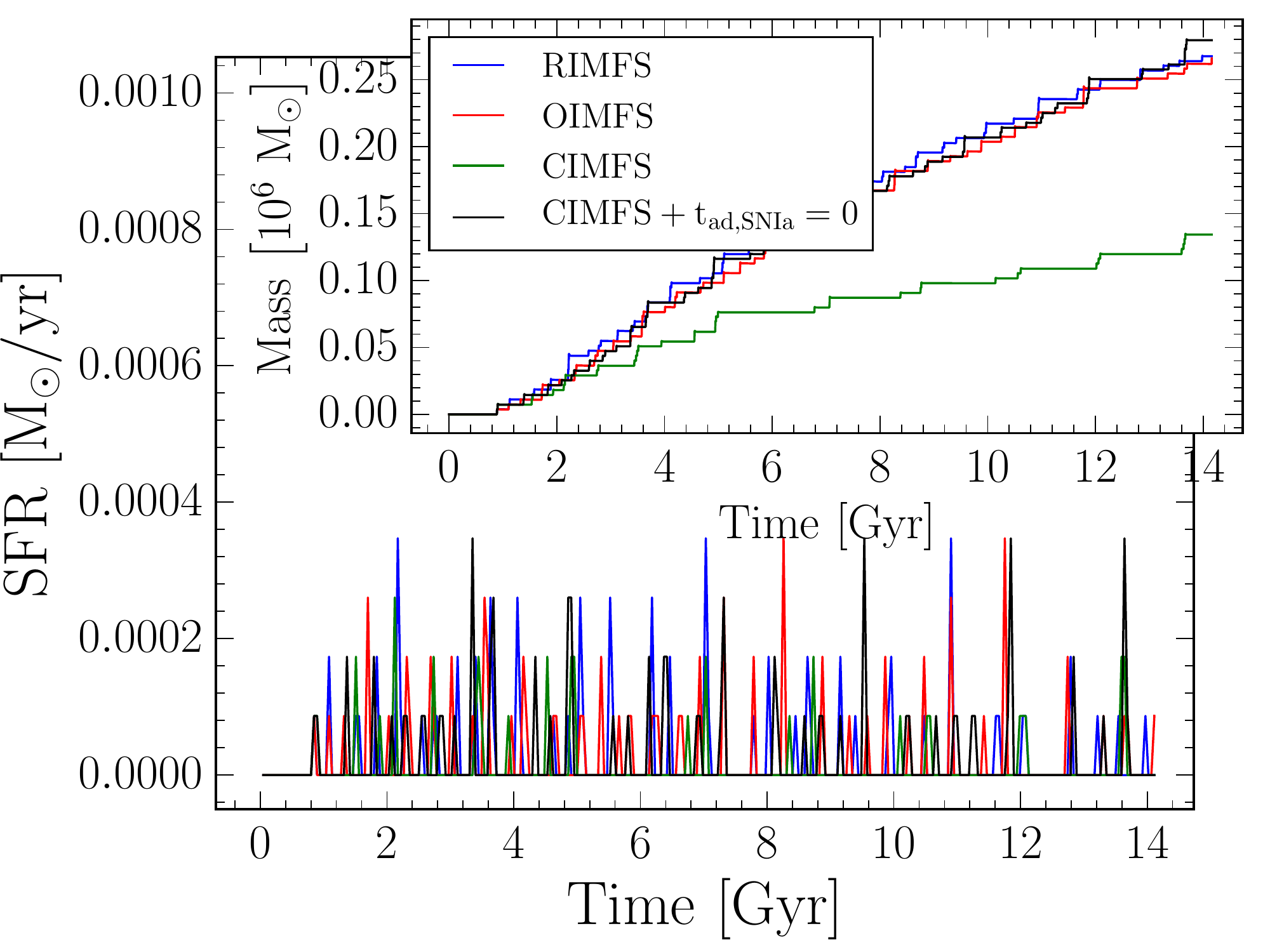}}}
  \subfigure[Sextans :$\mathfrak{r}=5, m_{\star}=2048\,\rm{M_{\odot}}$]{\resizebox{0.33\hsize}{!}{\includegraphics[angle=0]{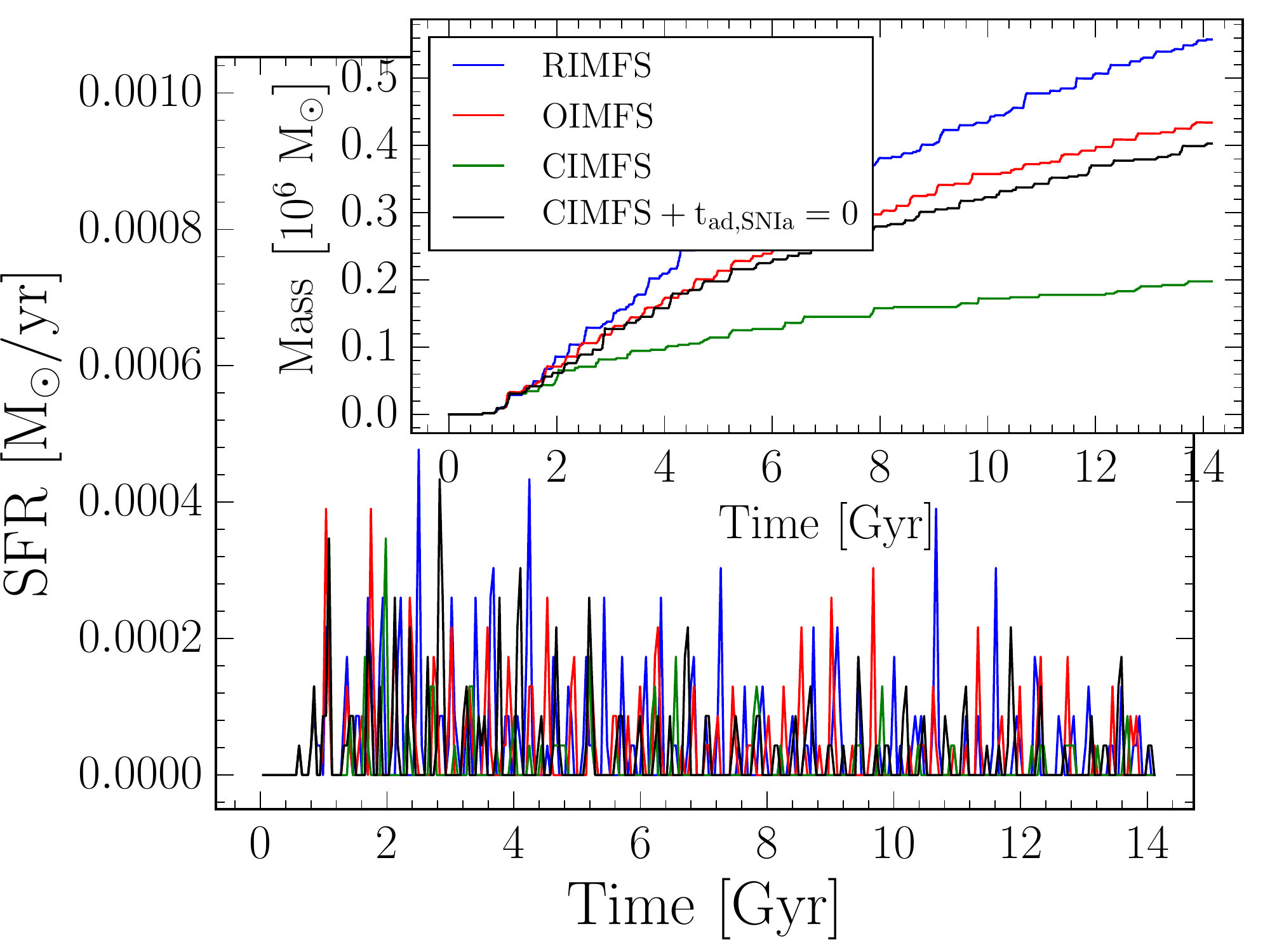}}}
  \subfigure[Sextans :$\mathfrak{r}=6, m_{\star}=1024\,\rm{M_{\odot}}$]{\resizebox{0.33\hsize}{!}{\includegraphics[angle=0]{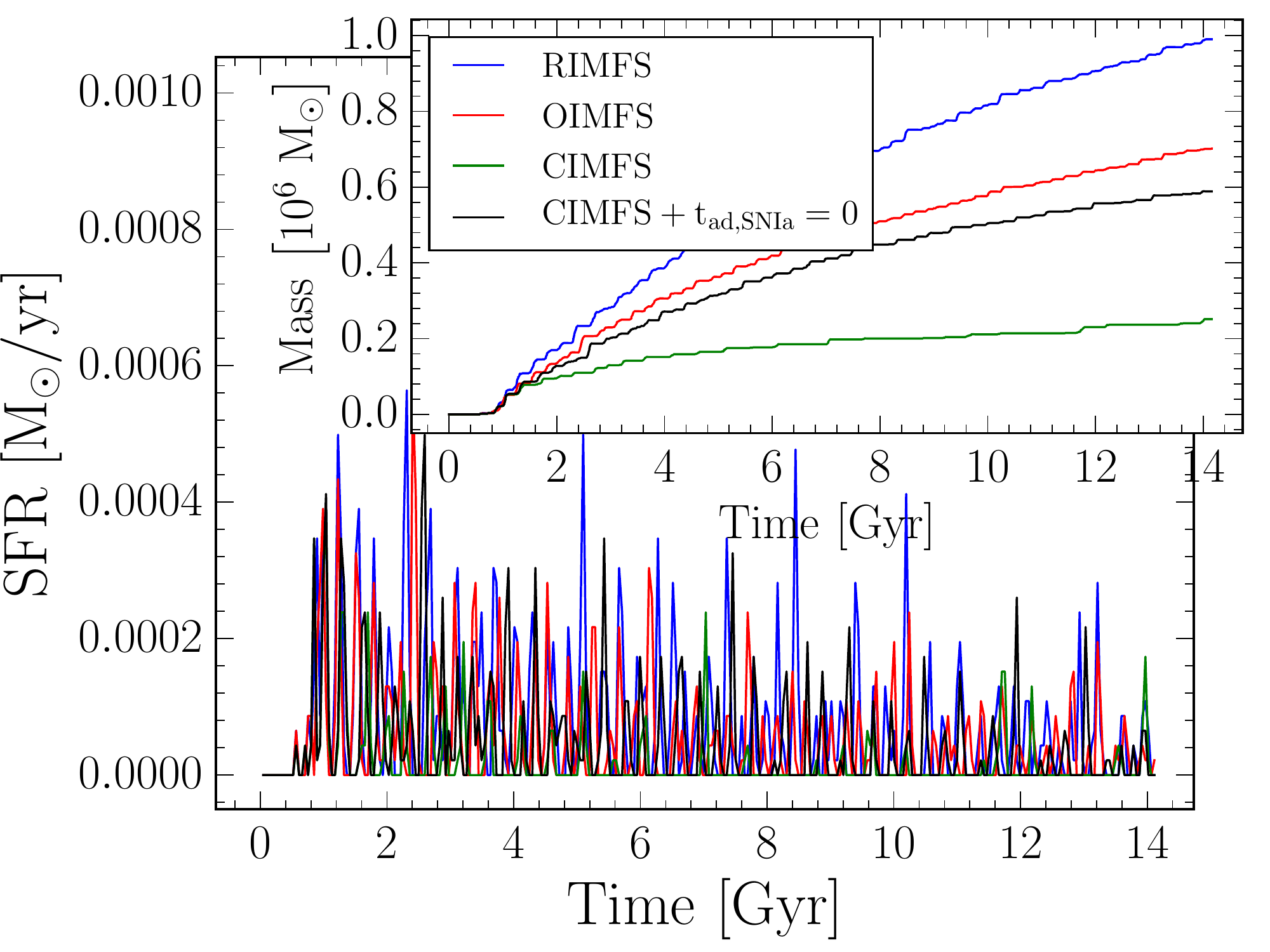}}}

\caption{Comparison of the star formation history and cumulative stellar mass evolution between the different models of Fornax and Sextans in four cases: random IMF sampling (RIMFS), optimal IMF sampling (OIMFS),
  and continuous IMF sampling with an adiabatic time $t_{\rm{ad,SNIa}}=5\,\rm{Myr}$ (CIMFS) and with $t_{\rm{ad,SNIa}}=0$ ($\rm{CIMFS}+t_{\rm{ad,SNIa}}=0$).
  Each panel corresponds to a different resolution.}
  \label{fig:IMFsfr}
\end{figure*}
In Figure~\ref{fig:IMFsfr}, we delineate the star formation history and evolution of the stellar mass for the three different schemes as a function
of the resolution from $\mathfrak{r}=4\,(m_{\star}=4096\,\rm{M_{\odot}})$ to $\mathfrak{r}=6\,(m_{\star}=1024\,\rm{M_{\odot}})$.
We analyse and compare these simulations in detail below.

\subsubsection{CIMFS and the SNIa adiabatic time}

The first and foremost feature observed in the CIMFS method is the quenching of star formation after about $3\,\rm{Gyr}$, 
independent of the resolution or the dSph model. 
Between $2$ and $10$ times fewer stars are generated in the case of the CIMFS which drastically impacts the final metallicity and abundances.

This quenching is a direct consequence of the heating of IMF by the SNIa and may be understood as follows.
%The number of SNIa issued, during a period of $10\,\rm{Gyr}$, from a stellar population of $10^6\,\rm{M_{\odot}}$ (corresponding to roughly the 
%amount of stars generated by the first peaks in the Fornax case), is about 1000. As pointed out in Section~\ref{continuous_sampling}, in this case, when considering 
%the CIMFS approach, at every times step a tiny fractions of SNIa will explode, generating about 1000 events divided by the number of time steps used during $10\,\rm{Gyr}$
%which is of the order of $10^6$. 
For the $\mathfrak{r}=4$ resolution runs ($m_{\star}=4096\,\rm{M_{\odot}}$), the typical timestep is $0.1\,\rm{Myr}$. According to the Kroupa IMF \citep{kroupa2001} and the \citet{kobayashi2000} SNIa model, 
the number of SNIa exploding during this timestep is about $10^{-4}$. This means that small fractions of SNIa continuously explode over a long time interval,
forcing the adiabatic time switch to be always active. Consequently, the gas cooling is artificially quenched along with the star formation rate. 
The problem worsens with increasing resolution.

In contrast, only full SNe explode in the RIMFS and CIMFS approaches, respecting the physical delay in time between all of them (see Figure~\ref{fig:NSNvsTime}). 
As $t_{\rm{ad}}=5\,\rm{Myr}$ is small compared to the mean time between two SNe, the cooling is still effective in maintaining a continuous star formation rate.

In a fourth model ($\rm{CIMFS}+t_{\rm{ad,SNIa}}=0$), we tested the suppression of the adiabatic period only after the SNIa explosions, since it is sometimes used in galaxy formation simulations.
Setting the adiabatic time for the SNIa to zero in the CIMFS generates a star formation rate akin to the RIMFS and OIMFS.
However,  we still have the problem that SNIa dilute the ejection of metals in time affecting 
the chemical evolution of the system.
As in this case, at every timestep all stellar particles experimenting SNe explosions expel at least a fraction of elements, and the mixing is artificially
boosted.
However, it is clear that CIMFS with $t_{\rm{ad,SNIa}}\ne 0$ does not reproduce any feature accurately and consequently, we will no longer consider those models in the future.

\subsubsection{The effect of the resolution}

The second clear feature on Fig.~\ref{fig:IMFsfr} is the important effect of the resolution on the star formation rate.
As long as the resolution is low ($\mathfrak{r}\le 4$, $m_{\star}\ge 4096\,\rm{M_{\odot}}$) all three sampling methods 
(RIMFS, OIMFS and $\rm{CIMFS}+t_{\rm{ad,SNIa}}=0$) nicely converge. 

As resolution increases the cumulative stellar mass begins to diverge, at the $\mathfrak{r}= 5$ resolution ($m_{\star}= 2048\,\rm{M_{\odot}}$), the RIMFS cumulative stellar mass deviates of about 10\% from the OIMFS and $\rm{CIMFS}+t_{\rm{ad,SNIa}}=0$.
%This tendency increases with the resolution.
For Sextans, with $\mathfrak{r}= 6$ ($m_{\star}= 1024\,\rm{M_{\odot}}$),
the RIMFS produces 40\% more stars than the $\rm{CIMFS}+t_{\rm{ad,SNIa}}=0$ and 30\% more that the OIMFS.
That the OIMFS is closer to the CIMFS because it represents a better fit to the continuous solution, as seen in Fig~\ref{fig:NSNvsTime}.

The deviation of the three methods with increasing resolution is a direct consequence of the method chosen for IMF sampling, a major problem for high-resolution simulations.
As we see in what follows, all of these methods bias the predictions of the chemical abundances once a given resolution is reached.

\subsubsection{Impact on the chemical abundances}\label{abundances_impact}

In addition to its effect on the global star formation rate, we also want to address the impact of the IMF sampling scheme
on the final abundances. 

\begin{figure*}  
  \subfigure[RIMFS, CIMFS: $\mathfrak{r}=4$, $m_{\star}=4096\,\rm{M_{\odot}}$]{\resizebox{0.33\hsize}{!}{\includegraphics[angle=0]{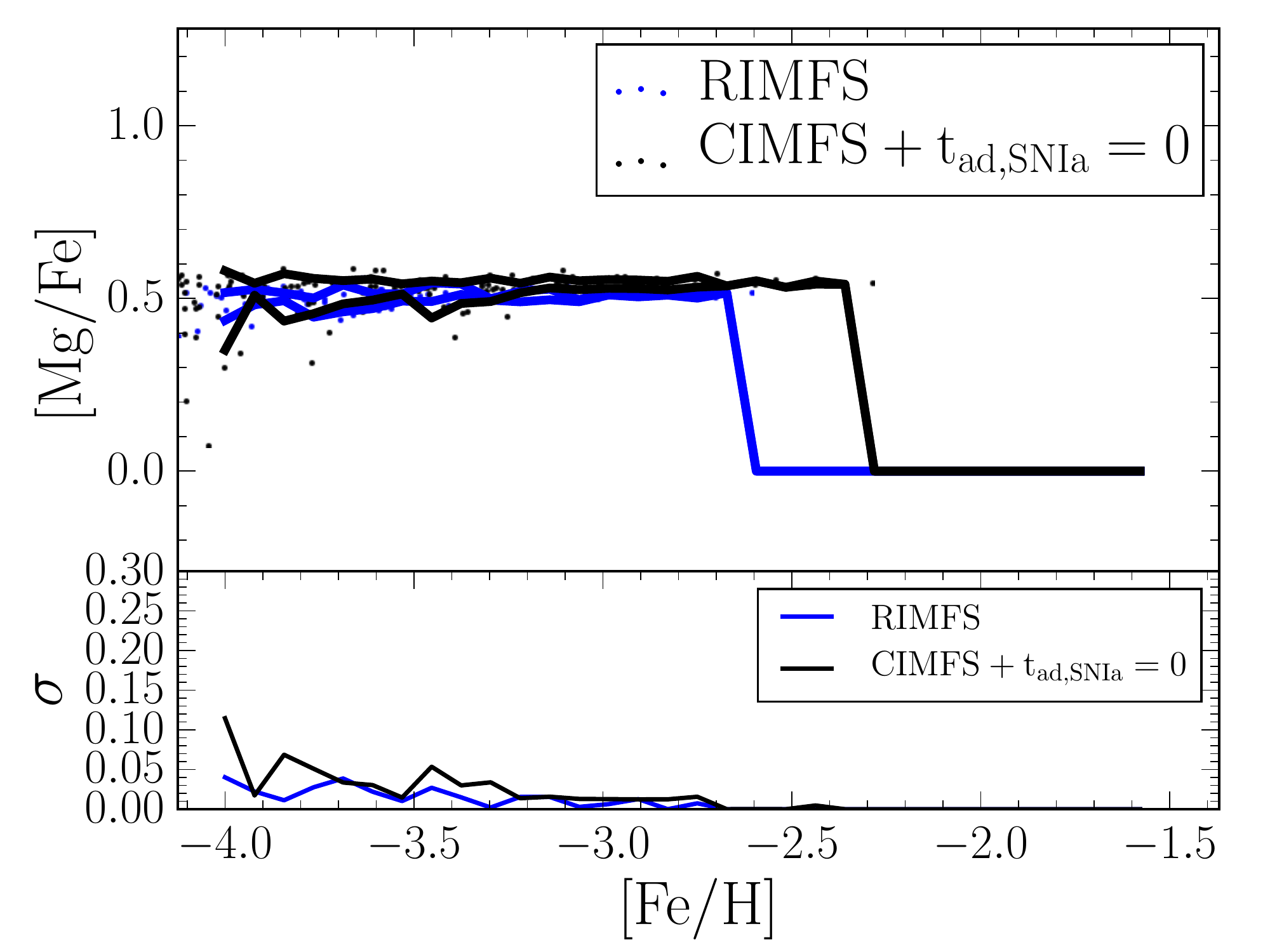}}}
  \subfigure[RIMFS, CIMFS: $\mathfrak{r}=5$, $m_{\star}=2048\,\rm{M_{\odot}}$]{\resizebox{0.33\hsize}{!}{\includegraphics[angle=0]{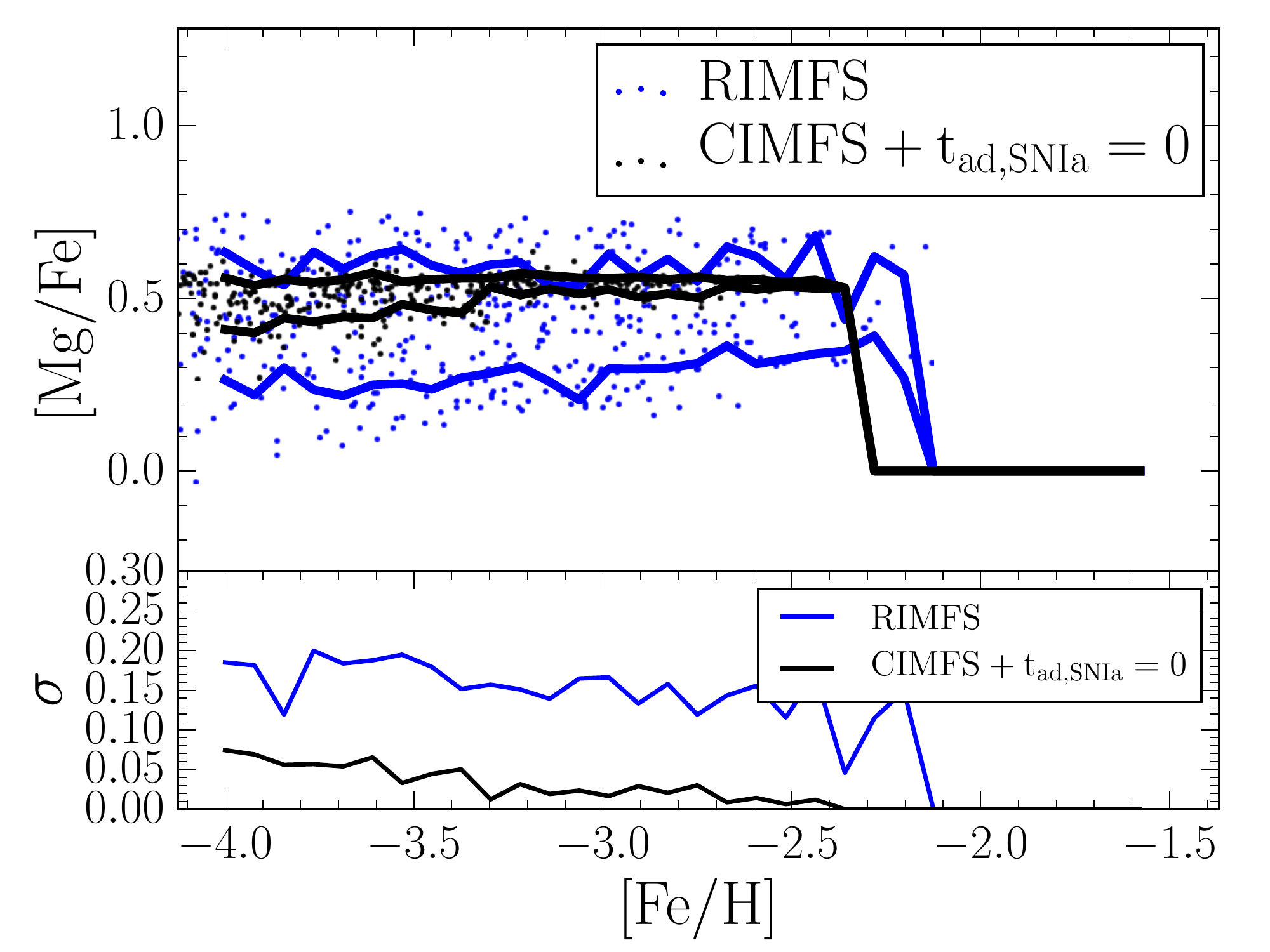}}}
  \subfigure[RIMFS, CIMFS: $\mathfrak{r}=6$, $m_{\star}=1024\,\rm{M_{\odot}}$]{\resizebox{0.33\hsize}{!}{\includegraphics[angle=0]{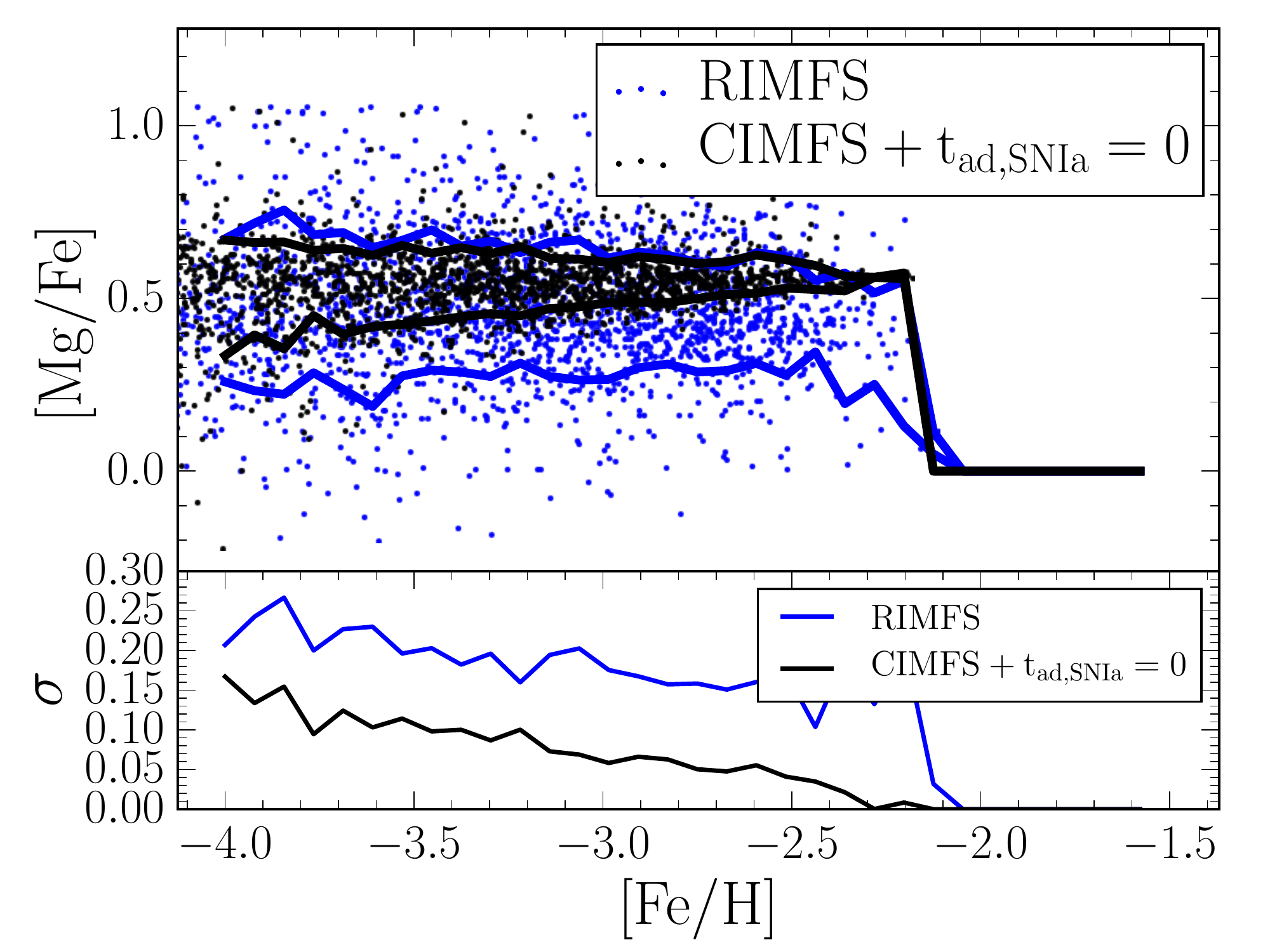}}}
  \subfigure[RIMFS, OIMFS: $\mathfrak{r}=4$, $m_{\star}=4096\,\rm{M_{\odot}}$]{\resizebox{0.33\hsize}{!}{\includegraphics[angle=0]{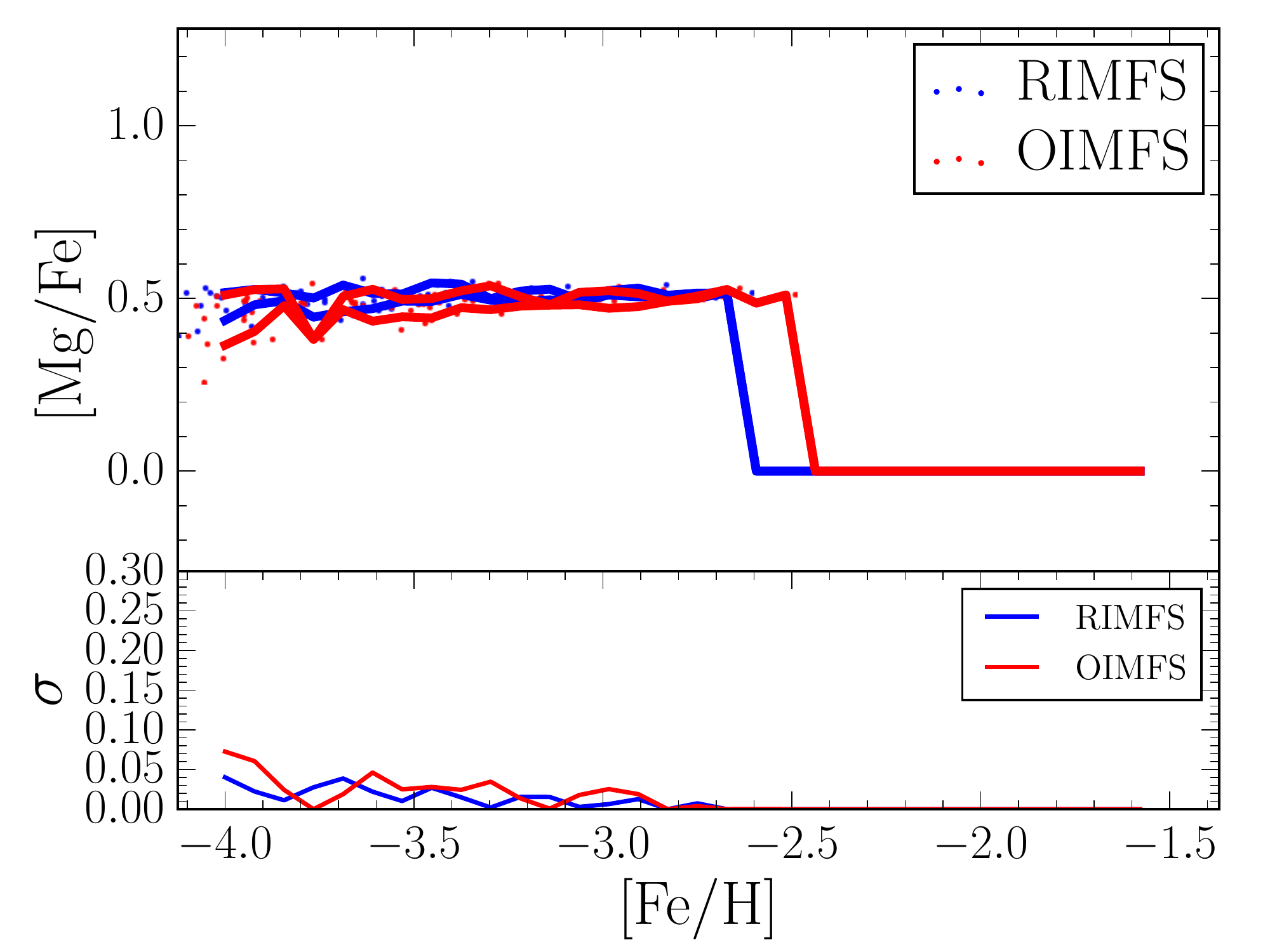}}}
  \subfigure[RIMFS, OIMFS: $\mathfrak{r}=5$, $m_{\star}=2048\,\rm{M_{\odot}}$]{\resizebox{0.33\hsize}{!}{\includegraphics[angle=0]{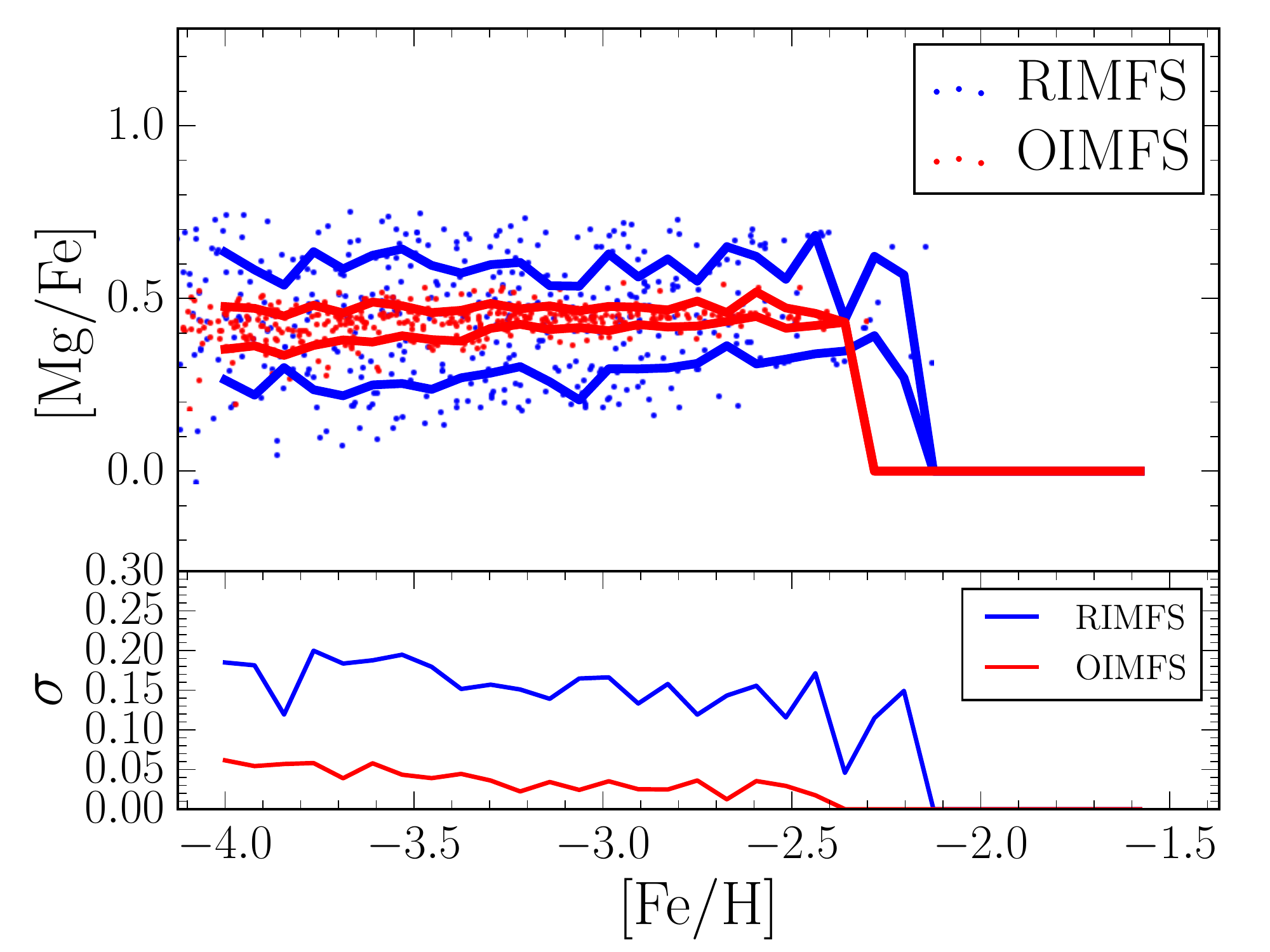}}}
  \subfigure[RIMFS, OIMFS: $\mathfrak{r}=6$, $m_{\star}=1024\,\rm{M_{\odot}}$]{\resizebox{0.33\hsize}{!}{\includegraphics[angle=0]{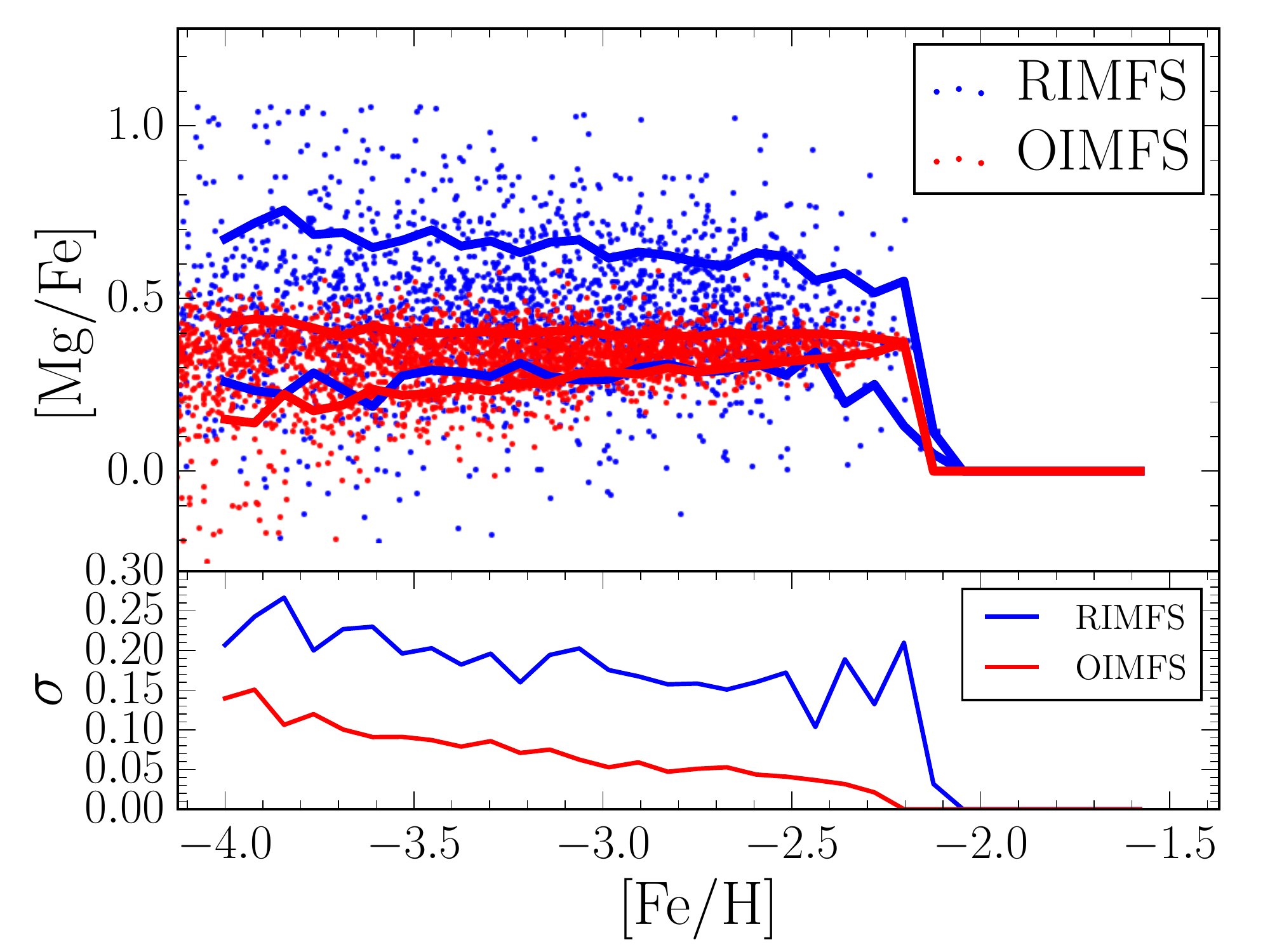}}}
\caption{Top: [Mg/Fe] for the gas component as a function of [Fe/H] for the Sextans model, at $t=1\,\rm{Gyr}$.
On top, the RIMFS scheme is compared to the CIMFS while on the bottom it is compared to the OIMFS scheme.
From left to right, the panels show the effect of the resolution increase. Bottom : the 1-$\sigma$ dispersion of the upper distributions.
  }
  \label{fig:MgFe_RIMFSvsCIMFS_Sex}
\end{figure*}

A reliable comparison between the different methods is not straightforward, as they all generate slightly different star formation histories that alter the final abundances.
We firstly compare the models at $1\,\rm{Gyr}$ when the amount of stars formed in the Sextans models are all similar, according to the bottom of Figure~\ref{fig:IMFsfr}.
In Figure~\ref{fig:MgFe_RIMFSvsCIMFS_Sex}, we show the magnesium abundance of the gas ([Mg/Fe]), our tracer of the $\alpha$-elements, as a function of the metallicity of the gas ([Fe/H]).
In Figure~\ref{fig:IMFsfr}, we indicate the three different IMF sampling methods and the three resolutions.
In each panel, a point corresponds
to one gaseous particle where the ratios [Mg/Fe] and [Fe/H] is computed with the smooth metallicity scheme (Eq.~\ref{XHsmoothed}). 
On top of that, the continuous and dashed black lines indicate the 1$\sigma$ dispersion of each sample. The values of the 1$\sigma$ dispersions are also reported on the bottom of the plot
for a direct comparison between the models. 

At low resolution ($\mathfrak{r}= 4$), all models display equivalent distributions with very low dispersions (1$\sigma<0.1$).
This arises because, at such low resolution, the neighbouring particles of a stellar particle cover a large volume, subsequently, ejecta are efficiently spread over the entire galaxy resulting in an excellent mixing. As expected, if the yields of all type II supernovae are mixed altogether, 
a mean value of $\rm{[Mg/Fe]}=0.5$ is obtained, forming a clear plateau. 

With increasing resolution, the dispersion increases and the differences between RIMFS, OIMFS and $\rm{CIMFS}+t_{\rm{ad,SNIa}}=0$ become clear.
Firstly, it is clear that in all cases, the dispersion decreases with increasing [Fe/H]. This is the direct consequence of the internal mixing that each particle experiments when it receives metals from successive supernovae.
Secondly, the plots clearly show that the RIMFS scheme always generates dispersions about a factor of two larger than both the OIMFS and the $\rm{CIMFS}+t_{\rm{ad,SNIa}}=0$ models.
The mixing is obviously stronger for the $\rm{CIMFS}+t_{\rm{ad,SNIa}}=0$ because, in this particular scheme, at every timestep,
each stellar particle that is older than the shortest star lifetime spreads ejecta.
A neighbouring gaseous particle is therefore more likely to receive ejecta from the whole range of stellar masses considered in the IMF.
This is in contrast with the RIMFS scheme where extreme cases, such as only a single $50\,\rm{M_{\odot}}$ SNe exploding for a stellar particle, may occur. In such a case, neighbouring gaseous particles are biased towards very high [Mg/Fe] ratios.
Similar cases such as low-mass supernovae exploding would bias the neighbouring particles towards low [Mg/Fe] ratios. The final dispersion is then larger for RIMFS than for the other schemes.

The OIMFS scheme, despite its discrete nature, aims to sample the IMF in a more continuous way, ensuring that a wide range of supernova masses are represented.
The dispersion obtained is indeed very similar to the $\rm{CIMFS}+t_{\rm{ad,SNIa}}=0$ dispersion.
However, this method suffers from an important bias. As described in Section~\ref{optimal_sampling} and \ref{implicit_limit}, for a given resolution ($m_{\star,\mathfrak{r}}=M_{\rm{SSP}}$),
an upper local stellar mass $m_{\rm{max}}$ exists, and stars with higher masses are ignored.
For the models considered here, $\mathfrak{r}=4,5,6$, the corresponding maximal masses are respectively $40.6$, $34.5$, and $26.9\,\rm{M_{\odot}}$.
Hence, with increasing resolution the OIMFS scheme misses the ejecta of
the most massive stars, which are those with a high $\alpha$-element ratio. As clearly seen for the
$\mathfrak{r}=6$ model, the low-metallicity plateau is offset towards lower [Mg/Fe], with respect
to the $\rm{CIMFS}+t_{\rm{ad,SNIa}}=0$ and RIMF schemes.

%

%% too dense and sfr less accurate
%\begin{figure*}  
%	\subfigure[RIMFS, CIMFS: $\mathfrak{r}=4$, $m_{\star}=4096\,\rm{M_{\odot}}$]{\resizebox{0.33\hsize}{!}{\includegraphics[angle=0]{./graphs/MgFe_RIMFSvsCIMFS/r2-RIMFS-CIMFS_0022-FNX.pdf}}}
%	\subfigure[RIMFS, CIMFS: $\mathfrak{r}=5$, $m_{\star}=2048\,\rm{M_{\odot}}$]{\resizebox{0.33\hsize}{!}{\includegraphics[angle=0]{./graphs/MgFe_RIMFSvsCIMFS/r4-RIMFS-CIMFS_0022-FNX.pdf}}}
%	\subfigure[RIMFS, CIMFS: $\mathfrak{r}=6$, $m_{\star}=1024\,\rm{M_{\odot}}$]{\resizebox{0.33\hsize}{!}{\includegraphics[angle=0]{./graphs/MgFe_RIMFSvsCIMFS/r8-RIMFS-CIMFS_0022-FNX.pdf}}}
%	\subfigure[RIMFS, OIMFS: $\mathfrak{r}=4$, $m_{\star}=4096\,\rm{M_{\odot}}$]{\resizebox{0.33\hsize}{!}{\includegraphics[angle=0]{./graphs/MgFe_RIMFSvsCIMFS/r2-RIMFS-OIMFS_0022-FNX.pdf}}}
%	\subfigure[RIMFS, OIMFS: $\mathfrak{r}=5$, $m_{\star}=2048\,\rm{M_{\odot}}$]{\resizebox{0.33\hsize}{!}{\includegraphics[angle=0]{./graphs/MgFe_RIMFSvsCIMFS/r4-RIMFS-OIMFS_0022-FNX.pdf}}}
%	\subfigure[RIMFS, OIMFS: $\mathfrak{r}=6$, $m_{\star}=1024\,\rm{M_{\odot}}$]{\resizebox{0.33\hsize}{!}{\includegraphics[angle=0]{./graphs/MgFe_RIMFSvsCIMFS/r8-RIMFS-OIMFS_0022-FNX.pdf}}}
%	\caption{Top: [Mg/Fe] for the gas component as a function of [Fe/H] for the Fornax model, at $t=1\,\rm{Gyr}$.
%		On top, the RIMFS scheme is compared to the CIMFS while on the bottom it is compared to the OIMFS one.
%		From left to right the panels show the effect of the resolution increase. Bottom : the 1-$\sigma$ dispersion of the upper distributions.
%	}
%	\label{fig:MgFe_RIMFSvsCIMFS_Fnx_Gas}
%\end{figure*}
%

\begin{figure*}  
  \subfigure[RIMFS, CIMFS: $\mathfrak{r}=4$, $m_{\star}=4096\,\rm{M_{\odot}}$]{\resizebox{0.33\hsize}{!}{\includegraphics[angle=0]{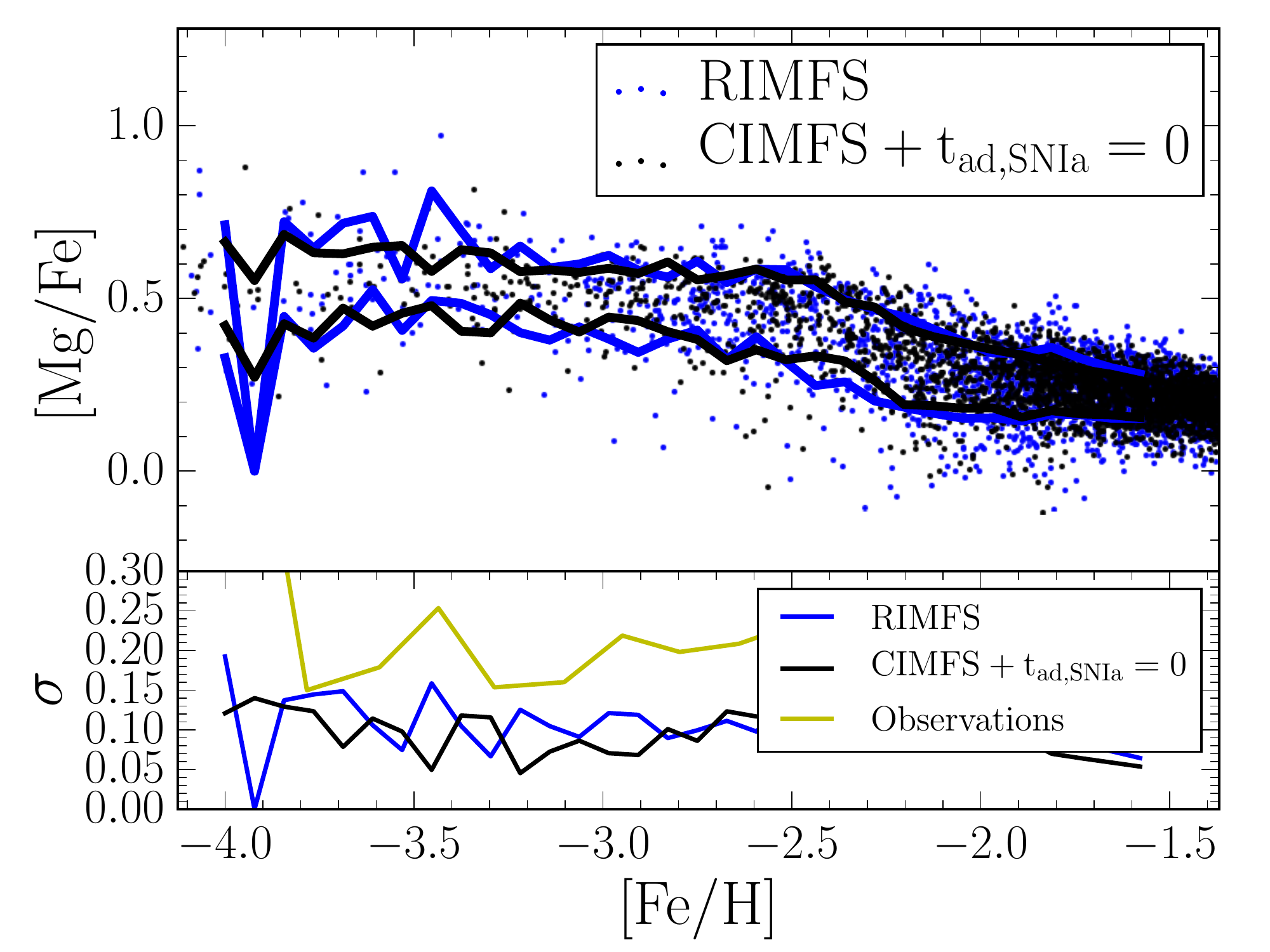}}}
  \subfigure[RIMFS, CIMFS: $\mathfrak{r}=5$, $m_{\star}=2048\,\rm{M_{\odot}}$]{\resizebox{0.33\hsize}{!}{\includegraphics[angle=0]{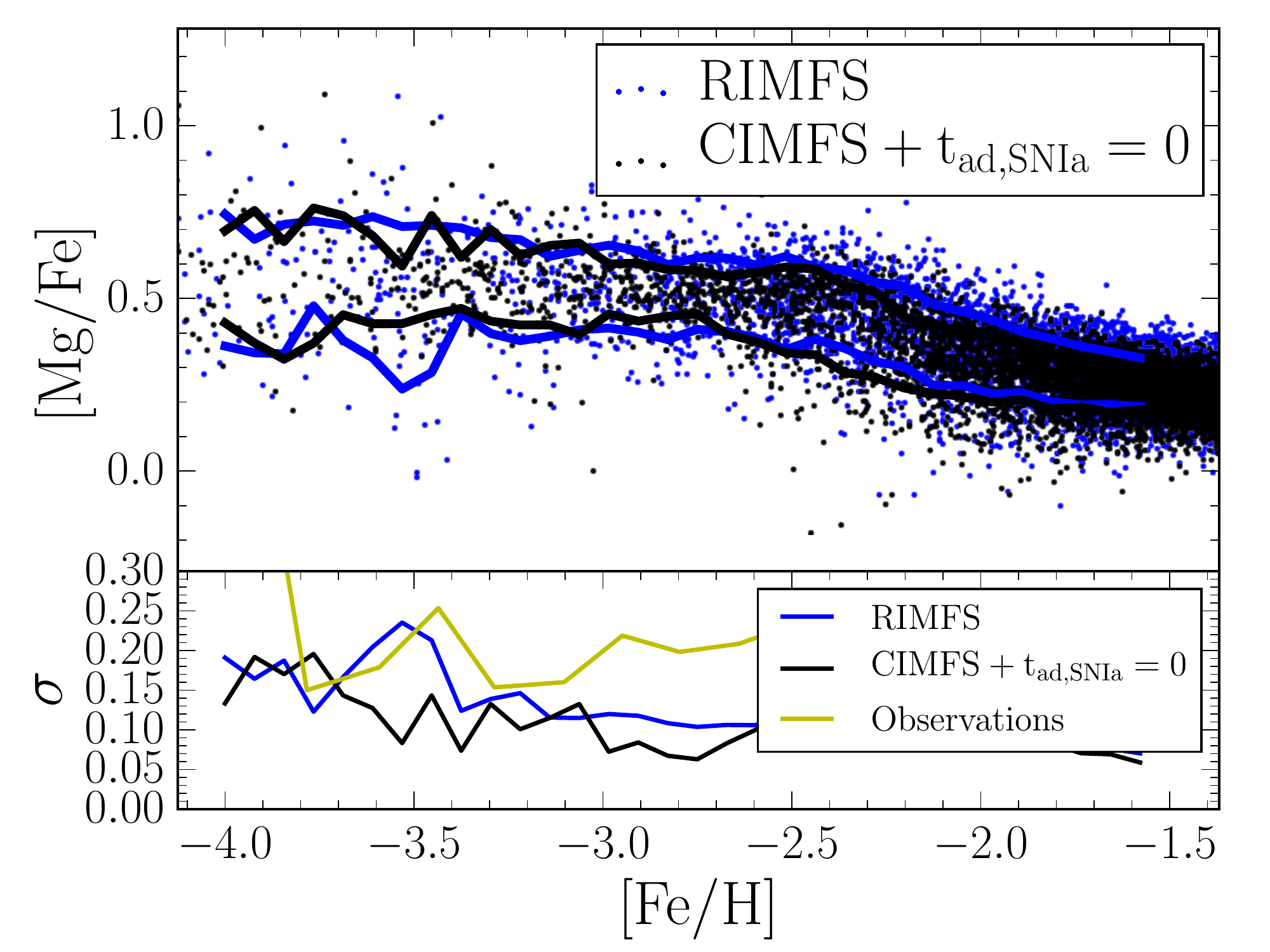}}}
  \subfigure[RIMFS, CIMFS: $\mathfrak{r}=6$, $m_{\star}=1024\,\rm{M_{\odot}}$]{\resizebox{0.33\hsize}{!}{\includegraphics[angle=0]{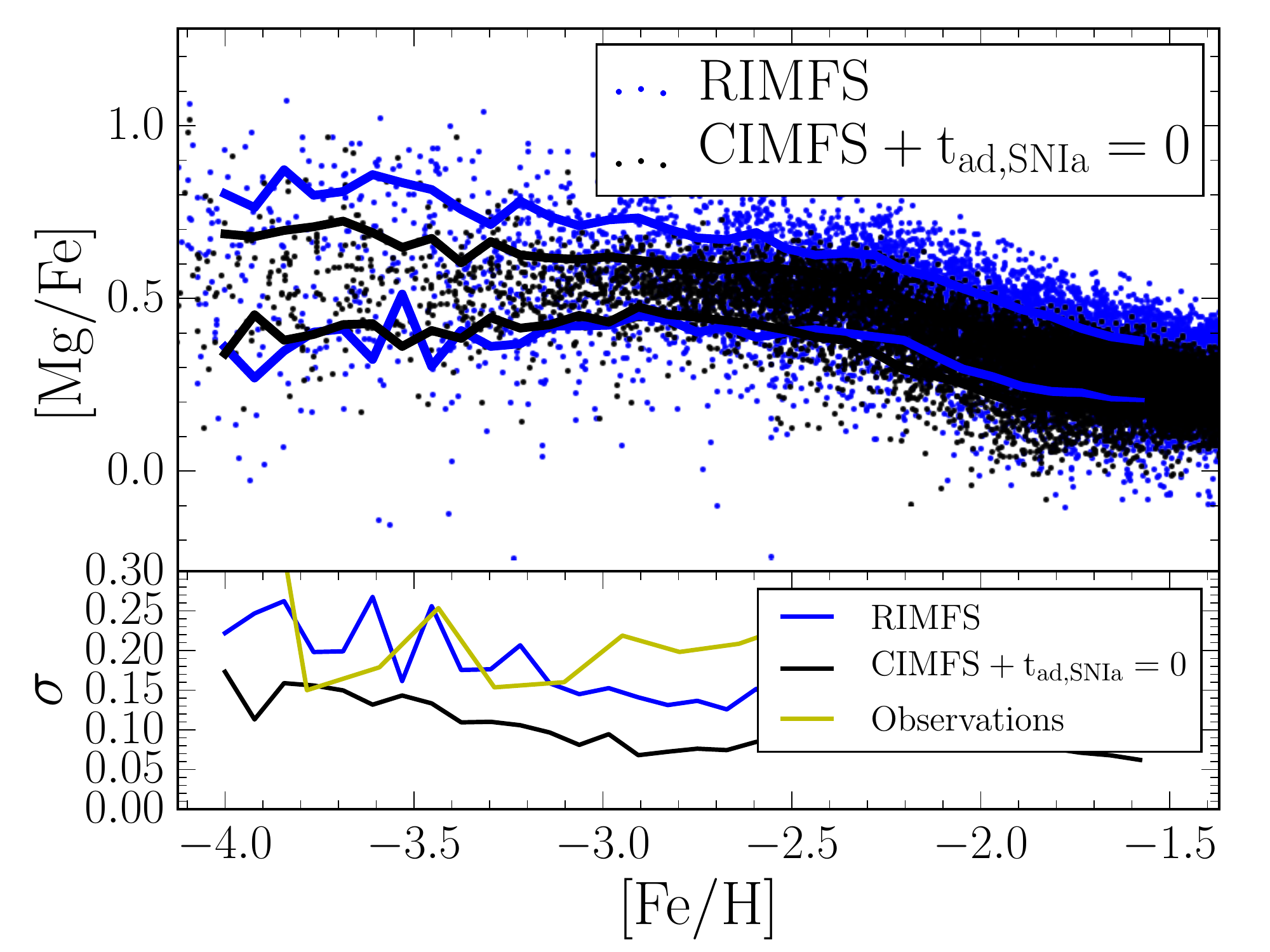}}}
  \subfigure[RIMFS, OIMFS: $\mathfrak{r}=4$, $m_{\star}=4096\,\rm{M_{\odot}}$]{\resizebox{0.33\hsize}{!}{\includegraphics[angle=0]{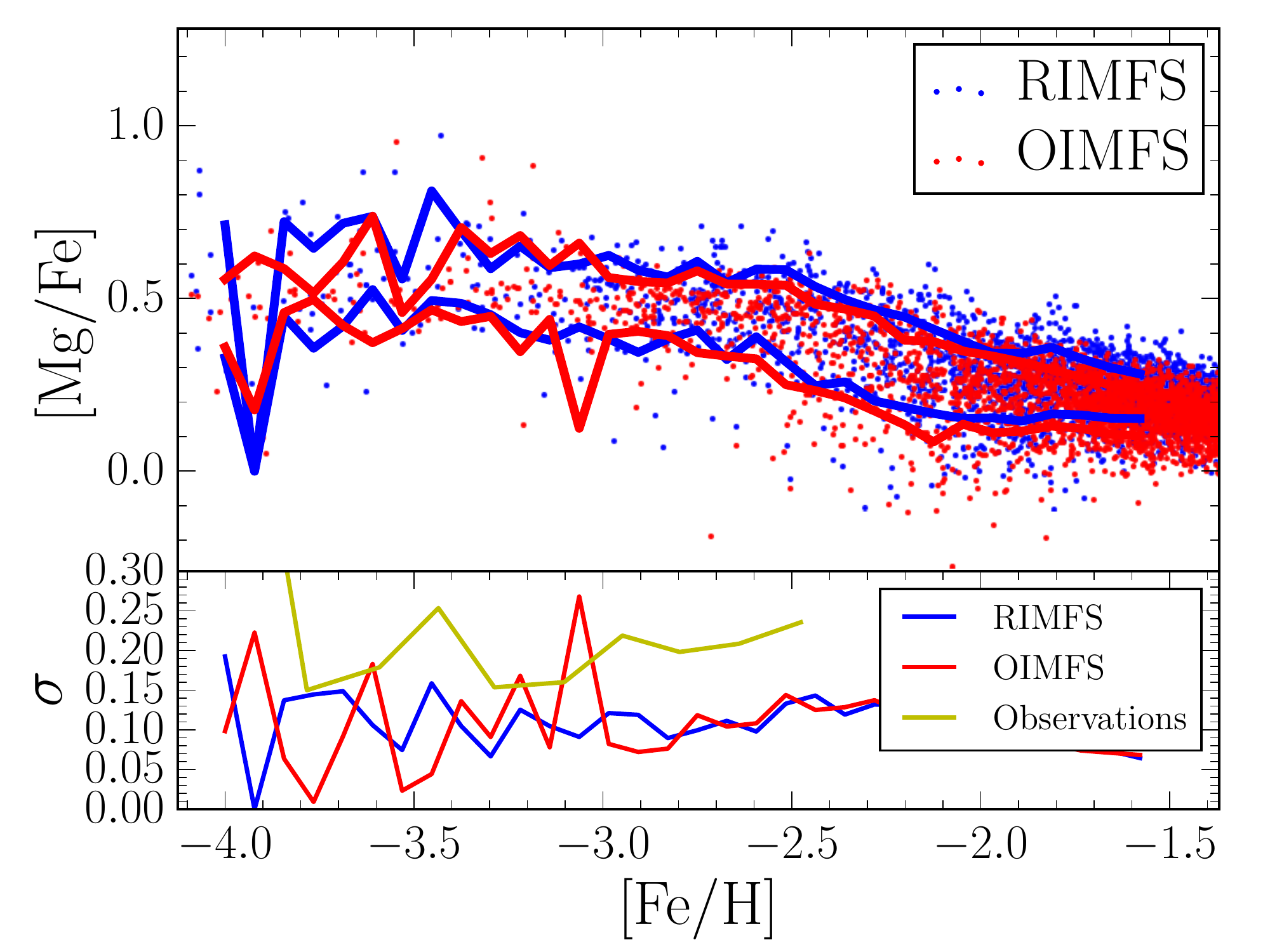}}}
  \subfigure[RIMFS, OIMFS: $\mathfrak{r}=5$, $m_{\star}=2048\,\rm{M_{\odot}}$]{\resizebox{0.33\hsize}{!}{\includegraphics[angle=0]{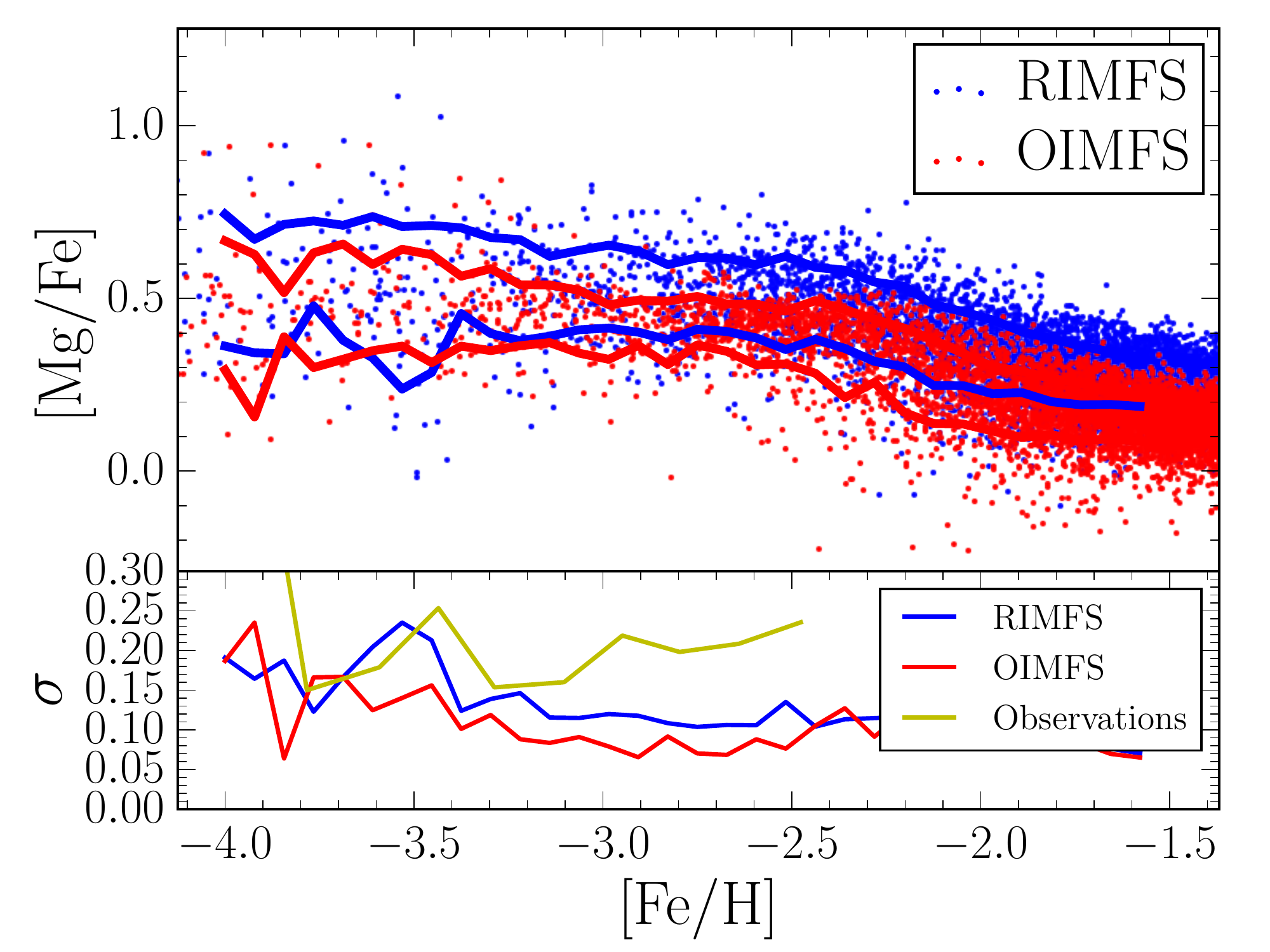}}}
  \subfigure[RIMFS, OIMFS: $\mathfrak{r}=6$, $m_{\star}=1024\,\rm{M_{\odot}}$]{\resizebox{0.33\hsize}{!}{\includegraphics[angle=0]{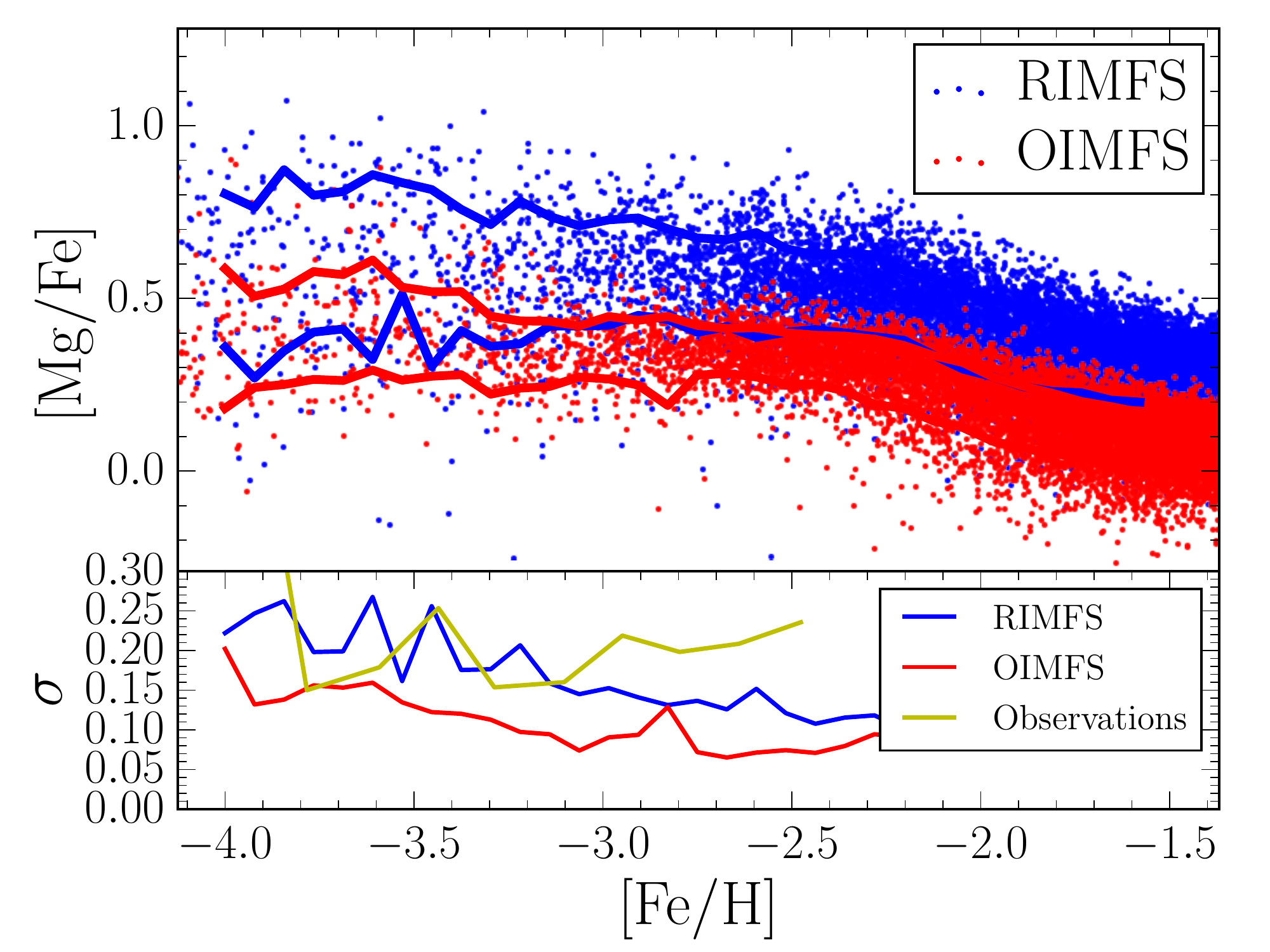}}}
\caption{Stars [Mg/Fe] as a function of [Fe/H] for the Fornax model, at $t=14\,\rm{Gyr}$.
On top, the RIMFS scheme is compared to the CIMFS, while on the bottom it is compared to the OIMFS scheme.
From left to right, the panels show the effect of the increase in resolution. The yellow curve shown on the bottom of each plot 
corresponds to the observed dispersion in [Mg/Fe] extracted from all stars of Figure~\ref{fig:EMPS}.
  }
  \label{fig:MgFe_RIMFSvsCIMFS_Fnx}
\end{figure*}

As observations only view the stars as they are today, we show in Figure~\ref{fig:MgFe_RIMFSvsCIMFS_Fnx} the [Mg/Fe] ratio for the Fornax models, at $t=14\,\rm{Gyr}$. 
Despite the fact that the quantity of stars formed in these models is different, owing to different star formation rates, 
these plots allow us to appreciate the effect of the different IMF sampling schemes on the final abundances. 
For comparison, the observed dispersion in [Mg/Fe] extracted from all stars of Figure~\ref{fig:EMPS} is shown as a yellow curve on the bottom of each plot.
While the trends discussed in the gas abundances are less pronounced in the stars, they all remain true:
(i) At low resolutions, the metallicity dispersion is low (even lower than that observed) and the three methods converge;
(ii) at higher resolutions, the dispersion in the RIMFS scheme is always larger than the $\rm{CIMFS}+t_{\rm{ad,SNIa}}=0$ and OIMFS schemes; and
(iii) the [Mg/Fe] plateau is offset owing to the lack of massive supernovae in the OIMFS case.

%

% lack of resolution
%\begin{figure*}  
%	\subfigure[RIMFS, CIMFS: $\mathfrak{r}=4$, $m_{\star}=4096\,\rm{M_{\odot}}$]{\resizebox{0.33\hsize}{!}{\includegraphics[angle=0]{./graphs/MgFe_RIMFSvsCIMFS/r2-RIMFS-CIMFS_0300-SEX.pdf}}}
%	\subfigure[RIMFS, CIMFS: $\mathfrak{r}=5$, $m_{\star}=2048\,\rm{M_{\odot}}$]{\resizebox{0.33\hsize}{!}{\includegraphics[angle=0]{./graphs/MgFe_RIMFSvsCIMFS/r4-RIMFS-CIMFS_0300-SEX.pdf}}}
%	\subfigure[RIMFS, CIMFS: $\mathfrak{r}=6$, $m_{\star}=1024\,\rm{M_{\odot}}$]{\resizebox{0.33\hsize}{!}{\includegraphics[angle=0]{./graphs/MgFe_RIMFSvsCIMFS/r8-RIMFS-CIMFS_0300-SEX.pdf}}}
%	\subfigure[RIMFS, OIMFS: $\mathfrak{r}=4$, $m_{\star}=4096\,\rm{M_{\odot}}$]{\resizebox{0.33\hsize}{!}{\includegraphics[angle=0]{./graphs/MgFe_RIMFSvsCIMFS/r2-RIMFS-OIMFS_0300-SEX.pdf}}}
%	\subfigure[RIMFS, OIMFS: $\mathfrak{r}=5$, $m_{\star}=2048\,\rm{M_{\odot}}$]{\resizebox{0.33\hsize}{!}{\includegraphics[angle=0]{./graphs/MgFe_RIMFSvsCIMFS/r4-RIMFS-OIMFS_0300-SEX.pdf}}}
%	\subfigure[RIMFS, OIMFS: $\mathfrak{r}=6$, $m_{\star}=1024\,\rm{M_{\odot}}$]{\resizebox{0.33\hsize}{!}{\includegraphics[angle=0]{./graphs/MgFe_RIMFSvsCIMFS/r8-RIMFS-OIMFS_0300-SEX.pdf}}}
%	\caption{The stars [Mg/Fe] as a function of [Fe/H] for the Sextans model, at $t=14\,\rm{Gyr}$.
%		On top, the RIMFS scheme is compared to the CIMFS while on the bottom it is compared to the OIMFS one.
%		From left to right the panels show the effect of the increase in resolution.
%	}
%	\label{fig:MgFe_RIMFSvsCIMFS_Sex_Stars}
%\end{figure*}
%%

\subsubsection{Convergence}\label{convergence}

\begin{figure}  
  \subfigure[RIMFS]{\resizebox{1.0\hsize}{!}{\includegraphics[angle=0]{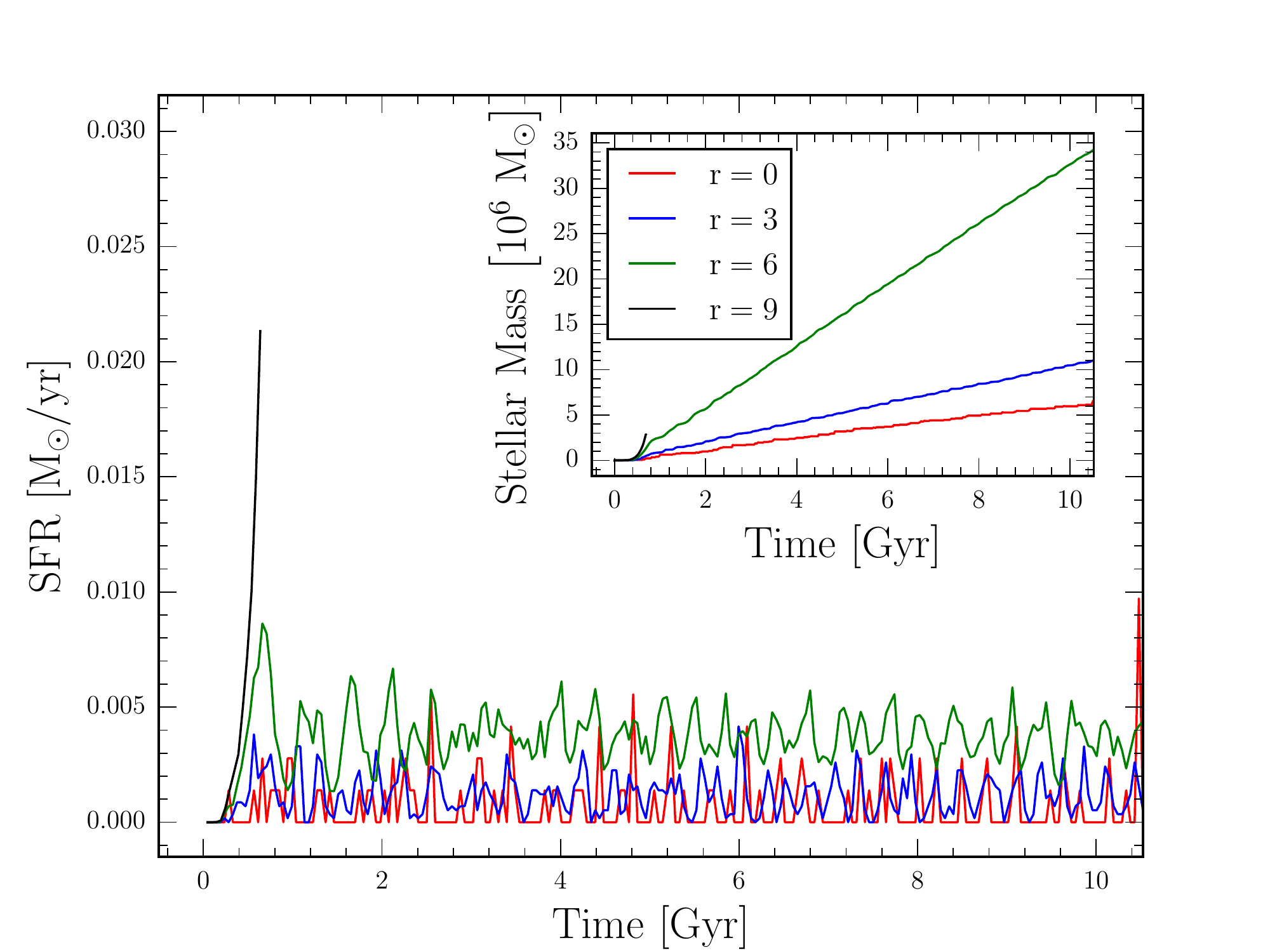}}}
  \subfigure[$\rm{CIMFS}+t_{\rm{ad,SNIa}}=0$]{\resizebox{1.0\hsize}{!}{\includegraphics[angle=0]{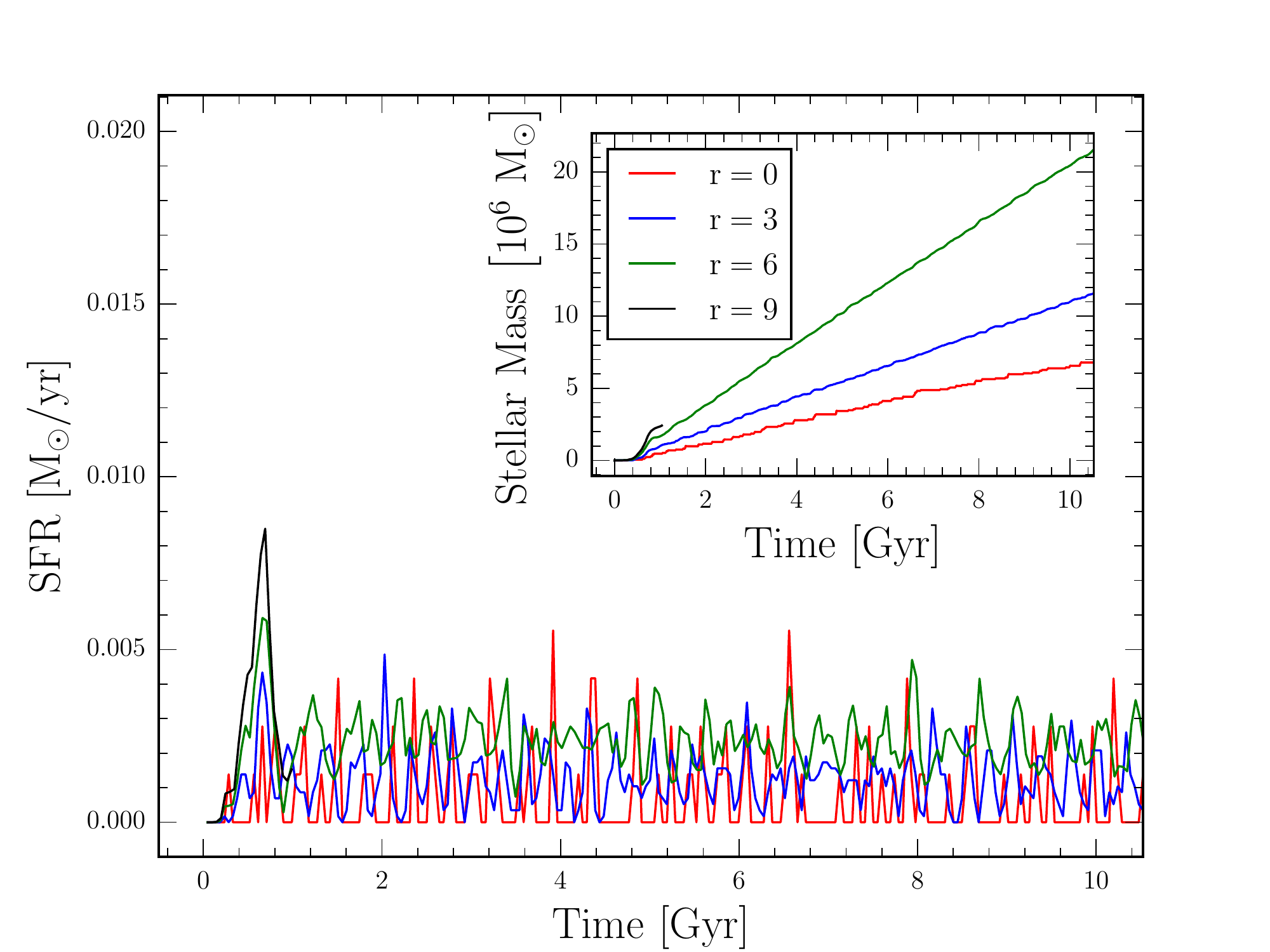}}}
\caption{Star formation and cumulative stellar mass as a function of time, for the Fornax models with different resolution, from 
  $\mathfrak{r}= 0$ to $9$. 
  }
  \label{fig:SFRConvergence}
\end{figure}

We briefly discuss here some convergence issues related to the resolution.
Figure~\ref{fig:SFRConvergence} shows the star formation and cumulative stellar mass for the Fornax model with resolutions increasing from $\mathfrak{r}= 0$ to $9$, i.e, covering two dex in terms of mass 
resolution. Here, no convergence is obtained with the RIMFS scheme\footnote{
As the divergence after a few hundreds of Myr is obvious and as $\mathfrak{r}= 9$ simulations are strongly CPUs consuming, 
we stopped the latter and did not run them up to $14\,\rm{Gyr}$.}, which primarily occurs for two reasons.

Firstly, for the high-resolution model ($\mathfrak{r}= 9$), the mass of one stellar particle is only $128\,\rm{M_{\odot}}$ inevitably
causing bias in the randomly generated IMF as described in Section~\ref{random_sampling}.
We found that the RIMFS simulation with a resolution lower that $\mathfrak{r}= 6$ generates precisely $87\times 10^{-4}$ SNeII per solar mass formed, in agreement with the choice
of the IMF. For the high-resolution model, this value can drop to $75\times 10^{-4}$ owing to incomplete sampling.
Consequently, the feedback associated with high-resolution simulations is smaller, fostering a higher
star formation rate.
The divergence between models is considerably smaller for the $\rm{CIMFS}+t_{\rm{ad,SNIa}}=0$ case. 
There, according to the definition of the CIMFS, the number of SNeII per solar mass formed is always precisely the same,
independent of the resolution.
In addition, energy is released at every timestep, preventing the runaway star formation.

Secondly, because the SPH radius is smaller for higher resolutions, energy is deposited into a smaller volume of gas, preventing the (temporary) shutdown in star formation over large sections of the galaxy seen in low-resolution models.

\subsection{Reproducing the scatter in [$\alpha$/Fe]}
%\subsubsection{Metal mixing}

A fundamental point when simulating the chemical evolution of a galaxy is to correctly reproduce the scatter in the element abundance ratios.
In this section we study the impact on the dSph chemical evolution of 
the elements spreading scheme discussed in Section~\ref{feedback} along with 
the two mixing methods introduced in Section~\ref{diffusion},
namely the smooth metallicity \citep{wiersma2009} and  diffusion process based on velocity dispersions \citep{greif2009}.
%and compare the case with no mixing processes.

\subsubsection{The scatter without mixing}\label{scatter_nomixing}

We first discuss our models in the absence of any additional mixing, by focusing on the final stellar [Mg/Fe] vs [Fe/H] distributions for Fornax and Sextans.
%Here, and for the rest of the discussion, we will only use the RIMFS scheme as, according to the previous section, it is the most reliable sampling method
%for the resolution considered here ($\mathfrak{r}=6$, $m_{\star}=1024\,\rm{M_{\odot}}$). 
All three IMF sampling methods suffer from different issues at the resolution considered here ($\mathfrak{r}=6$, $m_{\star}=1024\,\rm{M_{\odot}}$). 
The CIMFS artificially boosts the mixing of elements, the OIMFS induces a spurious offset in the $\alpha$-elements abundances ratio, and the RIMFS is
subject to Poissonian noise. We however found the RIMFS scheme to be better suited to study the influence of different numerical techniques on the scatter of elements abundances ratio. Thus for the rest of the discussion, we only use the RIMFS scheme. 
The results are indicated in dark blue on panels (a) and (e) of Figure~\ref{fig:allMgFe}, with observed values overplotted in yellow (from Figure~\ref{fig:EMPS}).
We compare only the low-metallicity end of the stellar distribution ($[\rm{Fe/H}]<-2.5$), where both Milky Way and dSph stars sit on the same high $[\alpha/{\rm Fe}]$ plateau, before they diverge because of differing star formation histories.
This choice allows us to compare the scatter with simulations without running into complications arising from the small statistics in most dwarfs.

At low metallicity in the simulations, the scatter is above 0.4 dex, about 2--3 times higher than that observed in metal-poor stars. 
In particular, stars are found with extreme [Mg/Fe] values, above $0.85$ and below $-0.5$.
As seen by Figure~\ref{fig:EMPS} very few metal-poor stars have been found with subsolar [Mg/Fe]
and only two have values below $-0.5$ \citep{aoki2014,jablonka2015}.
At high metallicity, the scatter decreases as each star has had a higher number of supernovae enriching it, however, many stars are still found with very low $[\alpha/{\rm Fe}]$ ratios.
By examining the extreme stars we find that very low-metallictiy stars are enriched either purely from low-mass $M<20\,\rm{M_{\odot}}$ supernovae or, in rare cases, by a single SNIa.
Meanwhile $\alpha$-rich stars all contain at least the ejecta of one SNII more massive than $25\,\rm{M_{\odot}}$ and few if any below $20\,\rm{M_{\odot}}$.
At higher metallicity, the evolution is more complex but extremely low $\alpha$ abundances at higher metallicity ($\rm{[Fe/H]}>-2$) only occur because of the presence of SNIa ejecta.

In all cases, stars with extreme $\alpha$ abundances result from the pollution of an incomplete IMF.
This is supported by the mean number of SNII supernovae responsible for the abundances in the Sextans model stars: only 13 for low-metallicity stars
and 48 for the more metal-rich but extremely $\alpha$-depleted stars, while the mean values for all stars is about 70. 
%With the exception of the star SDSS J0018-0939 discussed in \citet{aoki2014}, stars with $\alpha$-abundances as low as $-0.5$ have never 
%been observed. 
%
%The existence of those stars and the excessive scatter speaks in favour 
%of a lack of mixing inherent to the SPH technique. 
%%It is finally worse noting that when decreasing the resolution, from $\mathfrak{r}=6$ to $\mathfrak{r}=2$ the scatter increases. 
%The effect of adding some kind of mixing is discussed below.
%
The frequency of these stars in simulations and the excessive scatter point to limitations in the metal ejection scheme and possibly insufficient mixing in the standard SPH technique. 
We examine some proposed improvements of the method below.

\subsubsection{Metal spreading schemes}
\label{dsph_elt_spread}

Simple experiments involving two supernovae (Section~\ref{TwoSNs}) have shown how the scatter in element abundance ratios in the gas may be influenced
by the method used to distribute metals throughout the nearby gas.
Here, we apply the same recipes to our dSph models and check their impact on the final abundances of stars.

With the first scheme explored (the step function distribution $\Gamma$), all neighbouring particles receive the same amount of material, independent of their distance, avoiding any gradient related
to the kernel ($w_{ij}=\Gamma_{ij}$, Figure~\ref{fig:twoSNs} (b)).
The effect on our Fornax simulation is shown in panel (b) of Fig~\ref{fig:allMgFe}.
In contrast to what would be expected from the double supernovae experiments, the scatter is not decreased.
This lack of decrease suggests the scatter does not originate in the kernel, but by stochastic supernovae explosions having different masses and yields,
as discussed in Section~\ref{scatter_nomixing}.
At high metallicity ($\rm{[Fe/H]}>-2.5$), where the stochastic effect is less important, the scatter is slightly lowered.

Panels (c) and (d) show the impact of fixing the ejection radius to either a constant value ($R_{\rm{ej}}=0.125\,\rm{kpc}$) or to the
blast radius ($R_{\rm{ej}}=R_{\rm{E}}$).
None of these two recipes is able to decrease the scatter, which is in agreement that the scatter does not arise because of the choice of the kernel.
In the blast radius case, the effect is even more dramatic, and the scatter is increased at all metallicities.  
The constant ejected radius allows a better mixing compared to the blast radius.
The resulting scatter is similar to the default case, except above $\rm{[Fe/H]}>-2.0$, where it is slightly higher.
In both cases, the mean value of the [Mg/Fe] low-metallicity plateau is too high, lying well above the mean observed value of around 0.5. 

One can understand this through an examination of the supernova test run with a fixed radius ($r=0.125\,\rm{kpc}$) and no mixing (Figure~\ref{fig:twoSNs}). 
Using a fixed ejecta radius, the second supernova is only able to affect particles that still lie in $R_{\rm{ej}}$.
The most distant particles touched by the first massive supernova have moved outwards and are only polluted by $\alpha$-enhanced ejecta. Consequently, these particles form an $\alpha$-enhanced plateau at low metallicity.

\subsubsection{Smooth metallicity}

The smooth metallicity technique is a simple and natural way of mixing elements inside the SPH smoothing kernel.
Its effect, in the context of dSphs, is shown in Figure~\ref{fig:allMgFe} (a) and (e)  for the Fornax and Sextans models, respectively, where the [Mg/Fe] vs [Fe/H] distribution of the stars (in light blue) is compared to the case where no mixing is applied (in dark blue).
In these simulations, the star formation histories are similar, and the final stellar masses are within $1\%$ in the Fornax simulations and to within $3\%$ in the Sextans simulations.
The reason for this small impact is because only the local cooling function of the gas is impacted and its impact on the global dynamics and star formation history is minimal.
This similar star formation history allows us to reliably compare the simulations.
As seen in Figure~\ref{fig:allMgFe} the smooth metallicity scheme, strongly decreases the scatter by a factor of 2 to 3, becoming consistent with the scatter observed in stars (shown in yellow).
In particular, most of the stars with extreme abundances are forced towards the mean, reaching realistic values.

The final metallicity distribution function (MDF) is only slightly modified by the smooth metallicity scheme.
With the most metal-rich particles losing metals and the metal poorest gaining, and this results in a metal-rich cut-off in the MDF, an effect more pronounced in Sextans because 
of the small number of stars formed.
%%
%\begin{figure*}
%  \subfigure[Fornax]{\resizebox{0.5\hsize}{!}{\includegraphics[angle=0]{./graphs/MgFe_FiducialvsSmooth/MgFe_FiducialvsSmooth-FNX.pdf}}}
%  \subfigure[Sextans]{\resizebox{0.5\hsize}{!}{\includegraphics[angle=0]{./graphs/MgFe_FiducialvsSmooth/MgFe_FiducialvsSmooth-SEX.pdf}}}
%  \caption{
%  Comparison of the final [Mg/Fe] dispersion of the stars in the Fornax and Sextans ($\mathfrak{r}=6$) models run with the smooth metallicity scheme (first panel) 
%  and without it (second panel). As in Figure~\ref{fig:MgFe_RIMFSvsCIMFS_Sex} and \ref{fig:MgFe_RIMFSvsCIMFS_Fnx}, the black curves display the $1-\sigma$ dispersion around
%  the mean and is also reported in the third panel. The fourth panel compare the metallicity distribution function.
%  }
%  \label{fig:smooth}
%\end{figure*}
%%

\subsubsection{Metal diffusion}

\begin{figure*}
	
	\subfigure[smooth metallicity (Fornax)]           {\resizebox{0.247\hsize}{!}{\includegraphics[angle=0]{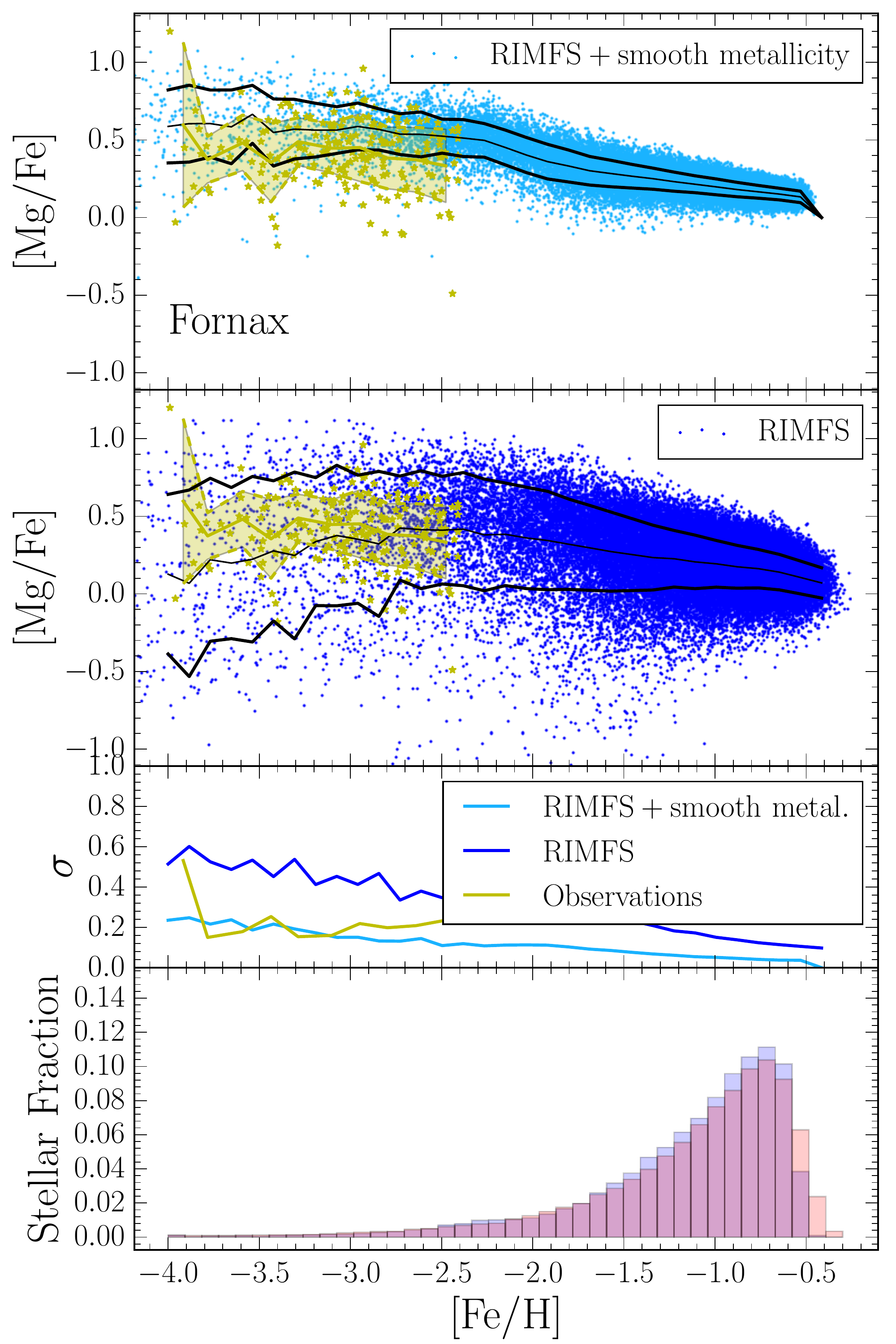}}}
	\subfigure[$w_{ij}=\Gamma_{ij}$ (Fornax)]            {\resizebox{0.247\hsize}{!}{\includegraphics[angle=0]{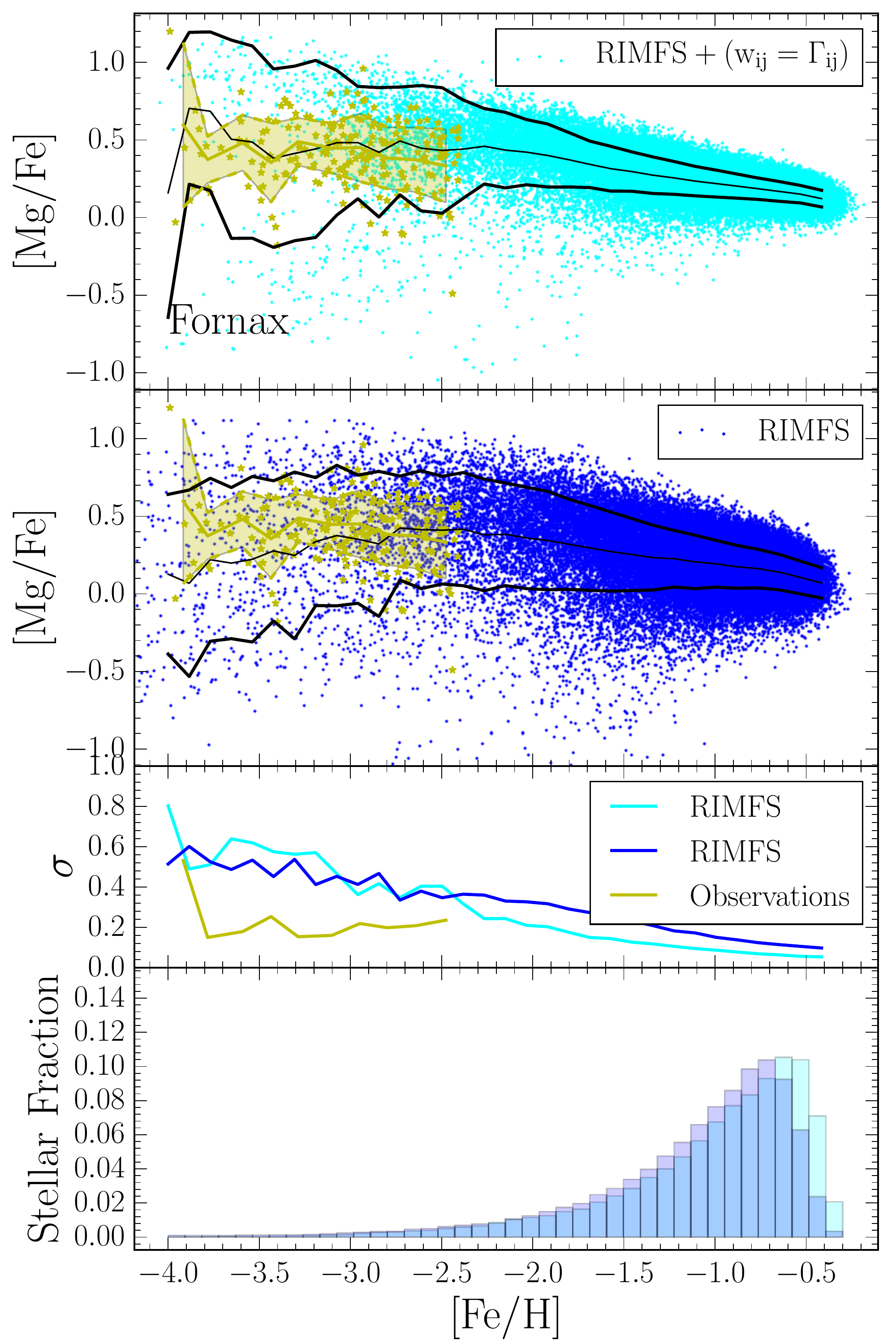}}}
	\subfigure[$R_{\rm{ej}}=0.125\,\rm{kpc}$ (Fornax)]{\resizebox{0.247\hsize}{!}{\includegraphics[angle=0]{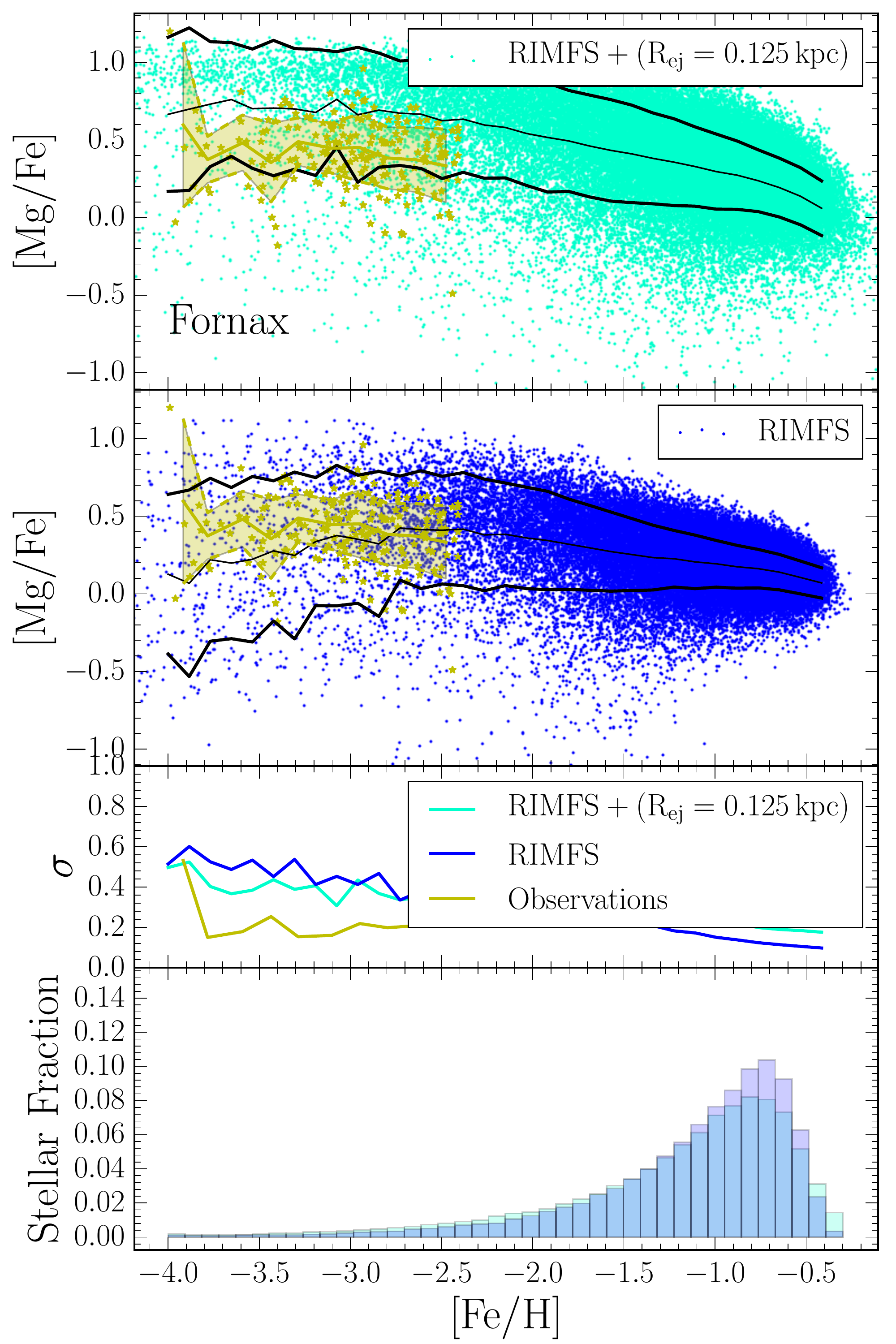}}}
	\subfigure[$R_{\rm{ej}}=R_{\rm{E}}$ (Fornax)]     {\resizebox{0.247\hsize}{!}{\includegraphics[angle=0]{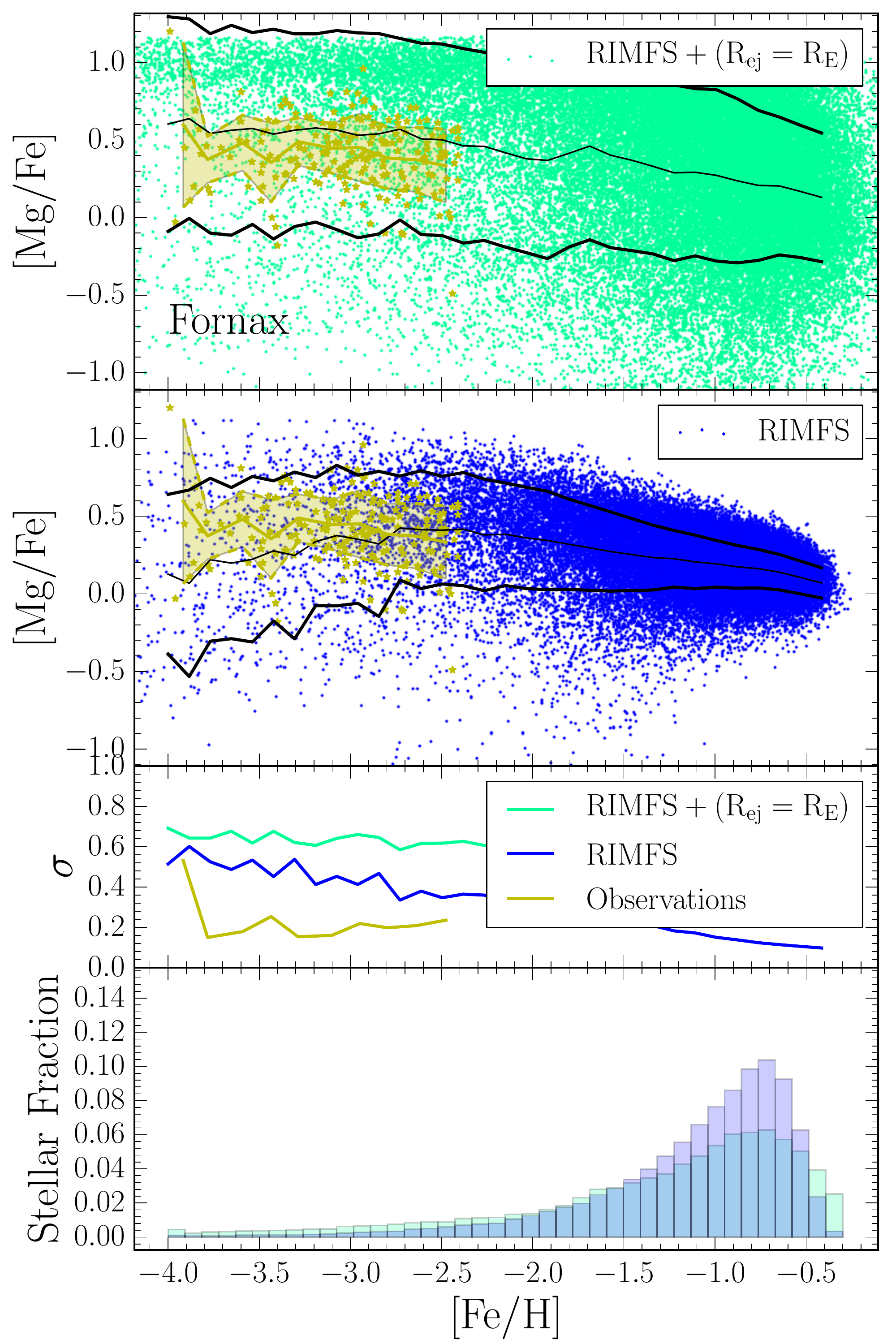}}}	
	\subfigure[smooth metallicity (Sextans)]           {\resizebox{0.247\hsize}{!}{\includegraphics[angle=0]{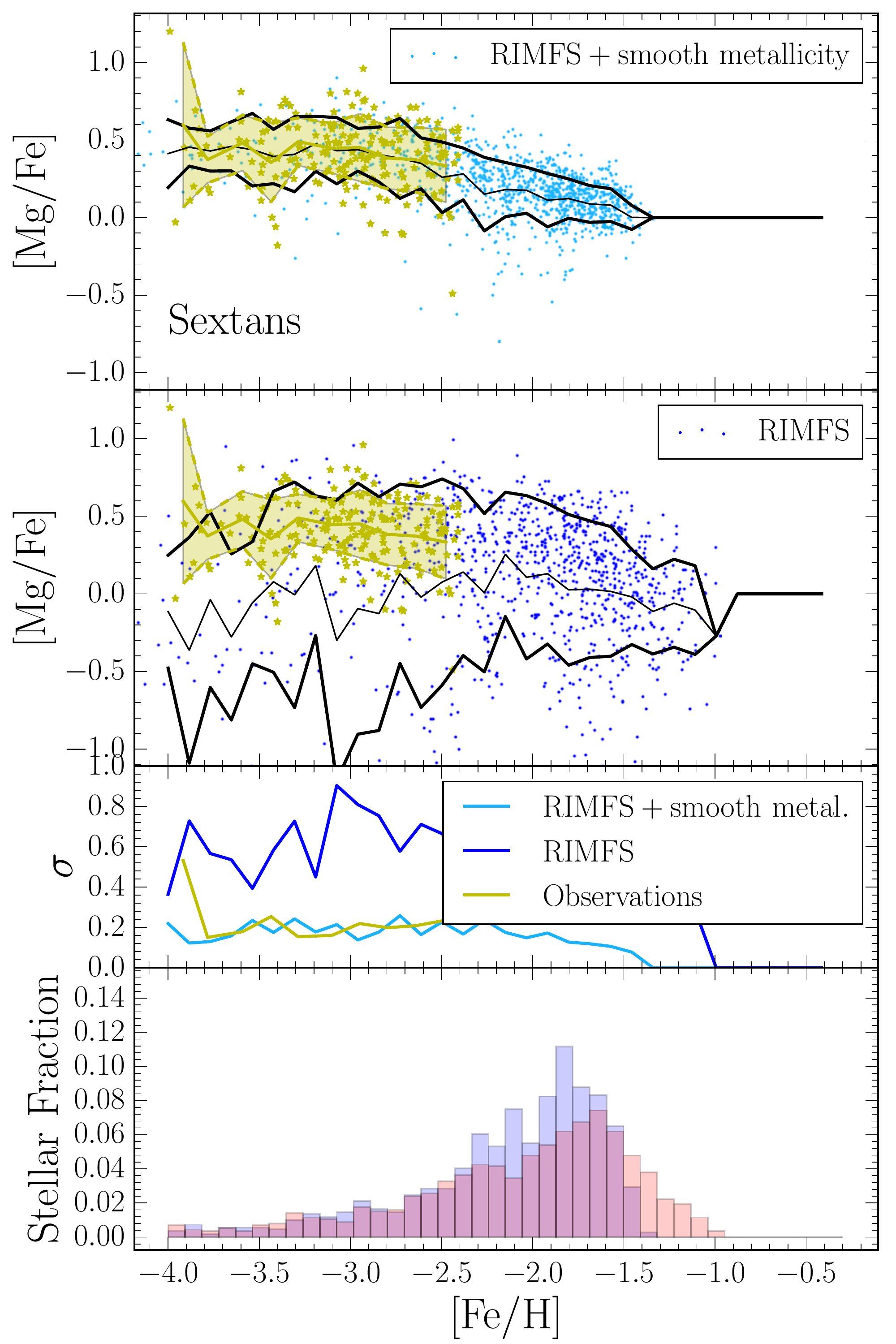}}}	
	\subfigure[diffusion : $d=0.003$ (Fornax)]        {\resizebox{0.247\hsize}{!}{\includegraphics[angle=0]{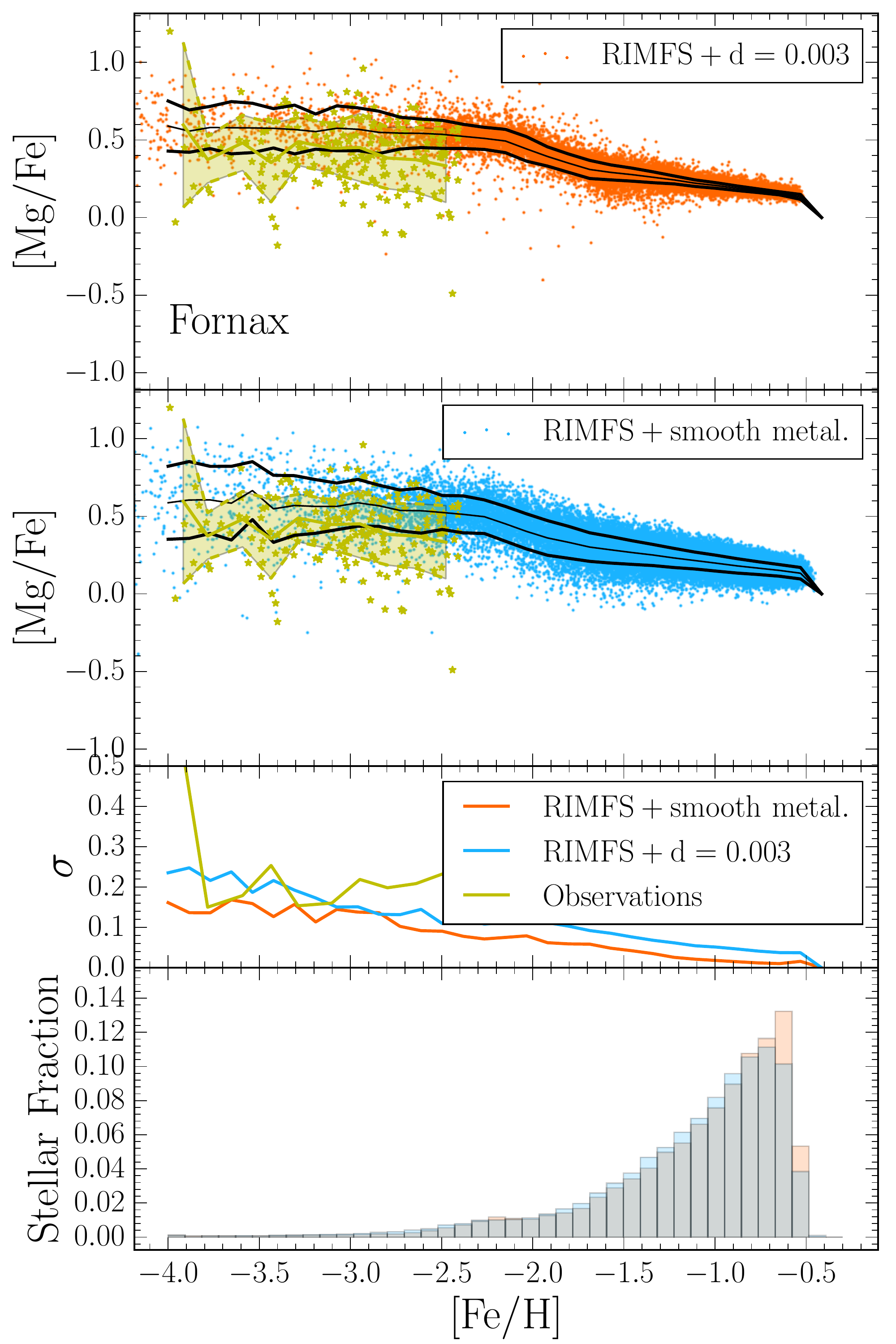}}}
	\subfigure[diffusion : $d=0.001$ (Fornax)]        {\resizebox{0.247\hsize}{!}{\includegraphics[angle=0]{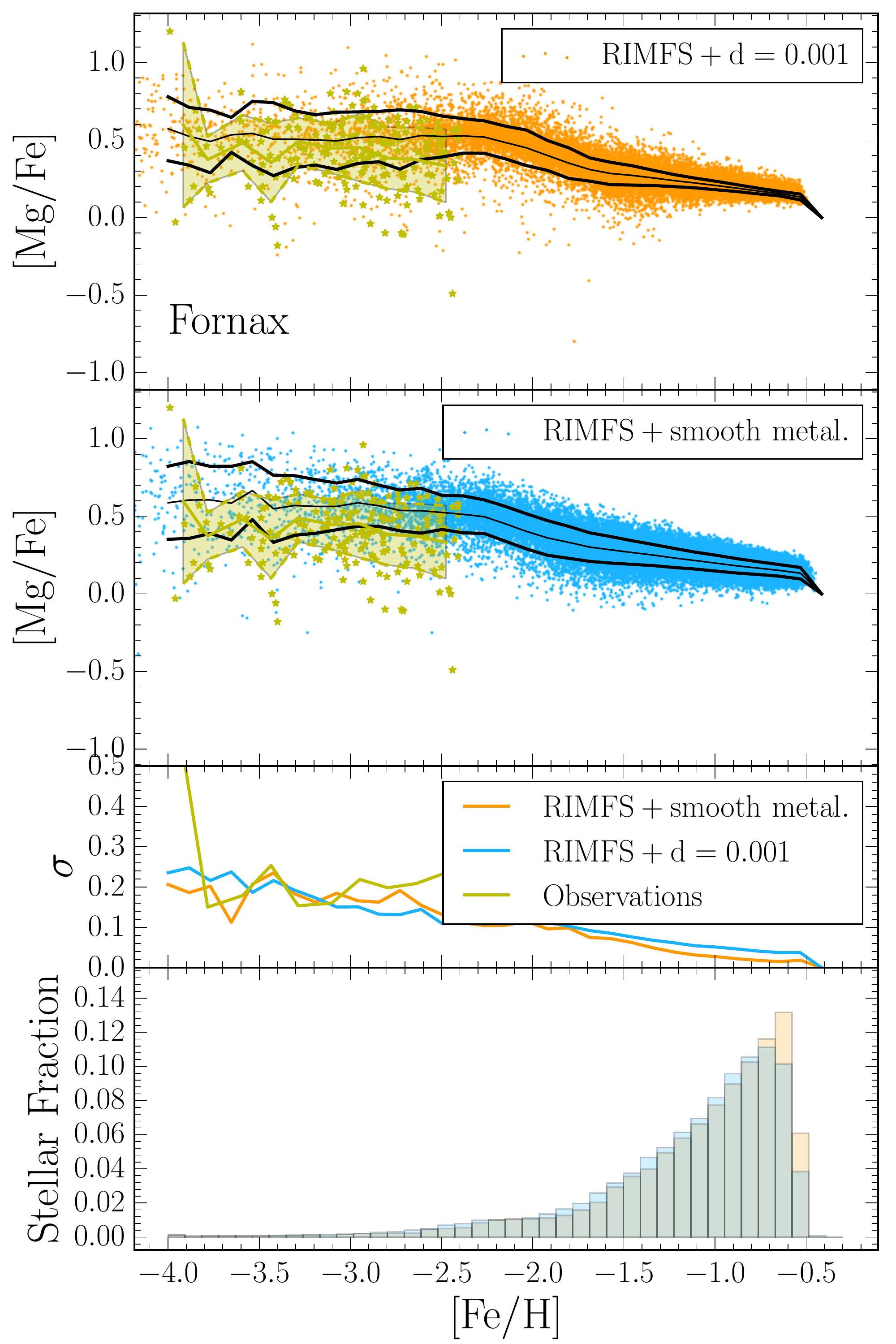}}}
	\subfigure[diffusion : $d=0.0003$ (Fornax)]       {\resizebox{0.247\hsize}{!}{\includegraphics[angle=0]{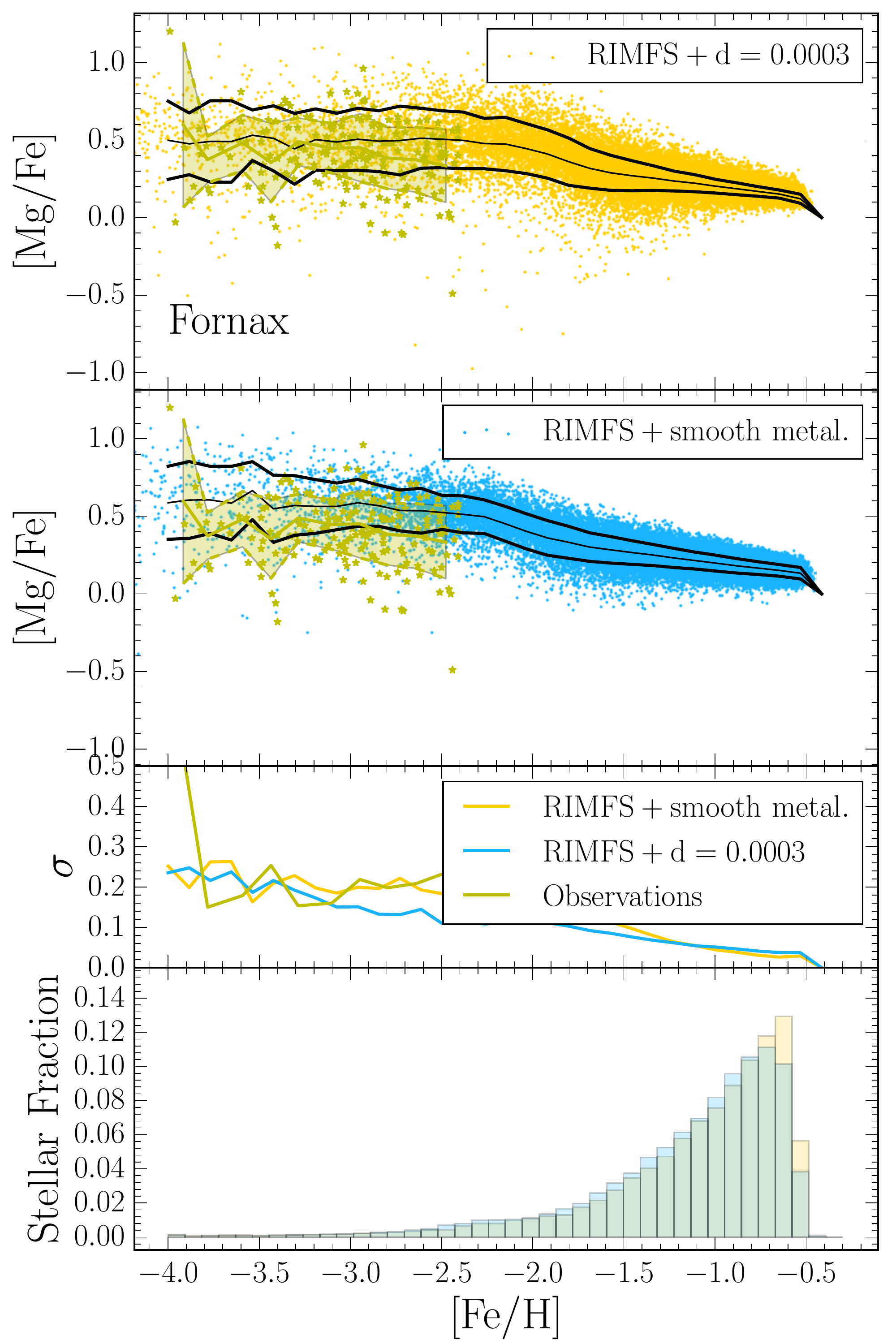}}}  
	\caption{
		Comparison of the final [Mg/Fe] stellar dispersion of the Fornax and Sextans ($\mathfrak{r}=6$) models run with different schemes.	
		As in Figure~\ref{fig:MgFe_RIMFSvsCIMFS_Sex} and \ref{fig:MgFe_RIMFSvsCIMFS_Fnx}, black curves show the 1$\sigma$ dispersion around
		the mean and is also reported in the third section of each panel. The comparison of the metallicity distribution function is shown on the bottom of each panel.				
		For comparison, observations at low metallicity ($[\rm{Mg}/\rm{Fe}]<-2.5$) of individual stars taken from Figure~\ref{fig:EMPS}, are shown in yellow. In the [Mg/Fe] vs [Fe/H] plot,
	    the shaded region corresponds to the 1$\sigma$ dispersion around the mean. Individual measurements are shown with small stars.
		Panels (a) and (e) show the effect of the smooth metallicity scheme on Fornax and Sextans.
		The model run with elements spreading independent of the distance to the source ($w_{ij}=\Gamma_{ij}$) is indicated on panel (b).
		The effect of imposing a constant ejection radius ($R_{\rm{ej}}=0.125\,\rm{kpc}$) or fixing it to the blast radius ($R_{\rm{ej}}=R_{\rm{E}}$) are
		shown on panels (c) and (d) respectively.
		The diffusion effect is dhown on panels (f) to (h). In those latter plots, the dispersion is compared to the fiducial model run with the smooth metallicity scheme,
		corresponding to the top of panel (a).
	}
	\label{fig:allMgFe}
\end{figure*}

Panels (f), (g), and (h) of Figure~\ref{fig:allMgFe} show the impact on the stellar [Mg/Fe] abundances of the
metal diffusion for diffusion coefficients of $d=0.003$, $d=0.001$, and $d=0.0003$  compared to the smooth metallicity scheme represented in light blue.
We also run a simulation with $d=0.0001$, however in this case the effect of the diffusion is very low compared to the fiducial case and, hence, is not discussed here.

It is clear that the diffusion plays a similar role to the smooth metallicity, and for the choice of $50$ neighbours a nice agreement is obtained for a diffusion coefficient of $d=0.001$.
Higher values of the diffusion coefficient are able to reduce the scatter even more, while lower values slightly increase the scatter.
This similarity is also confirmed by the MDF shown at the bottom of each panel.
A similar result is found in the case of the Sextans simulations (Figure~\ref{fig:Diffusion_Sex}).

%%
%\begin{figure*}  
%  \subfigure[$d=0.003$]{\resizebox{0.33\hsize}{!}{\includegraphics[angle=0]{./graphs/MgFe_FiducialvsDiffusion/RIMFS-D1-Fnx.pdf}}}
%  \subfigure[$d=0.001$]{\resizebox{0.33\hsize}{!}{\includegraphics[angle=0]{./graphs/MgFe_FiducialvsDiffusion/RIMFS-D2-Fnx.pdf}}}
%  \subfigure[$d=0.0003$]{\resizebox{0.33\hsize}{!}{\includegraphics[angle=0]{./graphs/MgFe_FiducialvsDiffusion/RIMFS-D3-Fnx.pdf}}}
%  \caption{Effect of the coefficient diffusion on the final [Mg/Fe] dispersion of the Fornax ($\mathfrak{r}=6$) models. Each panel correspond to a different coefficient. 
%  The dispersion is compared to the fiducial model run with the smooth metallicity scheme (top panel).
%  As in Fig.~\ref{fig:MgFe_RIMFSvsCIMFS_Sex}, \ref{fig:MgFe_RIMFSvsCIMFS_Fnx} and Fig.~\ref{fig:allMgFe} (a), the black curves display the $1-\sigma$ dispersion around
%  the mean and is also reported in the third panel. The fourth panel compares the metallicity distribution function.
%  }
%  \label{fig:Diffusion_Fnx}
%\end{figure*}
%%

%

\subsubsection{Discussion}

\begin{figure}  
  \subfigure[Smooth metallicity]       {\resizebox{1\hsize}{!}{\includegraphics[angle=0]{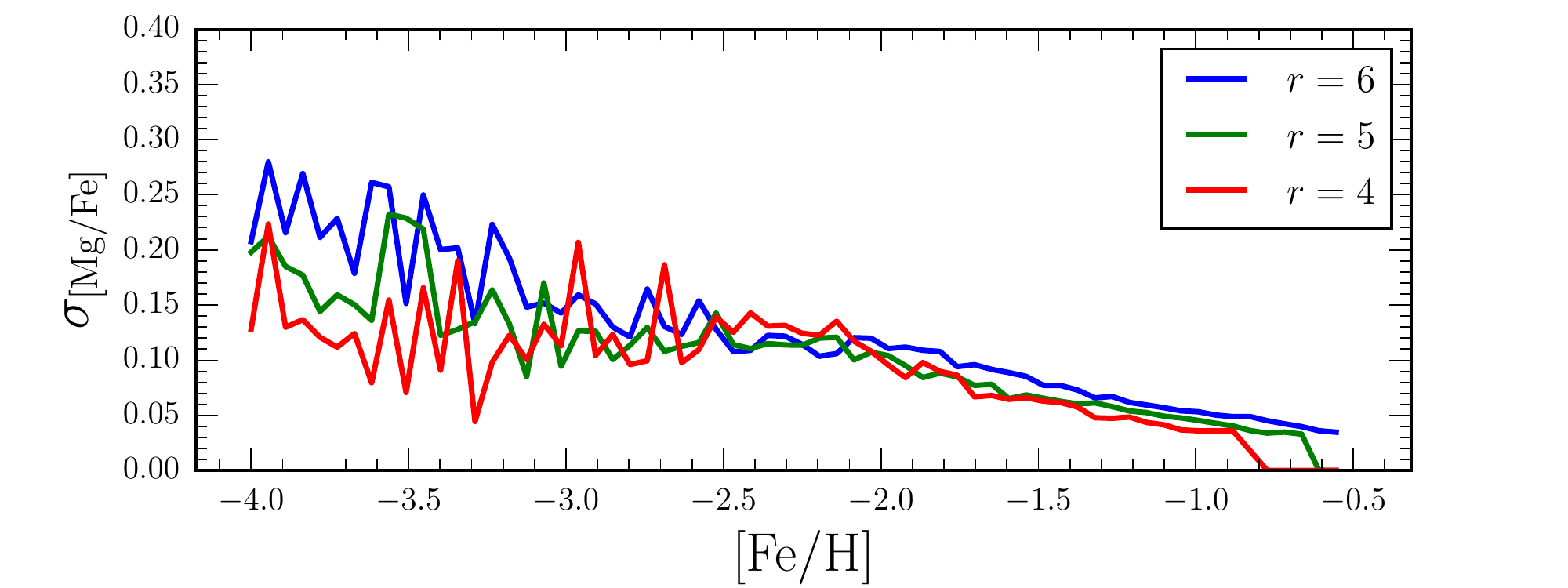}}}
  \subfigure[Diffusion with $d=0.001$] {\resizebox{1\hsize}{!}{\includegraphics[angle=0]{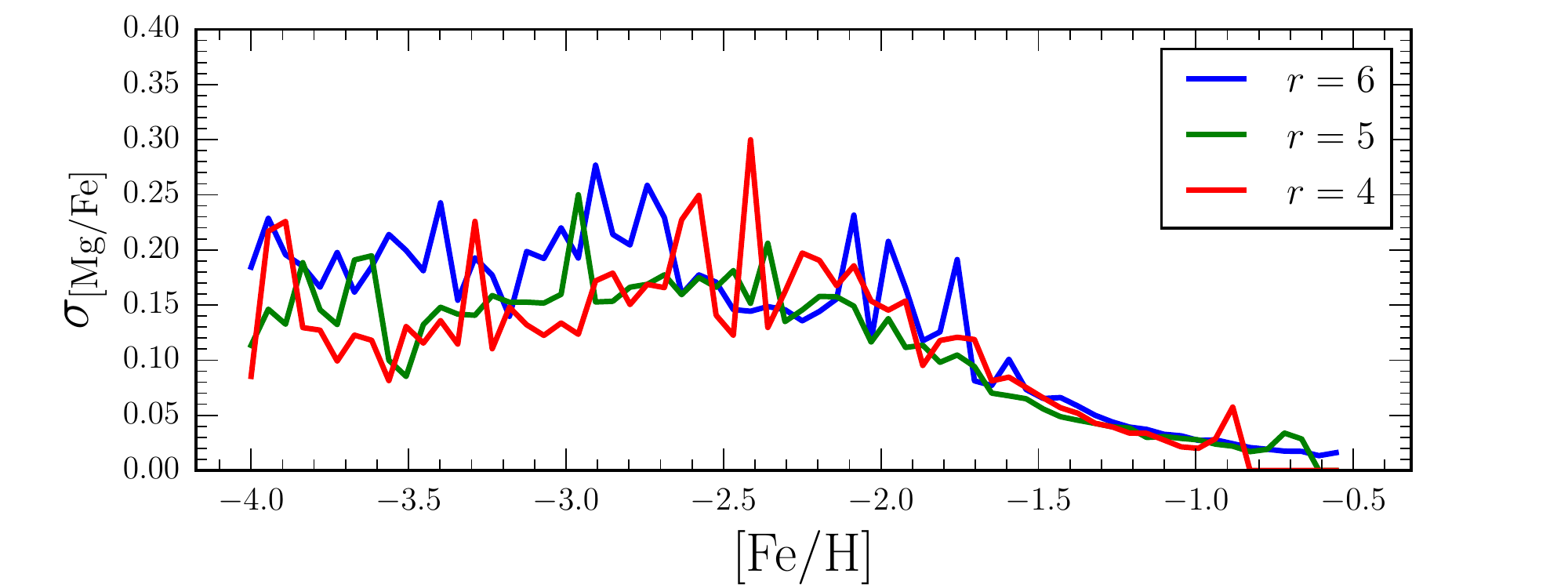}}}
\caption{Dependence of the scatter in abundance ratio [Mg/Fe] as a function of the model resolution. Panel (a) corresponds to the Fornax
model run with the smooth metallicity scheme, while (b) is the same model using the diffusion with a coefficient $d=0.001$.
  }
  \label{fig:Diffusion_Fnx_resolution}
\end{figure}

As both the smooth metallicity and diffusion produce similar results it is difficult to have a clear preference for one method over the other.
The disadvantage of the diffusion approach is the need for an additional parameter.
Its value may be approximately derived with physical arguments, but at the galactic scale numerical errors and 
resolution make this computation obsolete.
This requires the value of this coefficient to be calibrated as presented here.
On the other hand, the smooth metallicity approach appears more natural for an SPH technique, where the information stops
at the scale of the SPH smoothing radius.
%
%An important drawback of the smooth metallicity scheme is its approximate metal mass conservation.
%Indeed, the total mass in particle $i$ of element $X$ is defined through the ratio of the element density $\rho_i^{\rm{X}}$ 
%to the hydrogen density $\rho_i^{\rm{H}}$, both depending on the particles distribution (see Eq.\ref{rhoiX}).
%Thus, even when no supernovae eject new metals, the gas metallicity may vary with time. 
%In addition, in the presence of burst, the total content of metal is usually underestimated compared to the non smooth case.
%This arises because the supernovae explosion pollutes a small fraction of gas particles and the smoothing may take into account 
%particles with a very low metallicity content, biasing the estimation to lower values.
%However, from a direct comparison of our simulations the discrepancy of the total mass of Fe between the smoothed and non-smoothed models,
%is limited to about 5\%.
It is important to point out that both methods are sensitive to the resolution. This is illustrated in Figure~\ref{fig:Diffusion_Fnx_resolution}; when the number of particles is multiplied by 4, the scatter increases by about 0.05 dex at low metallicity.
Finally, it is also important to emphasise that the CPU time required for the two methods is identical 
and may not be used as an argument that one has an advantage over the other.

\section{Simulations of spiral galaxies}\label{spiral_galaxies}

%%%%%%%%%%%%%%%%%%%%%%%%%%%%%%%%%%%%%%%%%%%%%%%%%%%%%%%%%%%%%%%%%%%%%%%%%%%%%%%%%%%%%%%%%%%%%%%%%%%%%%%%%%%

A mixing scheme should work across multiple resolutions and galaxy sizes. As such, we test the effect of the smooth metallicity and metal diffusion in much more massive Milky-Way like spiral galaxies.

\subsection{Initial conditions}

The initial conditions of our models are inspired from the works of \cite{bullock2001} and \cite{kaufmann2006}, which consist of
an isolated halo containing a rotationally supported gas component that gently flows towards the centre and forms a rotating disk.
Initially, both the gas and the dark matter follow an NFW density profile with a concentration 
parameter $c$ equal to 8. The baryonic fraction is set to 10\%.
The total angular momentum of the system is given by
$\lambda=0.12$ according to the notation of \cite{bullock2001}.
Its radial distribution is such that the halo initially starts with a solid body rotation around the z-axis.
Both dark matter and gas are truncated at the virial radius. The total mass is is $M = 9.14 \times 10^{11} \rm{M_{\odot}}$.
We run the models using a single resolution where the stellar mass is 
$10^5 \, \rm{M_{\odot}}$.

We have explored three models. The first model is run without any kind of mixing, 
the second uses the smooth metallicity scheme and the last model uses the metal diffusion scheme with a
diffusion parameter of $d=10^{-3}$, corresponding to the optimal value according to the dSph simulations (Section~\ref{dsph}).
All of the models use the RIMFS scheme, which is, equivalent to the OIMFS and the CIMFS schemes at this resolution (see Section~\ref{effect_of_IMFS}).
The star formation parameter $c_{\star}$ is fixed to $0.7$, the supernova efficiency $\epsilon_{\rm{SN}}$ is $0.08$ and, as in
the dSph simulations, an adiabatic time of $5\,\rm{Myr}$ is used.

% \begin{table}
%         \begin{center}
%         \begin{tabular}{| l | c | }
%                 \hline
%                 \hline 
%                 $t_{\rm{ad,SNII}}$ [Myr] & 5  \\
%                 \hline
%                 $t_{\rm{ad,SNIa}}$ [Myr] & 0  \\
%                 \hline
%                 $\epsilon$ [$\%$] & 30  \\
%                 \hline 
%                 $T_{\rm{SFR,max}}$ [K] & $3\cdot10^4$  \\
%                 \hline
%                 $c_{\star}$ [-] & 1 \\
%                 \hline          
%         \end{tabular}
%         \caption{{\color{red}petit tableau que je sais pas trop ou mettre avec les paramètres des simus MW...} $t_{\rm{ad,SNIa/SNII}}$ represents the adiabatic time of the SNe. $\epsilon$ is the fraction of the total SNe energy converted into thermal energy when the SNe blows up. $T_{\rm{SFR,max}}$ and $c_{\star}$ are parameters related to the stellar formation {\color{red} en fait on s'en fiche pour ce papier !? }}
%         \end{center}
%         \label{tab:mw_param}
% \end{table}

\subsection{Results}

%
%\begin{figure*}	
%	\subfigure[Gas surface density]{\resizebox{0.19\hsize}{!}{\includegraphics{./graphs/Map_Spiral/qSPH_gsd.pdf}}}
%	\subfigure[Gas temperature]{\resizebox{0.19\hsize}{!}{\includegraphics{./graphs/Map_Spiral/qSPH_gt.pdf}}}
%	\subfigure[Gas metallicity]{\resizebox{0.19\hsize}{!}{\includegraphics{./graphs/Map_Spiral/qSPH_gfe.pdf}}}
%	\subfigure[Stars surface density]{\resizebox{0.19\hsize}{!}{\includegraphics{./graphs/Map_Spiral/qSPH_ssd.pdf}}} 
%	\subfigure[Gas and Stars (ege-on)]{\resizebox{0.19\hsize}{!}{\includegraphics{./graphs/Map_Spiral/qSPH_eo.pdf}}} 
%	\caption{
%		Morphological and physical properties of our Milky Way model run with the smooth metallicity scheme, at $t=14\,\rm{Gyr}$. 
%		The metallcitity and temperature maps are obtained in computing the physical quantities ([Fe/H] or T) in the $z=0$ plane, using the SPH technique.
%		The gas and stars surface density maps are simple mass projection after particles being convolved by the SPH kernel.
%	}
%	\label{fig:xMWMaps}
%\end{figure*}
%
\begin{figure*}	
	\subfigure[Gas surface density]{\resizebox{0.19\hsize}{!}{\includegraphics{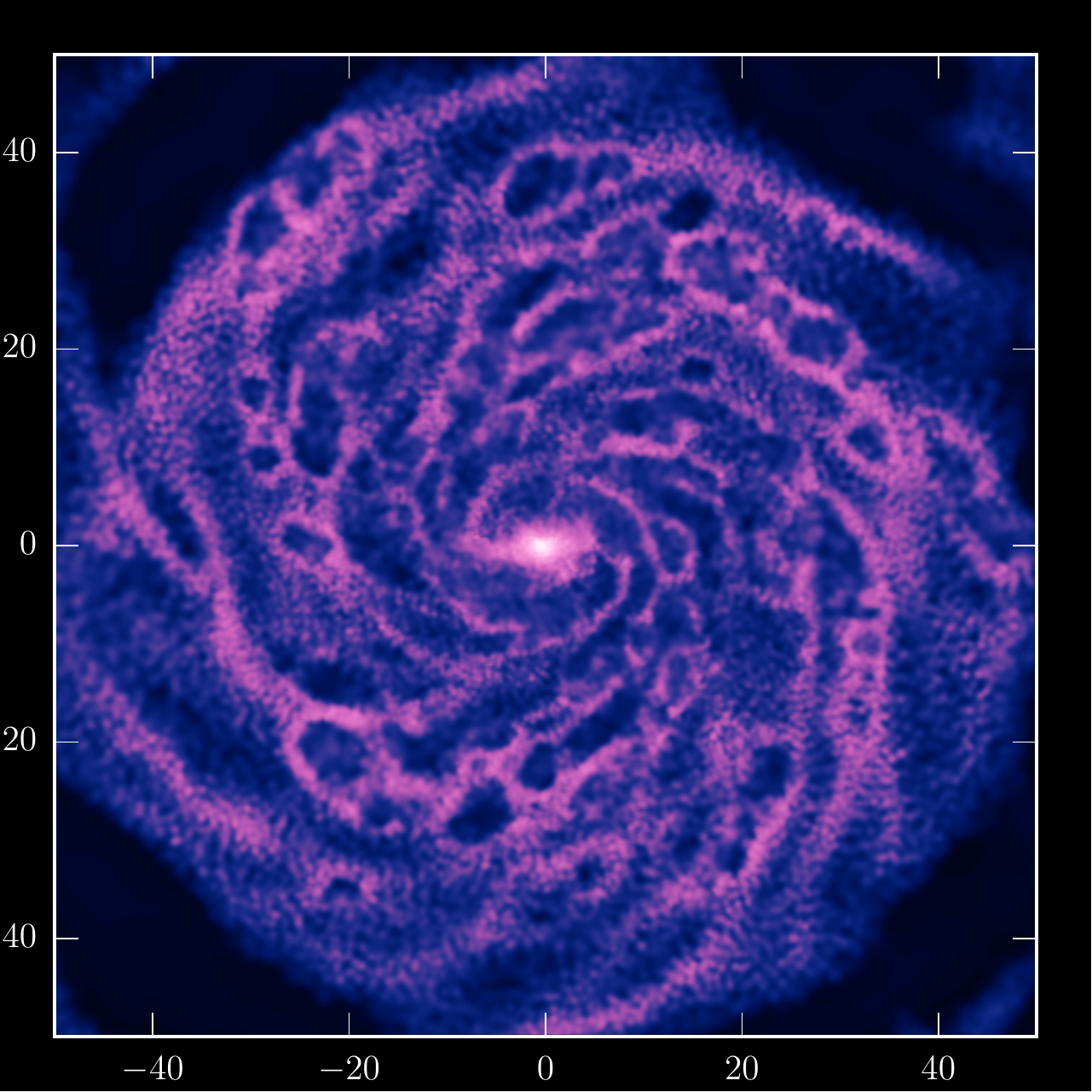}}}
	\subfigure[Gas temperature]{\resizebox{0.19\hsize}{!}{\includegraphics{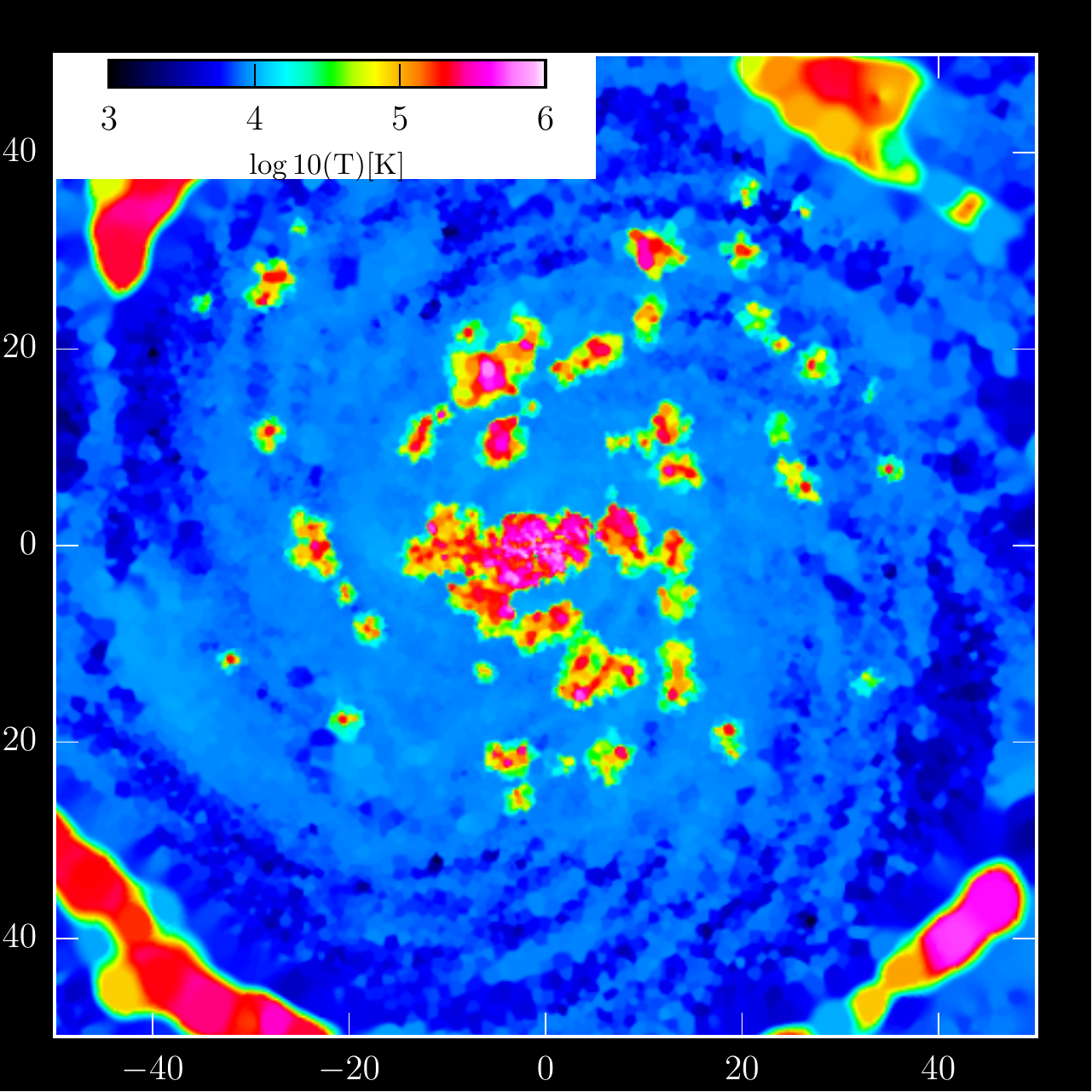}}}
	\subfigure[Gas metallicity]{\resizebox{0.19\hsize}{!}{\includegraphics{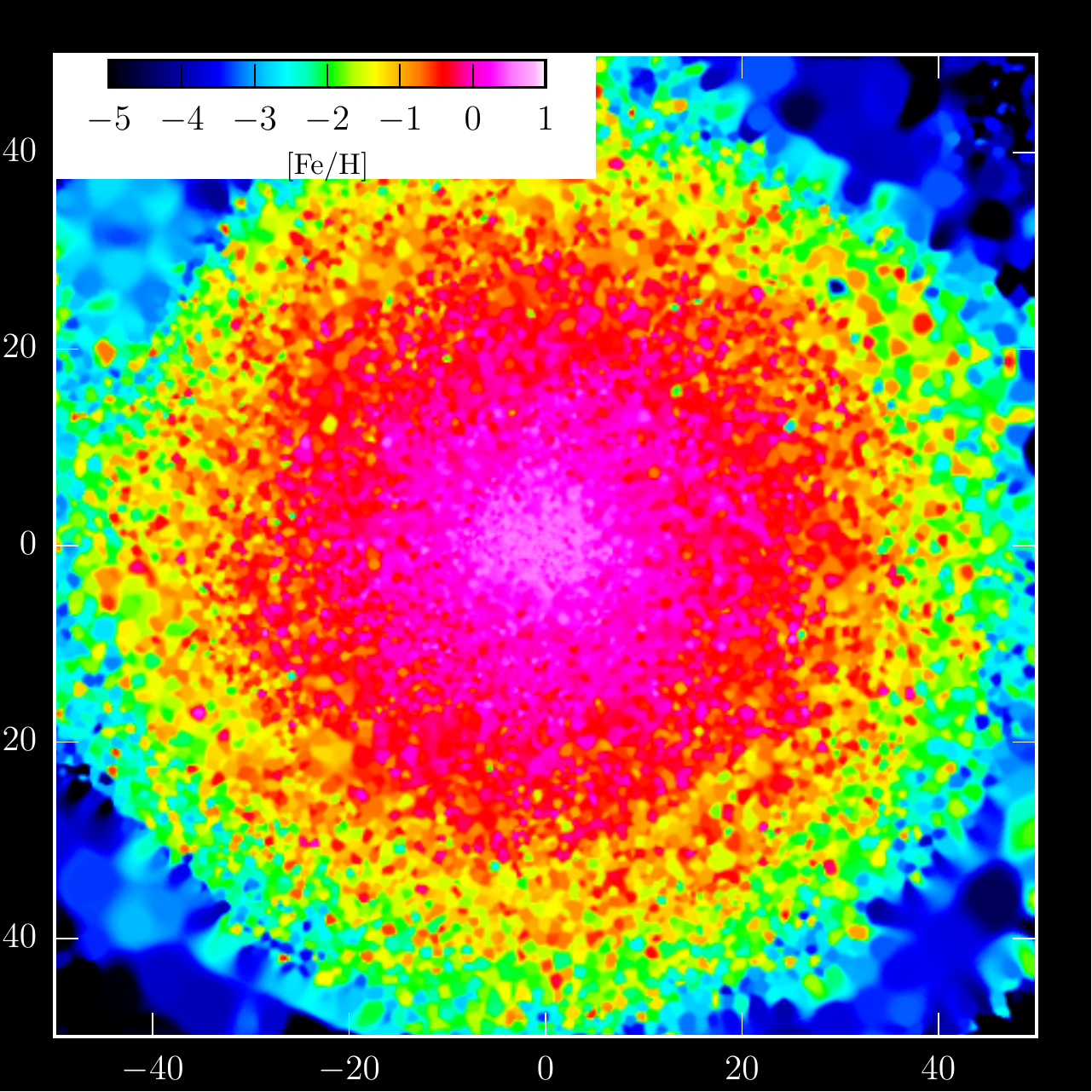}}}
	\subfigure[Stellar surface density]{\resizebox{0.19\hsize}{!}{\includegraphics{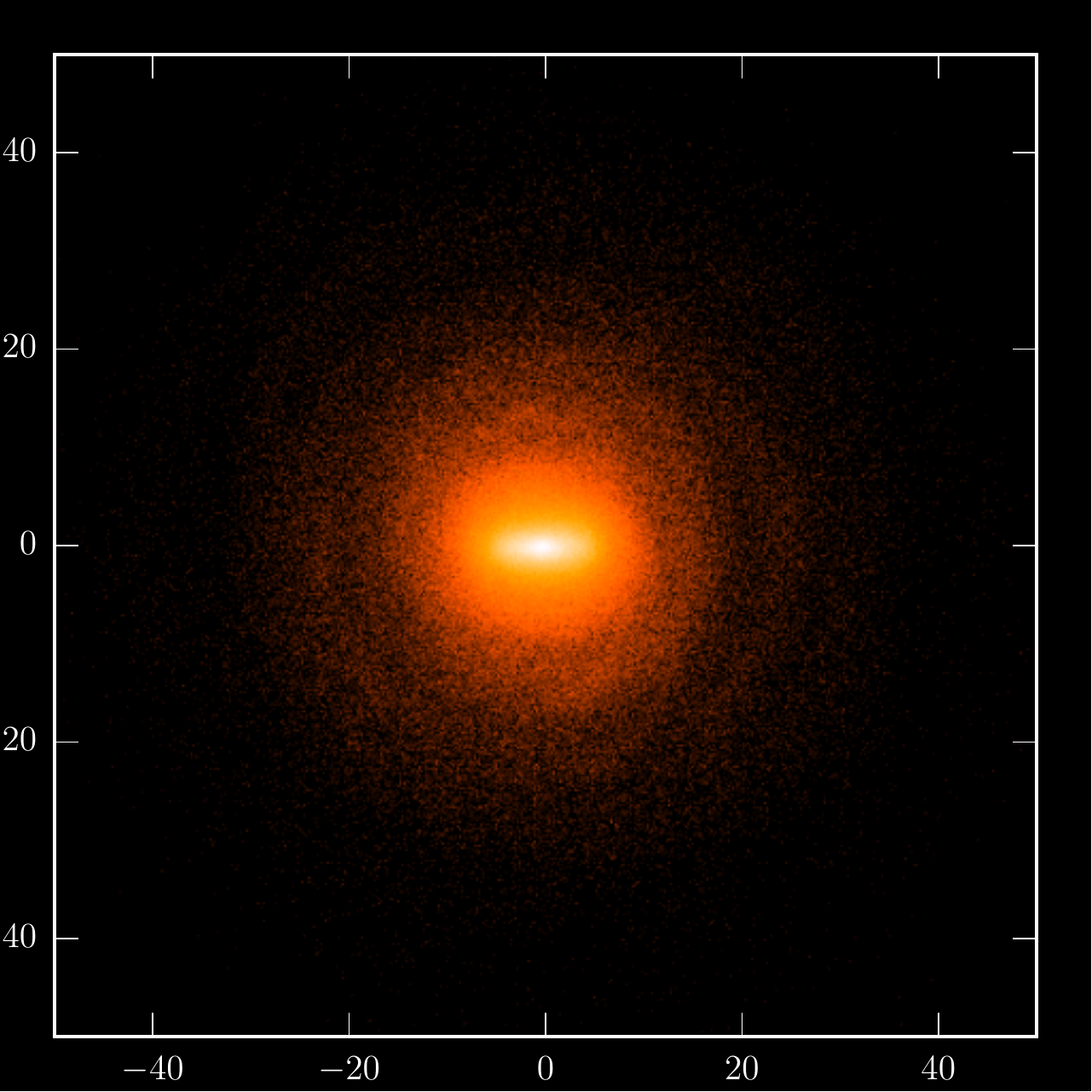}}} 
	\subfigure[Gas and Stars (edge-on)]{\resizebox{0.19\hsize}{!}{\includegraphics{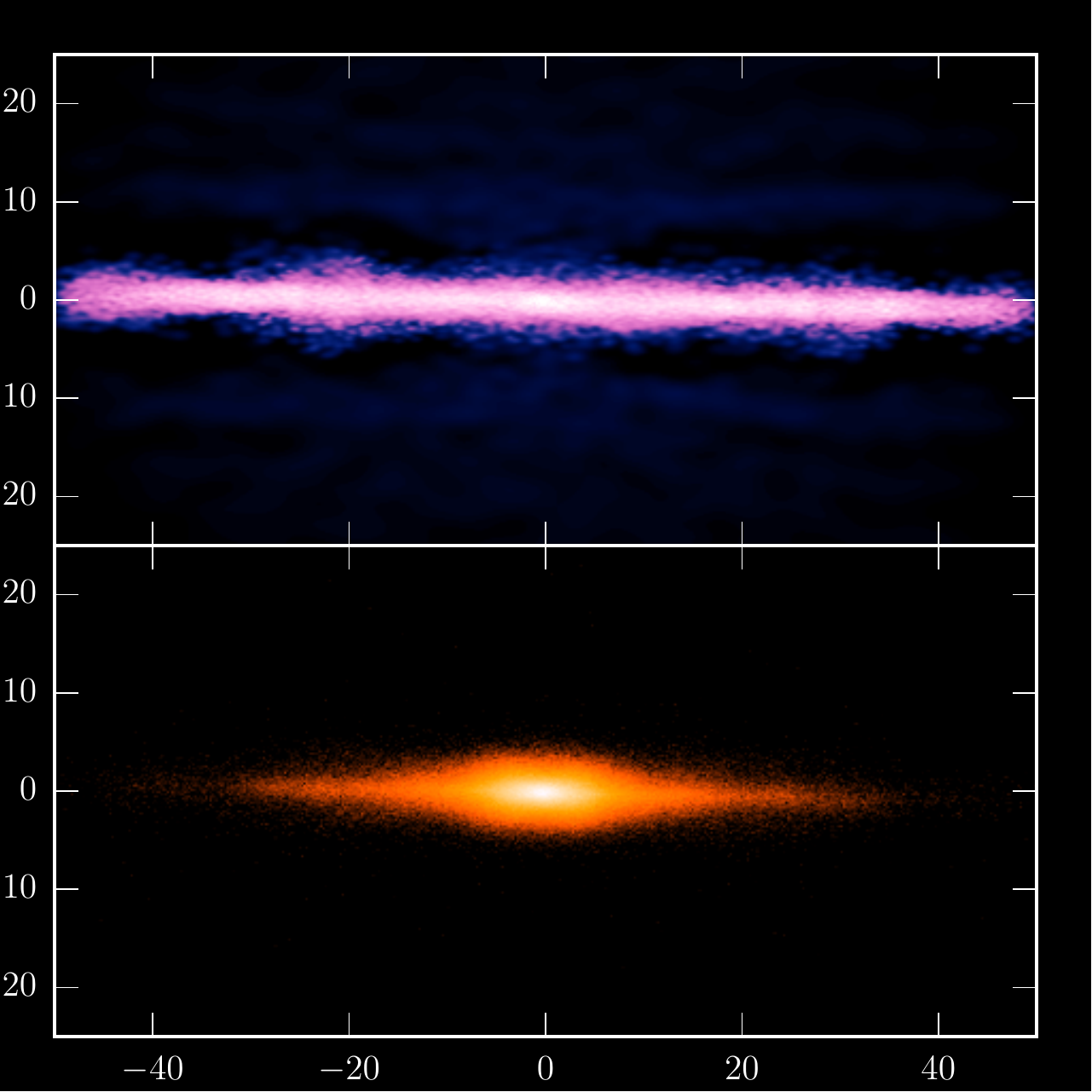}}} 
	\caption{
		Morphological and physical properties of our Milky Way model run with the smooth metallicity scheme, at $t=14\,\rm{Gyr}$. 
		The metallicity and temperature maps are obtained in computing the physical quantities ([Fe/H] or T) in the $z=0$ plane.
		The gas and stars surface density maps are simple mass projections after particles have been convolved with the SPH kernel.
	}
	\label{fig:xMWMaps}
\end{figure*}

All three models (no mixing scheme, dispersion scheme, and smooth metallicity scheme) present very similar morphological and physical features. 

\begin{figure}[h]
	\centering
	\leavevmode   
	\includegraphics[width=9cm]{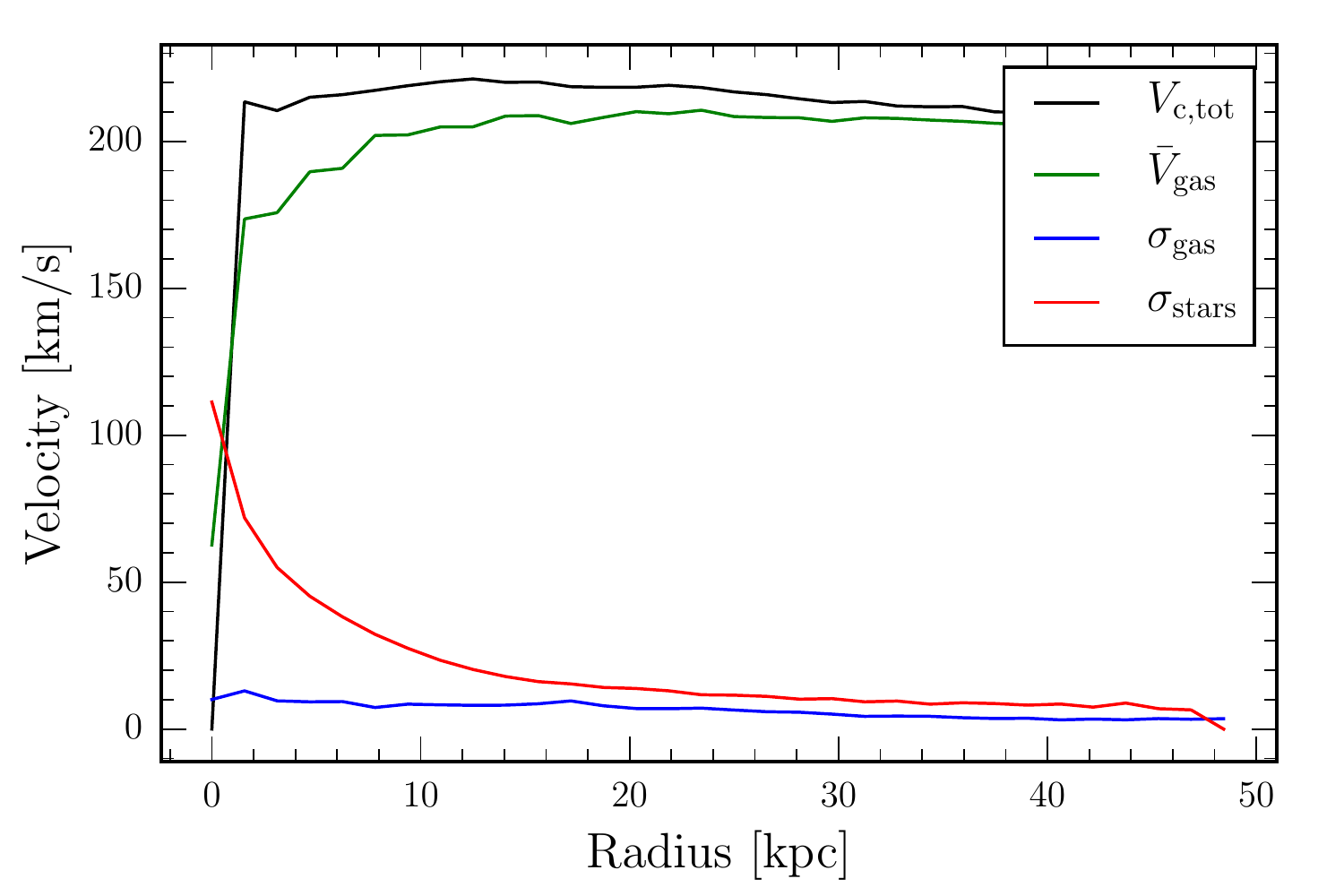}
	\caption{Rotation curve and velocity dispersions of a Milky Way model run with the smooth metallicity scheme. The black line corresponds to the total
		rotation curve while the green line is the mean azimuthal rotation of the gas. The velocity dispersions of the gas and 
		the stars are given by blue and red curves, respectively.}
	\label{fig:RC}
\end{figure}

\begin{figure}[h]
	\centering
	\leavevmode   
	\includegraphics[width=9cm]{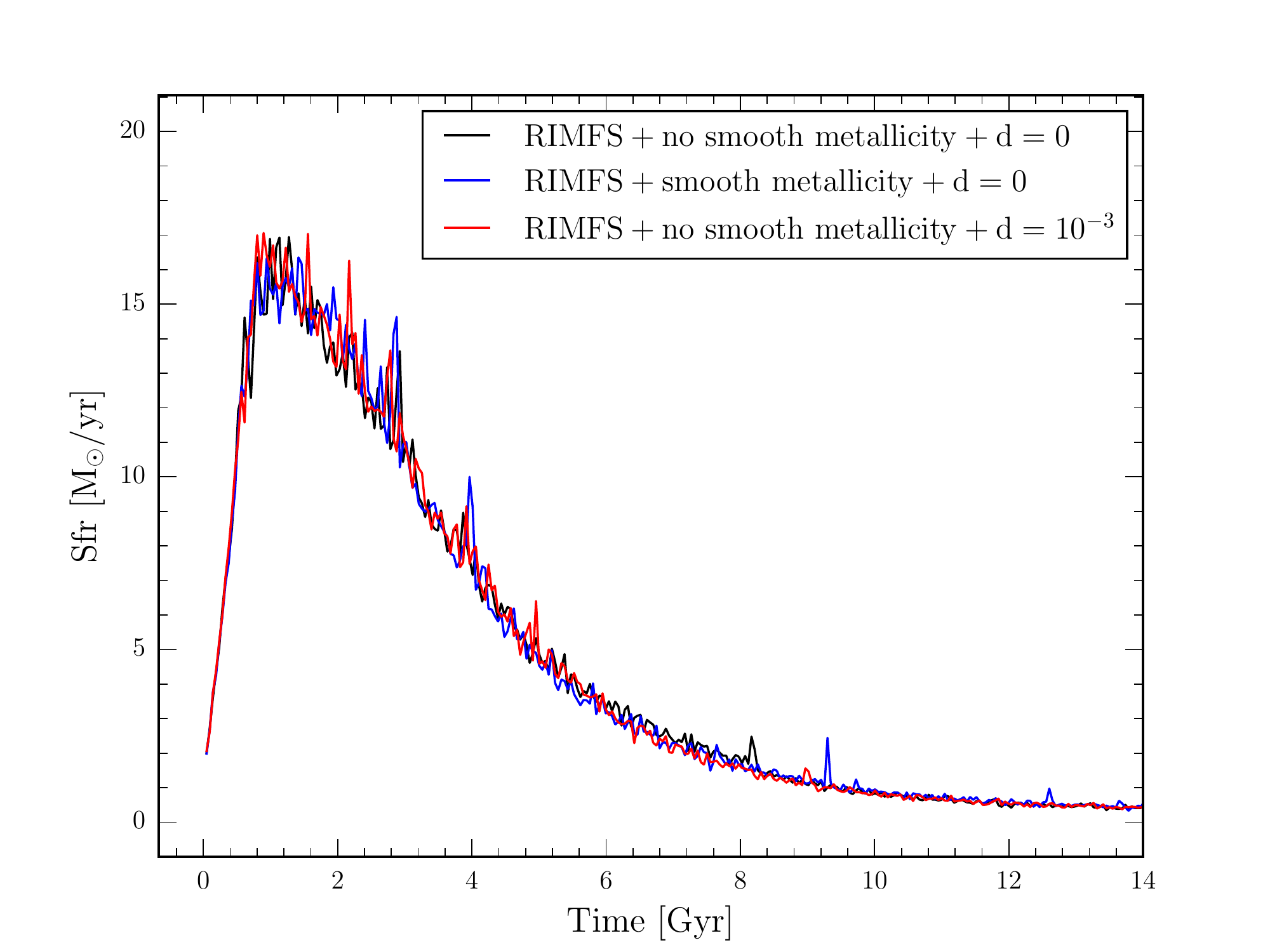}
	\caption{Comparison of the star formation rate of the three Milky Way-like models.
		The black curve corresponds to the model run with no additional mixing scheme; blue,
		with the smooth metallicity switched on; and red, with a diffusion scheme where the
		diffusion coefficient is $d=0.001$.
		All models are run using the RIMFS method.}
	\label{fig:Sfr_Spiral}
\end{figure}

\begin{figure}[h]
  \centering
  \leavevmode   
   \includegraphics[width=9cm]{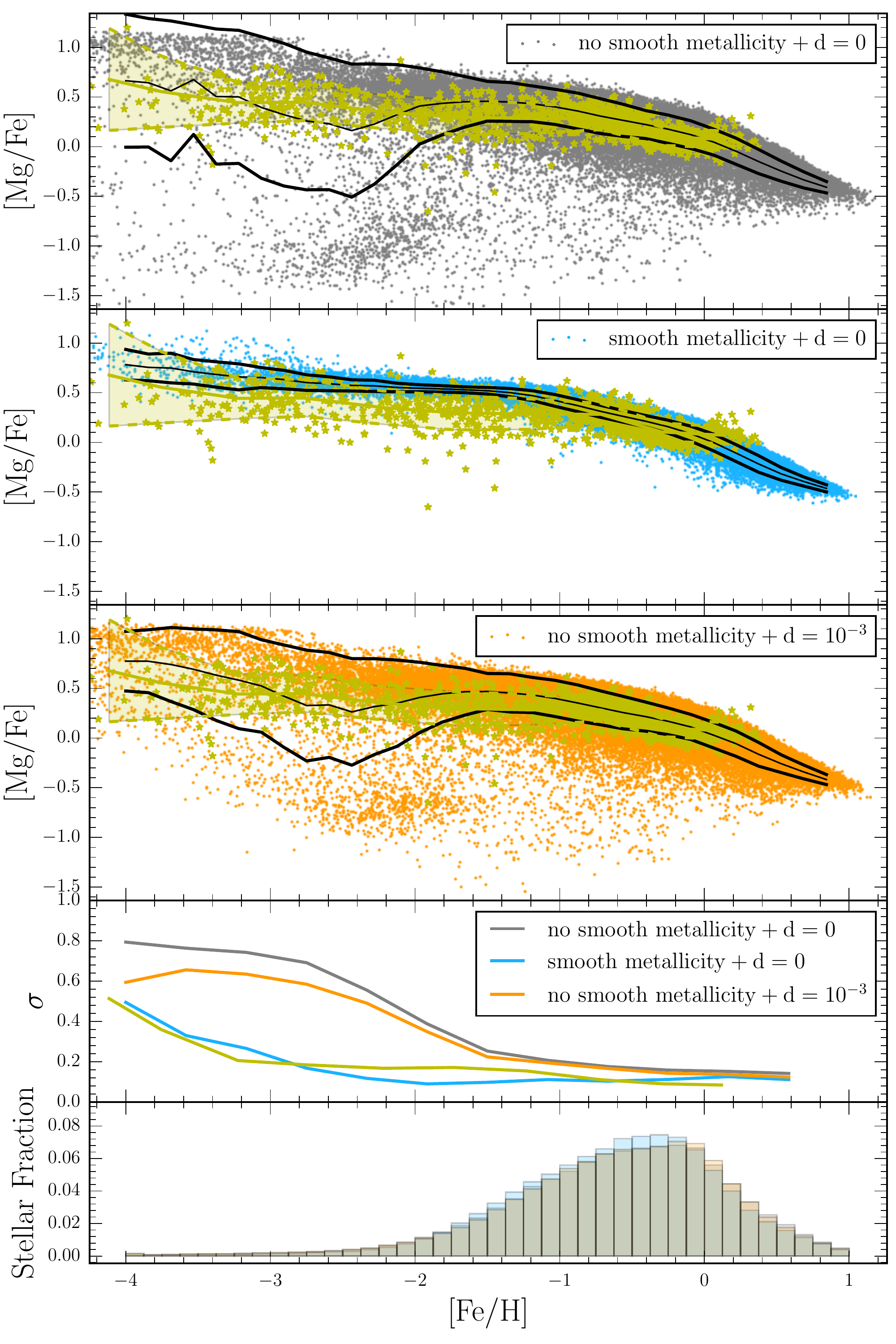}
  \caption{Comparison of the final [Mg/Fe] stellar dispersion of the three Milky Way-like models.
  In the top three panels, the points correspond to the [Mg/Fe] and [Fe/H] values of each stellar particle.
  The black curves show the 1$\sigma$ dispersion around the mean [Mg/Fe]. 
  Those values are also reported in the fourth panel for a direct comparison.
  The bottom panel shows the metallicity distribution function of the three models. 
  In each plot, the yellow points and curves correspond to the observational values of Milky Way stars only, taken from Figure~\ref{fig:EMPS}
  \citep{cayrel2004,gratton03,venn2004,gehren06,reddy06,andrievsky2010,cohen2013}.
  The yellow dashed region corresponds to the observation 1$\sigma$ dispersion.}
  \label{fig:MgFe_Spiral}
\end{figure}

Hence, in Figure~\ref{fig:xMWMaps}, we only show the the smooth metallicity scheme to illustrate these properties, where the galaxy is shown after $14\,\rm{Gyr}$ of evolution.
The first panel (a) shows the extended gaseous disk together with its spiral structure. Its corresponding temperature, computed in
the plane $z=0$ is given in panel (b).
The green regions indicate a dominant quasi-isothermal gas component around $10^4\,\rm{K}$.
The red hot spots ($T>10^5\,\rm{K}$) trace the regions recently affected by supernovae explosions.
The metallicity map for the gas (c) exhibits a radial gradient similar to that observed in spirals.
The [Fe/H] ratio decreases from about 1 at the 
centre, down to -4 at about $50\,\rm{kpc}$. 
Panel (d) shows the exponential stellar disk with a bar clearly present at the centre.
The last panel is an edge-on view of the gaseous disk [top panel of (e)] as well as the stellar component [bottom panel of (e)].
Superheated low density bubbles are visible within the gaseous disk, a consequence of supernova explosions.
In addition to these features, the model presents realistic dynamical properties given in Figure~\ref{fig:RC}. The total rotation curve shown in
black is nearly constant around $200\,\rm{km/s}$.
The mean azimuthal gas velocity (in green) slightly decreases in the central region owing to the larger velocity dispersion (asymmetric drift effect).
The gas velocity dispersion is nearly constant between $10$ and $20\,\rm{km/s}$.
Figure~\ref{fig:Sfr_Spiral} shows the star formation rate of the three models, all of which are similar.
Furthermore, the star formation history is in good agreement with that recently proposed by \citet{snaith14},
which reproduces the chemical abundances of long-lived stars of the Milky Way; the majority of the stars
are formed between 0 and $5\,\rm{Gyr}$.
The resulting $\alpha$-abundances for the three models are shown in Figure~\ref{fig:MgFe_Spiral}.
The first three panels show the [Mg/Fe] vs [Fe/H] distribution of the stars at $t=14\,\rm{Gyr}$.
We selected the stellar particles located 6 to 10 kpc from the centre, in the disk plane,
to compare with observations that take place in the solar neighbourhood.
We compare the distribution with observational values (shown in yellow) obtained from high-resolution spectroscopy of Milky Way stars only, as in Figure~\ref{fig:EMPS}. 
Overplotted are lines corresponding to the mean [Mg/Fe] value, and the 1$\sigma$ deviation.
As can be clearly seen, the three models result in different levels of scatter. The fourth panel directly compares the scatter between
the models and observations.

The base model (run without smooth metallicity or diffusion) as well as the diffusion model are clearly inconsistent with the observations.
At very low metallicity $[\rm{Fe/H}]<-3$, most of the stars are found
with high [Mg/Fe]. In the range $-3 < [\rm{Fe/H}] < -1$, a low [Mg/Fe] tail is present with stars having [Mg/Fe]
as low as $-1.5$. These two features which are not observed are the direct result of a lack of gas mixing.
Gas particles initially located in these external regions are not affected by SNe explosions.
Eventually, such pristine particles may fall towards the galactic disk and be 
impacted as the $\alpha$-rich ejecta of a massive supernova ($M>30\,\rm{M_{\odot}}$).
As the star formation rate is high during the first few Gyr, the polluted gas particles may quickly be transformed into stars.
The opposite process can happen at later time (and thus slightly higher [Fe/H]), when lower mass supernovae explode, 
impacting the gas with low [Mg/Fe].
At low metallicity, the resulting scatter is up to four times the observed scatter of about $0.2\,\rm{dex}$.

When smoothing the abundance on a scale corresponding to the SPH resolution scale (the smooth metallicity scheme), 
all these features are washed out and the scatter becomes very similar to the observed scatter. 
The model still tends to be slightly too metal rich, mainly due to an over-active stellar formation in the last giga-years.

The diffusion scheme, while lowering the scatter compared to the base model, is not sufficient to smooth out these features and results in an exaggerated scatter.

%

%\begin{figure}[ht]
%  	\centering
%  	\leavevmode
%	\includegraphics[width=9cm]{images/MgFe-smooth-MW.png}
%	\caption{MgFe ratio for our MW simulations. The red star symbols represent the observational values from \cite{gratton03,gehren06,reddy06,andrievsky2010}. We selected the stellar particles in the simulation at the solar neighbourhood, i.e. closer than 1kpc from the disk plane and between 6 and 10 kpc from its center. The tuning of the parameters is indicated in the text and in Tab.\ref{tab:mw_param}. {\color{red} ou bien est-ce que j'en reparle ici? Une fois qu'on est d'accord sur quelle figure présenter je la refait en suivant les codes couleurs des plots d'alexis}}
%	\label{fig:mgfe_mw}
%\end{figure}

%As for the dSph models, the final metallicity function displayed at the bottom of Fig.~\ref{fig:MgFe_Spiral}} is only weakly influenced by the different scheme used.
%It is worse noting that in all cases, the models tend to be slightly too metallic. This
%is mainly due to an overactive stellar formation in the last Gyr.

%%%%%%%%%%%%%%%%%%%%%%%%%%%%%%%%%%%%%%%%%%%%%%%%%%%%%%%%%%%%%%%%%%%%%%%%%%%%%%%%%%%%%%%%%%%%%%%%%%%%%%%%%%%

\section{Conclusions}

Using self-consistent, N-body, chemo-dynamical models of dSphs and Milky Way-like galaxies, 
we have investigated the impact of different numerical schemes
on the star formation history of galaxies and the chemical properties of stars therein;
these schemes are the 
IMF sampling schemes (Section~\ref{imf}), elemental spreading schemes (Section~\ref{feedback}), and metal mixing schemes (Section~\ref{diffusion}).
We focused in particular on the scatter in [$\alpha$/Fe] observed in metal-poor stars both in dSphs and Milky Way halo.

As feedback processes (both element and energy injection) are at the heart of this work, 
we ensured the reliability of our code \texttt{GEAR} in a simple case, where one supernova
explodes in an homogeneous medium. 
We nicely reproduce the Sedov-Taylor solution at very high resolution
when the individual and adaptive timestep scheme improvements of \citet{durier2012} are used.
Here we found that the artificial viscosity formulation of \cite{monaghan83} produces superior results to that of \citet{monaghan1997}
(the default \texttt{Gadget-2} viscosity formulation).
Additional simple experiments involving two supernovae led us to understand the effect of different metal spreading schemes,
that are essential to the understanding of pollution in complex galactic systems.

\subsection{The IMF sampling schemes}

We have tested the influence of different IMF sampling methods: a continuous scheme (CIMFS), an ``optimal scheme" (OIMFS) and a
random scheme (RIMFS). 

The most important result is that all schemes impose, at different levels, a limit to the
stellar mass resolution of the simulations. 
From theoretical considerations, we have found that:
\begin{itemize}

\item As the CIMFS continuously ejects metals and energy, below a stellar mass resolution of about $10^5\,\rm{M_\odot}$ the products of SNeIa are diluted
  over unrealistic long periods of time, generating an artificial mixing. When coupled with the adiabatic time method \citep{stinson2006}, 
the cooling is strongly underestimated, resulting in a severe quenching of star formation.
Removing the adiabatic time within the CIMFS scheme prevents this anomalous quenching, but the dilution problem remains.

\item The OIMFS is designed to reduce the noise in the IMF compared to the random sampling. 
Unfortunately, it does not allow the use of a low particle mass (and subsequently high resolution). Below about $10^4\,\rm{M_\odot}$
the maximal stellar mass is limited to about $30\,\rm{M_\odot}$. Ignoring the yields of the most massive stars induces a bias
that results in an offset in the $\alpha$-element abundance plateau at low metallicity. 

\item Because of its stochastic nature, the RIMFS induces noise in the IMF when 
the stellar mass resolution reaches below about $10^4\,\rm{M_\odot}$.
Assuming that the IMF is distributed among neighbouring particles,
the stellar mass resolution may be decreased down to about $10^3\,\rm{M_\odot}$.
The IMF may no longer be considered complete for lower masses, with pockets of enriched gas showing unrealistically high scatter.

\end{itemize}

In practice, based on simulations of Fornax and Sextans dSph, we have found 
that for a low resolution (where the mass of stellar particles is greater or equal to  $4096\,\rm{M_\odot}$),
the number of stars inside a single stellar particle is large enough to allow for a good sampling of the IMF regardless of the scheme used.
This occurs with all three schemes, which results in similar star formations and abundance patterns.
However, the different schemes severely diverge in our tests performed at even higher resolution.
At a resolution of $1024\,\rm{M_\odot}$, the RIMFS is the most reliable sampling method. At higher resolution 
($128\,\rm{M_\odot}$), the RIMFS shows a lack of convergence owing to incomplete IMF sampling.
This incomplete sampling fails to inject enough energy across the galaxy to counterbalance the cooling and effectively regulate star formation.

Consequently, at very high resolution ($<1024\,\rm{M_\odot}$), no method is reliable. 
This is a severe limitation for the forthcoming simulations of galaxies that approach or exceed this resolution.
Indeed, it is now well established that supernova feedback plays a major role during the process of galaxy formation. 
However, the precise supernova rate is directly dependent on the IMF sampling.
This is also true for gas cooling which directly depends on its metal enrichment.
Accurately reproducing the feedback and consequently star formation and chemical enrichment at these low masses will require 
the development of novel methods to sample the IMF.

\subsection{The scatter in abundance ratios}

Focusing on the RIMFS scheme at the maximal reliable resolution of $1024\,\rm{M_\odot}$, 
we have demonstrated that the scatter in stellar abundance ratios is about three times larger
than that observed in the metal-poor stars of dSph galaxies. 
This scatter is due to the stochastic explosions of supernovae imprinting small pockets of gas with markedly different and sometimes extreme abundances.
In a first attempt to reduce this scatter we have tested the impact of the scheme used to spread metals into the ISM. We found that

\begin{itemize}
	
\item no statistically distinguishable differences exist when combining a classical SPH kernel
with either a volume or particle mass-weighting;

\item replacing the classical SPH kernel with a step function (gas particles receive ejecta independent of their distance 
to the exploding supernova) does not reduce the scatter, which is still dominated by stochastic explosions of supernovae; and

\item using a constant radius (either fixed or equal to the blast radius) greatly worsens the problem.
\end{itemize}

In a second attempt, we tested the impact of introducing a mixing scheme.
We implemented both the smooth metallicity technique \citep{wiersma2009}  and metal diffusion \citep{greif2009}.
Both are able to reduce the scatter to a realistic value inside dwarf spheroidal galaxies. 
The smooth metallicity scheme has the advantage that it is independent of any additional parameters and very natural with regard to the SPH method.
Metal diffusion needs the introduction of a diffusion coefficient, a free parameter set that reproduces the observations ($d=10^{-3}$).
%Both methods are however resolution dependent.

As a final test, we simulated a Milky Way-like galaxy that agrees with the star formation history deduced by \citet{snaith14}.
The abundance of $\alpha$-elements is only reproduced when using the smooth metallicity technique.
The diffusion scheme with the same parameters calibrated in the context of dSphs is not able to adequately reduce the number of stars with extreme abundances.

We have shown that the current best practices in chemo-dynamical simulations delivers reliable prediction concerning the chemical properties of galaxies 
in restricted contexts only, where the stellar mass resolution is limited to value below (masses above)  $10^3\,\rm{M_\odot}$.
We found the smooth metallicity scheme, combined with a random initial mass sampling scheme (RIMFS) to be the best combination to reproduce the dispersion of abundances.
Increasing the mass resolution in future N-body chemo-dynamical modelling while avoiding any bias will require a drastic redesign of the classical star formation recipes presently used.

\appendix

\section{Diffusion}
\begin{figure*}  
  \subfigure[$d=0.003$]{\resizebox{0.33\hsize}{!}{\includegraphics[angle=0]{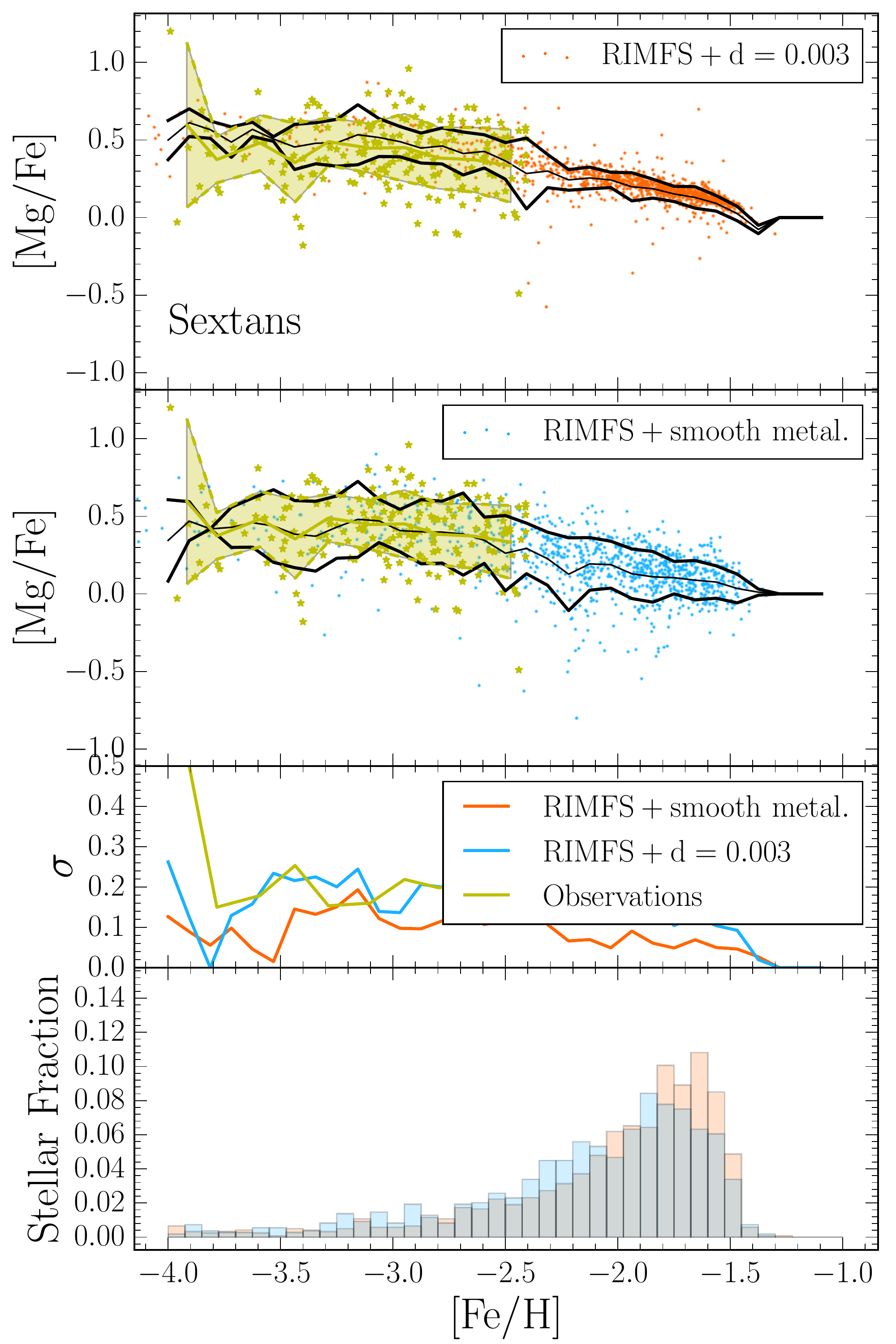}}}
  \subfigure[$d=0.001$]{\resizebox{0.33\hsize}{!}{\includegraphics[angle=0]{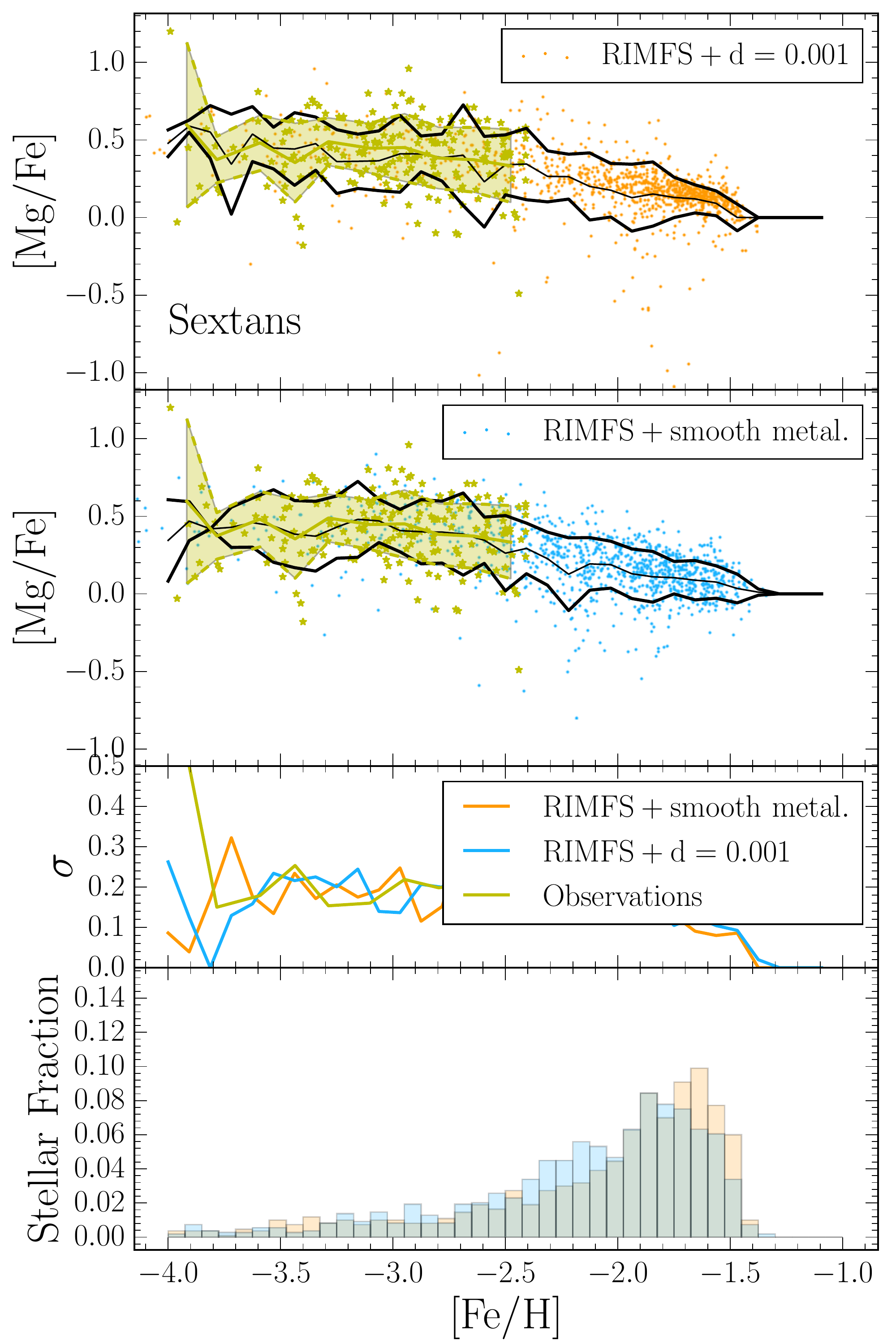}}}
  \subfigure[$d=0.0003$]{\resizebox{0.33\hsize}{!}{\includegraphics[angle=0]{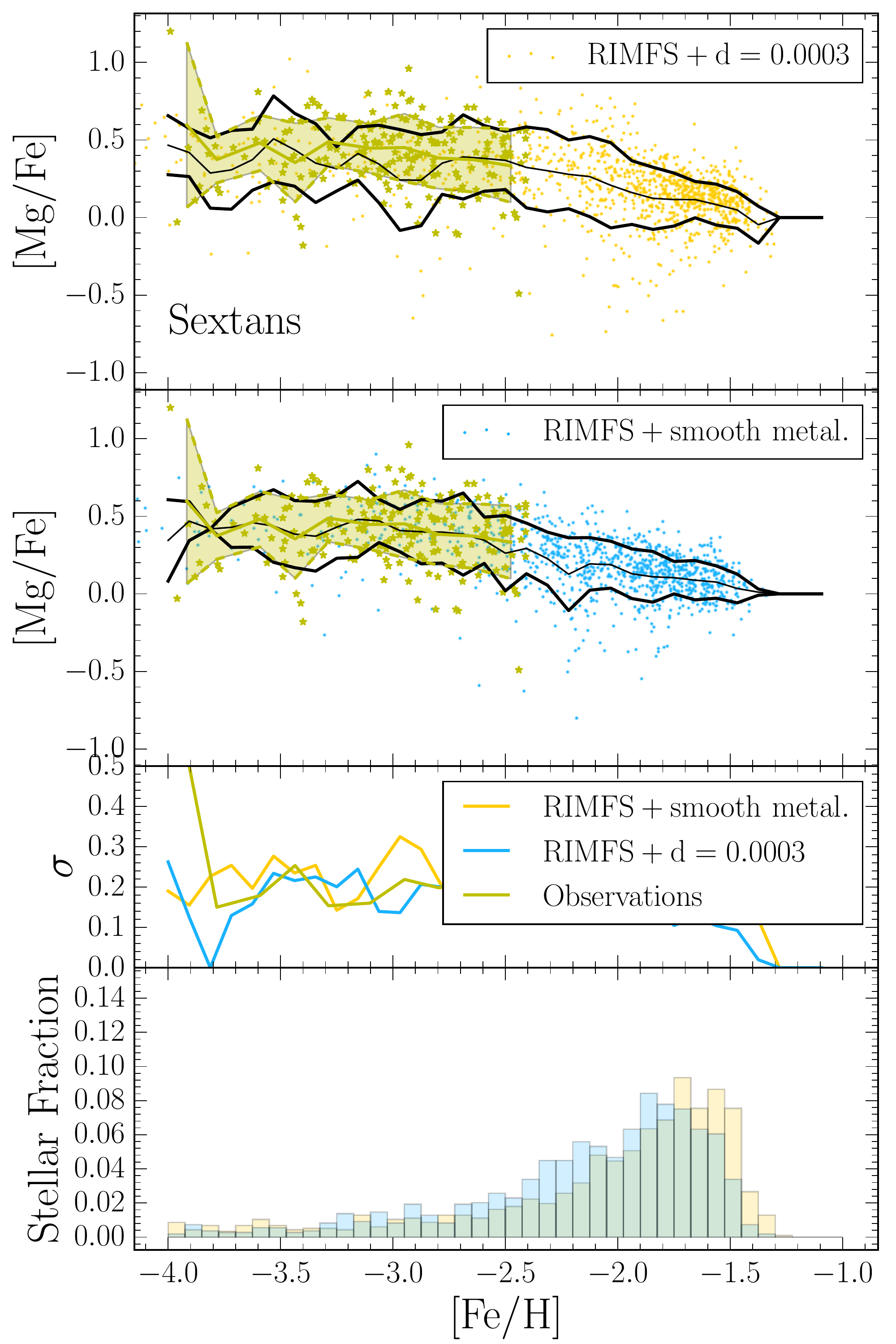}}}
  \caption{
  Effect of the coefficient diffusion on the final [Mg/Fe] dispersion of the Sextans ($\mathfrak{r}=6$) models. Each panel corresponds to a different coefficient. 
  The dispersion is compared to the fiducial model run with the smooth metallicity scheme.
  As in Figures~\ref{fig:MgFe_RIMFSvsCIMFS_Sex} and \ref{fig:MgFe_RIMFSvsCIMFS_Fnx}, the black curves show the 1$\sigma$ dispersion.
  For comparison, observations at low metallicity ($[\rm{Mg}/\rm{Fe}]<-2.5$) of individual stars taken from Figure~\ref{fig:EMPS} are shown in yellow. In each [Mg/Fe] vs [Fe/H] plots,
  the shaded region corresponds to the 1$\sigma$ dispersion around the mean. Individual measurements are shown with small stars.
  }
  \label{fig:Diffusion_Sex}
\end{figure*}

\begin{acknowledgements}

We thank the anonymous referee for his/her useful remarks which strongly improved the paper.
We also thank Fabrice Durier for providing his adaptive timestep routine
and Pierre North for precious suggestions and comments.
We are grateful to the International Space Science Institute (ISSI) at Bern
for their funding of the team ``The first stars in dwarf galaxies'', as well
as SCITAS, the scientific IT and application support of the Swiss Federal Institute of Technology in Lausanne (EPFL).
As for the simulations, data reduction and galaxy maps have been
performed using the parallelised Python \texttt{pNbody} package
(\texttt{http://lastro.epfl.ch/projects/pNbody/}).  This work was supported
by the Swiss National Science Foundation.

\end{acknowledgements}

\bibliographystyle{aa}
\bibliography{bibliography}
\nocite{*}

\end{document}